\documentclass{aa}
\usepackage[varg]{txfonts}
\usepackage{graphicx}
\usepackage{booktabs}
\graphicspath{{img/}}
\nocite{*}
\bibpunct{(}{)}{;}{a}{}{,}

\begin{document}
	
	\title{A new panchromatic classification of unclassified Burst Alert Telescope active galactic nuclei}
	\titlerunning{Classifying the unclassified BAT AGNs}
	\authorrunning{Giuliani et al.}
	\author{L. Giuliani\inst{1}\thanks{\email{luca.giuliani.mail@gmail.com}}
		\and G. Ghisellini\inst{2}
		\and T. Sbarrato\inst{1,2}
	}
	\institute{Dipartimento di Fisica G. Occhialini, Univ. Milano--Bicocca, P.za della Scienza 3, I--20126 Milano, Italy   
		\and INAF -- Osservatorio Astronomico di Brera, Via Bianchi 46, I--23807 Merate, Italy \\
	}
	
	\date{Received <date> /
		Accepted <date>}
	
	\abstract{
		We collect data at all frequencies for the new sources classified as unknown active galactic nuclei (AGNs) in the latest Burst Alert Telescope (BAT) all-sky hard X-ray catalog.
		Focusing on the 36 sources with measured redshift, we compute their spectral energy distribution (SED) from radio to $\gamma$-rays with the aim to classify these objects.
		
		We apply emission models that attempt to reproduce the obtained SEDs, including: i) a standard thin accretion disk together with an obscuring torus and a X-ray corona;  ii) a two temperature thick advection-dominated flow; iii) an obscured AGN model, accounting for absorption along the line of sight at kiloelectronvolt energies and in the optical band; and iv) a phenomenological model to describe the jet emission in blazar-like objects.
		We integrate the models with the SWIRE template libraries to account for the emission of the host galaxy.
		
		For every source we found a good agreement between data and our model.
		Considering that the sources were selected in the hard X-ray band, which is rather unaffected by absorption, we expected and found a large fraction of absorbed radio-quiet AGNs (31 out of 36) and some additional rare radio-loud sources (5 out of 36), since the jet emission in hard X-rays is important for aligned jets owing to the boost produced by the beaming effect.
		
		With our work we can confirm the hypothesis that a number of galaxies, whose optical spectra lack AGN emission features, host an obscured active nucleus.
		The approach we used proved to be efficient in rapidly identifying objects, which commonly used methods were not able to classify.}
	
	\maketitle      
	
	\section{Introduction}
	
	Active galactic nuclei (AGNs) are point-like sources, found in the nuclei of galaxies, which emit a very large luminosity (10$^{46}$-10$^{48}$ erg s$^{-1}$).
	According to the unified model \citep{uni_antonucci,uni_urry}, these sources are composed by an accreting supermassive black hole (10$^{6}$-10$^{10}$ M$_{\odot}$) that is surrounded by ionized plasma and dusty regions in the center of a host galaxy.
	About $10\%$ of AGNs, besides accreting matter, launch matter and radiation in two oppositely directed relativistic jets, whose emission is highly beamed and anisotropic.
	Sources with the jet axis pointing directly toward us are called blazars.
	
	The number density of AGNs subject to absorption with a hydrogen column density along the line-of-sight $N_H>10^{24}$ cm$^{-2}$ (Compton Thick AGNs) is a key parameter to understand the accretion history of the Universe and the cosmic X-ray background (CXB) \citep{agn_xrb}.
	Many observations imply their presence in large quantity in the local Universe \citep[e.g.,][]{obs_local}.
	Much of the peak intensity of the CXB at 30 keV is thought to be produced by these objects \citep{ueda_2014}.
	Hard X-ray surveys at energies above 10 keV provide an ideal opportunity to select this population of AGNs.
	The Swift satellite, launched in 2004, is carrying out an all-sky survey at 14-195 keV with its Burst Alert Telescope (BAT) instrument \citep{bat_art}.
	The last official data release, the 105 month Swift-BAT all-sky hard X-ray survey \citep{bat_survey}, was published in early 2018 and contains a total of 1632 sources.
	In this new catalog, there are a number of sources that are identified as AGN but of unknown type, that is, sources associated with an extended galaxy in the optical image whose optical spectra and type classification are not known or lack firm evidence of AGN optical emission diagnostics.
	Under this classification there are 73 new objects not found in the 70-month catalog \citep{bat_70}, of which 36 have redshift measurements.
	We aim to characterize these new sources.
	
	\section{Procedure}
	
	Considering that the sources are selected in the hard X-ray band, which is rather unaffected by absorption, we expected and found a relatively large fraction of absorbed radio-quiet AGNs and some more rare radio-loud sources. In fact, the jet emission in hard X-rays becomes important for aligned jets as a result of the boost produced by the beaming effect.
	
	In the literature, it is common to characterize heavily obscured AGNs by the properties of their soft-X-ray spectra  \citep[e.g.,][]{suzaku_obs} or their infrared to ultraviolet emission \citep[e.g.,][]{obs_ir}.
	Furthermore, there are a great number of studies that aim to classify Swift/BAT sources through optical spectroscopy from archival data or with new observation campaigns \cite[e.g.,][]{class:parisi,class:rojas}.
	The process of studying high-energy sources through the analysis of optical data is also used to recognize gamma-ray blazar candidates within the sources associated in the Fermi catalog \citep{class:gamma}.
	Also, there are precedents in the literature that attempt to identify possible blazars quantitatively by comparing a set of spectral energy distribution (SED) templates, representative of the various blazar classes to the observed multiwavelength flux, for example, the work of \cite{class:paiano} on Fermi $\gamma$-ray sources.
	
	In our study, instead, we chose to characterize unclassified BAT AGNs from the comparison of predicted emission models with the SED of the sources; this comparison is constructed using archival data at all frequencies.
	This is necessary because our source selection itself implies that the optical data are either unavailable or show no signs of the typical AGN emission.
	
	We verified the optical counterpart redshift in the literature for all the sources using the online archives NASA/IPAC Extragalactic Database\footnote{https://ned.ipac.caltech.edu/} (NED) and SIMBAD\footnote{http://simbad.u-strasbg.fr/simbad/}.
	To collect the spectral data for the SED of each source we used the SSDC Sky Explorer Web Tools\footnote{https://tools.ssdc.asi.it/}, which provides flux measurements taken from a large set of astronomical online databases and mission archives.
	Since the source identification by SSDC is based only on the position in the sky, we checked that each object was really included in the referenced archive, especially when several upper limits were present.
	
	To obtain soft X-ray data, we built the spectrum of each source using the Swift online tool\footnote{https://www.swift.ac.uk/} and fitted the flux with an absorbed power law using XSPEC in  the $0.3-2$ keV and $2-10$ keV bands as follows:
	\begin{equation}\label{alphaX}
		f(E)=Ke^{-N_H\sigma(E)}E^{-\alpha},
	\end{equation}
	where $\alpha$ is the photon index, $\sigma$ the photo-electric cross section, $K$ the normalization parameter at 1 keV, and $N_H$ the Galactic hydrogen column density to account for our galaxy absorption, which is obtained from the source position in the sky\footnote{https://www.swift.ac.uk/analysis/nhtot/index.php}. 
	We note that in this work we are not considering any absorption from the source galaxy, which is accounted for by our emission model.
	In this phase we are only interested in obtaining the observed flux and the observed power-law slope.
	In \tablename~\ref{fit_X} we show the XSPEC fit results in both the $0.3-2$ keV and $2-10$ keV bands, alongside the source redshift and the total exposure time for each spectra.
	
	\begin{table*}[tb]
		\centering
		\resizebox{\textwidth}{!}{
			\begin{tabular}{lrrrrrrrrrrrrrr}
				\toprule 
				Source & $z$ & $t_{tot}$    & $\alpha_{0.3-2}$ & $\alpha_{0.3-2}^{low}$     & $\alpha_{0.3-2}^{high}$     & $f_{0.3-2}$   & $f_{0.3-2}^{low}$       & $f_{0.3-2}^{high}$       & $\alpha_{2-10}$  & $\alpha_{2-10}^{low}$     & $\alpha_{2-10}^{high}$     & $f_{2-10}$        & $f_{2-10}^{low}$       & $f_{2-10}^{high}$ \\
				\midrule
				SWIFT J0000.5+3251 & 0.033 & 16.0 & 2.241 & 1.554 & 2.830 & 6.46E-14 & 5.21E-14 & 7.44E-14 & 1.306 & -0.274 & 2.923 & 1.35E-13 & - & 1.57E-13\\ 
				SWIFT J0043.9-5009 & 0.029 & 10.9 & 0.395 & -1.300 & 1.624 & 5.24E-14 & 4.37E-14 & 7.21E-14 & 0.796 & 0.341 & 1.256 & 1.30E-12 & 9.85E-13 & 1.42E-12\\ 
				SWIFT J0120.7-1444 & 0.040 & 16.9 & -0.496 & -1.009 & 0.827 & 1.86E-14 & 8.87E-15 & 2.51E-14 & 0.592 & -0.495 & 1.422 & 4.67E-13 & - & 5.42E-13\\ 
				SWIFT J0155.2-3048 & 0.134 & 14.7 & 0.892 & -0.880 & 1.795 & 5.68E-14 & 4.77E-14 & 7.00E-14 & 0.805 & 0.196 & 1.431 & 5.14E-13 & 3.17E-13 & 5.78E-13\\ 
				SWIFT J0201.5+5032 & 0.016 & 12.0 & -0.012 & -0.789 & 0.600 & 2.86E-13 & 2.63E-13 & 3.10E-13 & 1.362 & 0.993 & 1.738 & 1.67E-12 & 1.41E-12 & 1.78E-12\\ 
				SWIFT J0248.3+1202 & 0.034 & 2.7 & 0.111 & -1.067 & 0.969 & 4.10E-13 & 3.27E-13 & 4.62E-13 & 1.758 & 1.152 & 2.430 & 2.29E-12 & 1.54E-12 & 2.59E-12\\ 
				SWIFT J0302.2+6853 & 0.022 & 8.7 & 0.626 & -0.133 & 1.254 & 3.64E-13 & 3.28E-13 & 3.84E-13 & 2.331 & 1.984 & 2.696 & 9.96E-13 & 8.37E-13 & 1.21E-12\\ 
				SWIFT J0317.1+1542 & 0.174 & 0.9 & 2.189 & 0.819 & 3.294 & 9.38E-13 & 8.09E-13 & 1.05E-12 & 2.211 & 0.467 & 4.047 & 1.43E-12 & - & 1.54E-12\\ 
				SWIFT J0327.9-5301 & 0.060 & 5.1 & - & - & - & - & - & - & -0.508 & -2.679 & 1.245 & 7.05E-13 & - & 6.21E-13\\ 
				SWIFT J0329.4-2157 & 0.035 & 15.2 & 1.039 & 0.777 & 1.282 & 5.70E-13 & 5.42E-13 & 6.00E-13 & 1.931 & 1.587 & 2.287 & 1.33E-12 & 1.20E-12 & 1.43E-12\\ 
				SWIFT J0726.6-4634 & 0.030 & 12.0 & 3.892 & 2.279 & 5.300 & 2.79E-14 & 2.22E-14 & 4.08E-14 & 0.108 & -0.585 & 0.778 & 6.78E-13 & 2.46E-13 & 7.19E-13\\ 
				SWIFT J0749.8+3359 & 0.016 & 19.6 & -1.860 & -3.623 & -0.098 & 1.34E-15 & - & 2.00E-12 & -1.067 & -1.602 & -0.557 & 7.79E-13 & 4.80E-13 & 8.49E-13\\ 
				SWIFT J0943.1-4147 & 0.016 & 11.8 & - & - & - & - & - & - & -2.243 & -4.501 & -0.466 & 7.35E-13 & - & 6.97E-13\\ 
				SWIFT J1048.6-3901 & 0.045 & 9.0 & 2.502 & 2.343 & 2.658 & 1.64E-12 & 1.58E-12 & 1.71E-12 & 1.638 & 1.124 & 2.196 & 1.46E-12 & 1.15E-12 & 1.59E-12\\ 
				SWIFT J1121.1-0529 & 0.030 & 4.4 & - & - & - & - & - & - & 0.370 & -0.894 & 1.609 & 7.83E-13 & - & 8.03E-13\\ 
				SWIFT J1300.5-0759 & 0.026 & 9.6 & - & - & - & - & - & - & -0.502 & -1.896 & 0.856 & 6.43E-13 & 1.09E-14 & 6.46E-13\\ 
				SWIFT J1343.1+8038 & 0.045 & 14.8 & 0.832 & -0.929 & 1.840 & 3.37E-14 & 2.47E-14 & 4.21E-14 & 1.148 & 0.191 & 2.343 & 2.76E-13 & 1.35E-13 & 3.01E-13\\ 
				SWIFT J1534.5+6258 & 0.085 & 5.9 & 1.991 & 1.782 & 2.193 & 1.07E-12 & 1.02E-12 & 1.14E-12 & 2.195 & 1.498 & 2.908 & 1.20E-12 & 8.86E-13 & 1.30E-12\\ 
				SWIFT J1643.4+0986 & 0.047 & - & - & - & - & - & - & - & - & - & - & - & - & -\\ 
				SWIFT J1649.3-1739 & 0.023 & 10.1 & -1.839 & -2.588 & -0.756 & 5.66E-14 & 2.41E-14 & 6.17E-14 & 1.022 & 0.731 & 1.323 & 2.89E-12 & 2.57E-12 & 3.09E-12\\ 
				SWIFT J1651.2-0144 & 0.041 & 2.1 & - & - & - & - & - & - & -1.244 & -2.485 & 1.680 & 1.08E-12 & - & 1.11E-12\\ 
				SWIFT J1700.8+3602 & 0.113 & 4.9 & - & - & - & - & - & - & -1.452 & -2.905 & 0.349 & 7.36E-13 & - & 5.33E-13\\ 
				SWIFT J1725.7-4517 & 0.020 & 4.5 & 1.096 & 0.139 & 1.830 & 3.55E-13 & 3.06E-13 & 3.98E-13 & 2.110 & 1.397 & 2.864 & 1.44E-12 & 1.04E-12 & 1.59E-12\\ 
				SWIFT J1735.7+2045 & 0.024 & 16.2 & 4.512 & 2.999 & 6.355 & 2.38E-14 & 1.43E-14 & 2.97E-14 & -1.615 & -2.704 & -0.508 & 7.34E-13 & - & 6.94E-13\\ 
				SWIFT J1832.2+3146 & 0.044 & 6.9 & - & - & - & - & - & - & -0.228 & -0.561 & 5.663 & 3.90E-13 & - & 3.61E-13\\ 
				SWIFT J2004.6-1111 & 0.047 & 8.5 & 2.447 & 2.317 & 2.573 & 2.59E-12 & 2.50E-12 & 2.66E-12 & 1.842 & 1.580 & 2.120 & 3.22E-12 & 2.96E-12 & 3.40E-12\\ 
				SWIFT J2043.8-0958 & 0.142 & 20.6 & 0.194 & -0.173 & 0.518 & 2.27E-13 & 2.13E-13 & 2.43E-13 & 1.948 & 1.537 & 2.378 & 1.08E-12 & 9.25E-13 & 1.15E-12\\ 
				SWIFT J2112.4-4249 & 0.494 & 12.5 & 2.971 & 1.987 & 3.920 & 8.74E-14 & 7.18E-14 & 1.06E-13 & 1.821 & -1.315 & 5.416 & 6.03E-14 & - & 6.60E-14\\ 
				SWIFT J2117.7-0208 & 0.090 & 8.8 & 1.890 & 1.655 & 2.113 & 8.39E-13 & 8.01E-13 & 8.79E-13 & 2.085 & 1.541 & 2.697 & 9.26E-13 & 7.38E-13 & 1.02E-12\\ 
				SWIFT J2147.6-5360 & 0.031 & 9.7 & -2.823 & -5.646 & 2.568 & 1.73E-14 & - & 3.75E-14 & 0.564 & 0.016 & 1.135 & 1.13E-12 & 7.97E-13 & 1.23E-12\\ 
				SWIFT J2306.3-5147 & 0.096 & 8.2 & -1.651 & -2.797 & -0.384 & 5.92E-14 & 3.66E-15 & 7.44E-14 & 1.104 & 0.724 & 1.497 & 1.46E-12 & 1.19E-12 & 1.56E-12\\
				\bottomrule
		\end{tabular}}
		\caption{Fit results for the sources SWIFT soft-X spectra.
			The meanings of the parameters are $z$ the source redshift, $t_{tot}$ the total exposure time in ks, $\alpha_{X-Y}$ and $f_{X-Y}$ the photon index and the total flux (in erg cm$^{-2}$ s$^{-1}$) in the energy band $X-Y$ keV, each with their respective lower and upper limit.
			Fluxes and photon spectral indices are the observed values.
			The spectral indices $\alpha$ are defined as in Eq. \ref{alphaX}.}\label{fit_X}
	\end{table*}
	
	Then we applied emission models in an attempt to reproduce the obtained SEDs, including i) a standard thin accretion disk joined with an obscuring torus and a X-ray corona;  ii) a thick advection-dominated flow, emitting a synchrotron, Compton, and bremsstrahlung component; iii) an obscured AGN model, accounting for absorption along the line of sight both at kiloelectronvolt energies and in the optical band; and iv) a phenomenological model to describe the jet emission in blazar-like objects, composed of two smoothly broken power laws to represent the high- and low-energy bumps.
	Where needed, we integrated the models with the SWIRE template libraries \citep{swire_temp} to account for the emission of the host galaxy.
	
	In our model we have a large number of parameters, that is, up to 14 if all components are needed.
	In a relevant number of sources, these are much more numerous than the data points available.
	Hence we cannot apply automatic procedures via software, for example, a Levenberg-Marquardt algorithm to minimize the residuals.
	Moreover, AGNs are highly and fast variable sources, sometimes even by 2 or 3 orders of magnitude.
	Collecting data from different archives is meant to measure the source luminosity in different times and, probably, at different stages of their activity cycle.
	Therefore, the fitting procedures cannot give an exact value of the source physical properties.
	We emphasize that we are not trying to derive values, but we are inferring a range of parameters that may be compatible with the observed emission.  
	With our work we aim to understand if the measured mean luminosity, evaluated on a 0.1 logarithmic bin in frequency, is compatible with the applied model, without any intention to extract precise physical parameters.
	We only seek a coherent representation of the observed phenomenon consistent with the source classification type.  
	Therefore, we chose to perform a manual SED fit.
	
	\subsection{Standard accretion model}\label{AD}
	\subsubsection{Accretion disk}
	The simplest model for an accretion disk considers a steady state (all parameters are constant with time), geometrically thin (the characteristic height is much smaller than the linear size),  and optically thick (the emitted radiation is in thermodynamic equilibrium with the accreting plasma) disk.
	If we parameterize the viscosity in the disk with a parameter $\alpha$ ($0<\alpha<1$) the main physical parameters of the accretion disk, such as density, pressure, and radial velocity, can be determined as a function of $\alpha$ and its structure can be resolved analytically.
	Furthermore, the rate at which the energy is emitted at each radius is independent on the detailed local viscosity conditions.
	Moreover, the vertical optical thickness allows us to assume that each annulus emits radiation as a blackbody with a temperature $T(R)$ \citep{ss_acc_disk}.
	
	The amount of gravitational energy released from each ring of the disk is given by
	\begin{equation}
		F(R)=\frac{3}{8\pi}\left(\frac{R}{R_g}\right)^{-3}\left[1-\left(\frac{R}{R_{in}}\right)^{-1/2}\right]\frac{\dot{M}c^2}{R_g^2},
	\end{equation}
	where $R$ is the distance from the black hole, $R_g=GM_{BH}/c^2$ its gravitational radius, and $R_{in}$ is the inner radius of the disk.
	
	The assumption of large optical thickness implies that each ring of the disk emits blackbody radiation with temperature
	\begin{equation}
		T(R)=\left(\frac{F(R)}{\sigma}\right)^{1/4},
	\end{equation}
	where $\sigma$ is the Stefan–Boltzmann constant.
	Therefore, the emitted spectrum is a superposition of blackbody spectra and is written as
	\begin{equation}
		L_{\nu}=2\int_{R_{out}}^{R_{in}}2\pi R \pi B[\nu,T(R) dR],
	\end{equation}
	where $B[\nu,T] = \frac{2 h v^3}{c^2}\left(\exp\frac{h\nu}{kT}-1\right)^{-1}$ is the Plank function and $k$ the Boltzmann constant.
	
	\subsubsection{Obscuring torus}
	\begin{figure}[tb]
		\centering
		\includegraphics[width=0.7\columnwidth]{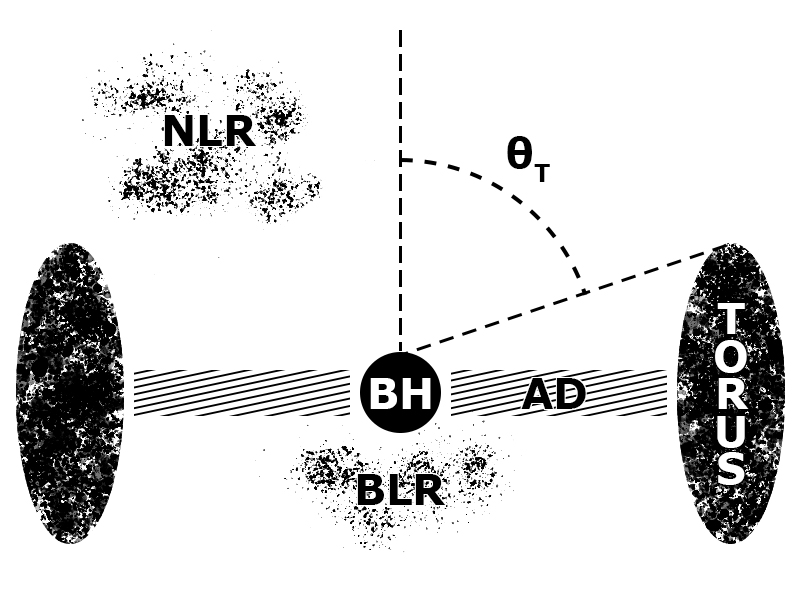}
		\caption{Scheme of the unified model of an AGN. The labeled parts are the supermassive black hole (BH), the accretion disk (AD), the narrow line region (NLR), the Broad Line Region (BLR), the obscuring torus, and the covering angle respect to the symmetry axis $\theta_{T}$.
		}\label{torusscheme}
	\end{figure}
	We consider the simplest case of a doughnut-shaped torus with opening angle $\theta_T$, measured from the symmetry axis (see \figurename~\ref{torusscheme}).
	For ease of calculation we can introduce the dimensionless parameters $x=R/R_{in}$ and $\eta=R_g/2R_{in}$.
	Setting $R_{in}=6R_{g}$ we can write the total emitted disk luminosity as
	\begin{equation}
		L_{disk}=\frac{1}{4}\dot{M}c^2\int_{r_{in}}^{r_{out}}x^{-2}-x^{-5/2}dx,
	\end{equation}
	where $\dot{M}$ is the accretion rate and $r=R/R_g$.
	
	The amount of disk radiation intercepted and re-emitted by the torus is written as
	\begin{equation}
		\frac{L_{torus}}{L_{disk}} = \frac{\int_{\theta_T}^{\frac{\pi}{2}} \sin\theta \cos\theta d\theta}{\int_{0}^{\frac{\pi}{2}} \sin\theta \cos\theta d\theta} = \cos^2\theta_T.
	\end{equation}
	Therefore the torus emits as a blackbody with temperature
	\begin{equation}\label{eq_T_torus}
		T=\left( \frac{L_{disk}\cos^2\theta_T}{\sigma S} \right)^{1/4},
	\end{equation}
	where $S$ is the emitting surface.
	
	We can define the torus covering factor $q$ as
	\begin{equation}
		q=\frac{\Omega_T}{4\pi}=\frac{2\times 2\pi \int_{\theta_T}^{\frac{\pi}{2}} \sin\theta d\theta}{4\pi}=\cos\theta_T.
	\end{equation}
	
	\subsubsection{Hot corona}
	The hot corona emission has been modeled with a simple power law and an exponential cut:
	\begin{equation}
		L_X(\nu)=D\nu^{-\alpha}e^{-\nu/\nu_{cut}}\qquad
		h\nu > kT_{\max},
	\end{equation}
	where $D$ is a scaling parameter, $T_{max}$ the accretion disk blackbody peak temperature, and $\nu_{cut}\sim300$ keV.
	In the literature, usually $\nu_{cut}$ varies between 100 keV and 400 keV and it has been shown to correlate with other spectral quantities (e.g., see the relationship between $\nu_{cut}$ and the spectral slope in Seyfert 1 galaxies in \citet{cutoff:perola}).
	However, after testing we found that our SED are not sufficiently resolved to appreciate differences as $\nu_{cut}$ varies within the same order of magnitude. Therefore we chose to adopt a fixed value for all the sources, thus reducing the number of free parameters in the fit.
	
	\subsection{Obscured AGN model}\label{obs_model}
	
	To reproduce the expected emission for a heavily obscured AGN we chose to start from the hot corona, since hard X-rays are less absorbed than other frequencies.
	Inspired by \citet{suzaku_obs}, we modeled the emission in the X-rays with an absorbed broken power law.
	The flux is written as
	\begin{equation}
		F(E)=e^{-\sigma(E) N_H^{gal}} \left[ fDE^{-\Gamma} e^{-\frac{E}{E_{cut}}} + e^{-\sigma(E)N_H}DE^{-\Gamma}e^{-\frac{E}{E_{cut}}}\right],
	\end{equation}
	where $\Gamma$ is the photon spectral index ($\Gamma = \gamma +1$), $e^{-\sigma(E) N_H^{gal}}$ is the absorption from the Galactic gas and dust ($N_H^{gal}\sim10^{20-21}$ cm$^{-2}$), $f$ is the fraction of emitted X-rays that is scattered, $e^{-\sigma(E)N_H}$ represents the obscuring torus absorption, $E_{cut}\sim 300$ keV is the cutoff energy and $\sigma(E)$ is the cross section of photoelectric absorption \citep{photoelectric} given by        \begin{equation}
		\sigma(E)=\frac{16\sqrt{2}}{3}\pi r_e^2 \alpha^4 Z^5 \left( \frac{m_e c^2}{E} \right)^{7/2}
		,\end{equation} 
	where $r_e$ and $m_e$ are the electron radius and mass, $\alpha$ the fine structure constant, and $Z\sim 1.4$ the average atomic number for galactic dust \citep{z_gal}.
	Unlike \citet{suzaku_obs}, we introduced the multiplicative exponential terms to account for the spectra cutoff above 100 keV and did not take into account the terms representing the reflection and iron K emission line, since our SED is not sufficiently resolved to observe these features.
	The scaling parameter $D$ is determined imposing the high-energy flux equal to the flux from BAT 105 at 52 keV.
	
	\subsubsection{Expected disk emission}
	We considered the BAT 105 flux as an indicator of the unabsorbed emission from the hot corona, since heavy absorption at hard X-ray energies is unlikely.
	Therefore, we assumed the spectral index from BAT 105 data as the index for the corona power law.
	To obtain the total emission of the corona, we integrated the unabsorbed power law from $\nu=10^{15}$ Hz to $\nu_{cut}$.
	
	Assuming an accretion efficiency of $\eta_{\mathit{eff}}=1/12$ and a constant mean ratio of 15 between the X-ray corona luminosity and the disk bolometric luminosity \citep{bat_vs_disk}, we can derive the accretion rate $\dot{M}$ as follows:
	\begin{equation}
		\dot{M}=\frac{15L_{X}}{\eta c^2}
		.\end{equation}
	From the correlation between the Eddington ratio to the X-ray luminosity in the BAT AGNs sample \citep{bat_edd} we can obtain the black hole mas:
	\begin{equation}
		\log\frac{L_{disk}}{L_{Edd}}=(0.92\pm 0.02)\log L_{BAT}+(-42\pm 3)
		,\end{equation}
	where $L_{\mathit{Edd}}=1.3\cdot 10^{47}\left(\frac{M_{BH}}{10^{9}M_\odot}\right)$ erg s$^{-1}$.
	We note that this method does not precisely weight the black hole, our purpose is to reproduce the entire SED without any intention to extract precise physical parameters.
	
	Therefore, having an indication on the black hole mass and accretion rate from the X-ray emission, interpreted as the hot corona, we can model an expected standard accretion disk as seen in Section~\ref{AD}.
	The values of $M_{BH}$ and $\dot{M}$ derived in this way are used as initial guess for our fitting procedure.
	
	\subsubsection{Optical absorption}
	The optical disk emission is then absorbed, to account for the contribution from silicate and graphite grains along the line of sight, according to \citet{abs_pei}.
	Since the absorption occurs in an area close to the disk, we calculated the absorption in the rest frame.
	
	According to the standard classification, an obscured AGN is defined through the absence of broad lines in the optical waveband; in this case we expect a typical extinction from dust for a Seyfert 2 source in the V band at 550 nm of at least $A_V\sim4-8$ mags \citep{sy:av}.
	
	\subsubsection{Infrared re-emission}
	The absorbed optical radiation is re-emitted by a toroidal dust structure up to 100 pc from the black hole.
	It is modeled by a blackbody spectra with temperature $T$ as in equation \ref{eq_T_torus}.
	To best fit the SED shape in all the sources with data in the millimeter to submillimeter band, it was necessary to introduce a second colder emitting structure at a larger distance (up to 1 kpc).
	
	\subsection{ADAF model}
	To describe the SED for an advection-dominated accretion flow (ADAF), we followed \citet{adaf_letter,adaf_article}.
	We developed a program to simulate the emission of an ADAF using the scaling laws from \citet{adaf_scaling}.
	Since the electrons are responsible for cooling, the temperature in these flows is determined by the energy balance equation for the electrons: the sum of the individual cooling processes (synchrotron, Compton, and bremsstrahlung) is equated to the total electrons heating for a given black hole mass, accretion ratio, viscosity parameter, and ratio of gas to total pressure; then the electron temperature is randomly varied until the equality is satisfied.
	Solving the emission equations for the electron temperature thereby fixes the slope of the Comptonized spectrum and the ADAF SED.
	Then we accounted for absorption in both the optical and the X-ray band as in section 2.2.
	We found that this model could not properly reproduce our sources emission.
	
	\subsection{Jet model}
	The mean nonthermal SED of blazars consists of two humps, peaking in the IR-X band and the MeV-TeV band, and a flat radio spectrum.
	In the most powerful sources, there is also often a third peak due to the accretion disk contribution.
	There is a major debate over the origin of the high-energy peak, involving leptonic, hadronic, or mixed emission mechanisms \citep[e.g., see][]{lepto:hadro}.
	In our work we chose a phenomenological description of the jet nonthermal emission.
	Our main focus is to understand whether a jet component is dominant or not, rather than derive specific jet parameters.
	Even though the analytic function is initially based on a leptonic jet model, this function is able to describe the characteristic two-bump blazar behavior without diving deeply into the jet nature debate.
	
	The simplest analytical function to approximate the SED is a single power law in the radio, connecting to two smoothly broken power laws to describe the high- and low-energy bumps \citep{blazar_sequence}.
	We note that this description is completely phenomenological.
	
	In the radio band, up to a frequency $\nu_t$ where the flat part ends, we have
	\begin{equation}
		L_R(\nu)=A\nu^{-\alpha_R}
		.\end{equation}
	The limit frequency $\nu_t$ can be interpreted as the self-absorbed frequency of the most compact jet emitting region.
	This power law connects with
	\begin{equation}
		L_{S+C}(\nu)=L_S(\nu)+L_C(\nu)
		.\end{equation}
	The low-energy part $L_S$, which can be associated with synchrotron flux, is assumed to be
	\begin{equation}
		L _ { \mathrm { S } } ( \nu ) = B \frac { \left( \nu / \nu _ { \mathrm { S } } \right) ^ { - \alpha _ { 1 } } } { 1 + \left( \nu / \nu _ { \mathrm { S } } \right) ^ { - \alpha _ { 1 } + \alpha _ { 2 } } } \exp \left( - \nu / \nu _ { \mathrm { cut } , \mathrm { S } } \right)
		,\end{equation}
	while the high-energy part, associated with the inverse Compton flux, is given by
	\begin{equation}
		L _ { \mathrm { C } } ( \nu ) = C \frac { \left( \nu / \nu _ { \mathrm { C } } \right) ^ { - \alpha _ { 3 } } } { 1 + \left( \nu / \nu _ { \mathrm { C } } \right) ^ { - \alpha _ { 3 } + \alpha _ { 2 } } } \exp \left( - \nu / \nu _ { \mathrm { cut }  , \mathrm { C }} \right)
		.\end{equation}
	The constants $A$, $B,$ and $C$ are obtained requiring the spectrum continuity and that, at each peak frequency, the emission is only due to the mechanism responsible for the peak.
	
	\section{Results}
	
\begin{figure}[tb]
		\centering
		\begin{minipage}{\columnwidth}
			\centering
			\includegraphics[width=.48\columnwidth]{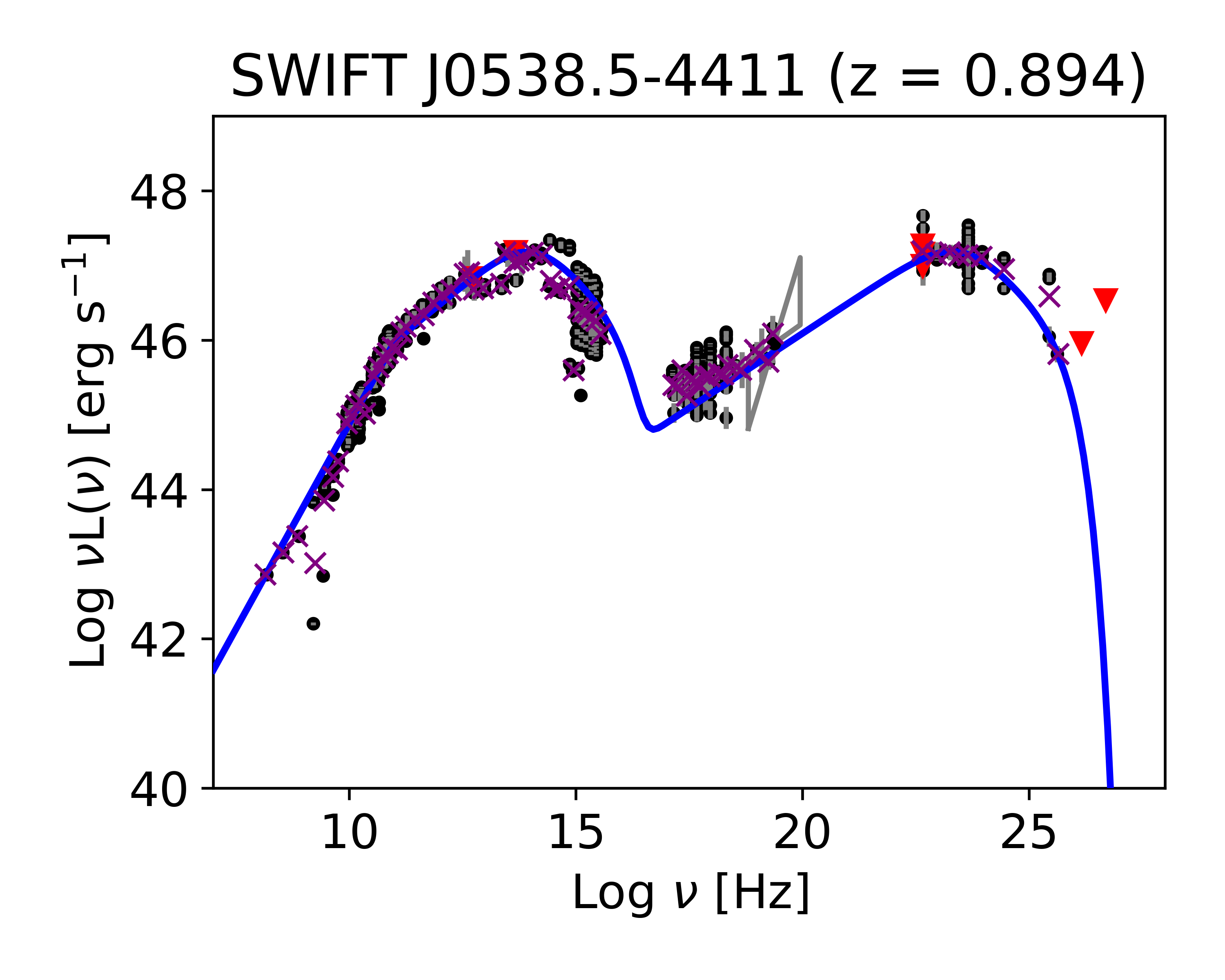}
			\includegraphics[width=.48\columnwidth]{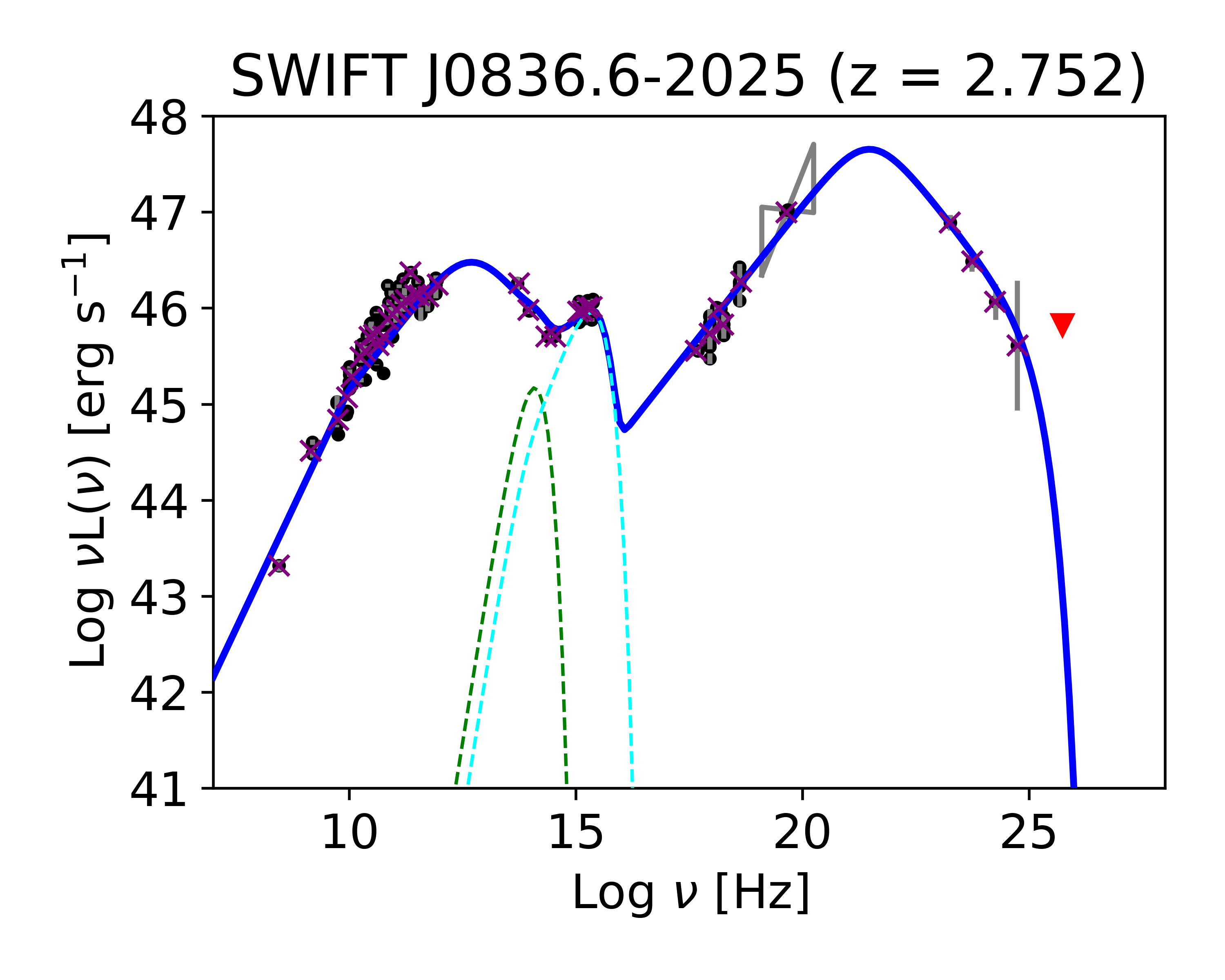}
		\end{minipage}
		\caption{Spectral energy distribution fit (blue line) for two beamed AGN sources.
			The black dots indicate archival data, the error bars are shown in gray, the red pointers represent the upper limits, and the purple crosses depict the mean flux in a 0.1 width logarithmic bin of frequency.
			The model considers a phenomenological description of the jet emission for blazar-like sources with the optional addition of an emitting thin accretion disk (cyan dashed line) and an obscuring torus (green dashed line).
		}\label{es_agn}
	\end{figure}

	To test our blazar-like model, we first studied the 31 new sources from BAT 105 beamed AGN type with redshift measurements.
	We obtained a good agreement between data and the phenomenological jet model for these sources, confirming their blazar nature. Two of these sources are shown
	in \figurename~\ref{es_agn}.
	We note that we can clearly see the accretion disk emission in SWIFT J0836.6-2025.
	
\begin{figure}[tb]
		\centering
		\begin{minipage}{\columnwidth}
			\centering
			\includegraphics[width=.48\columnwidth]{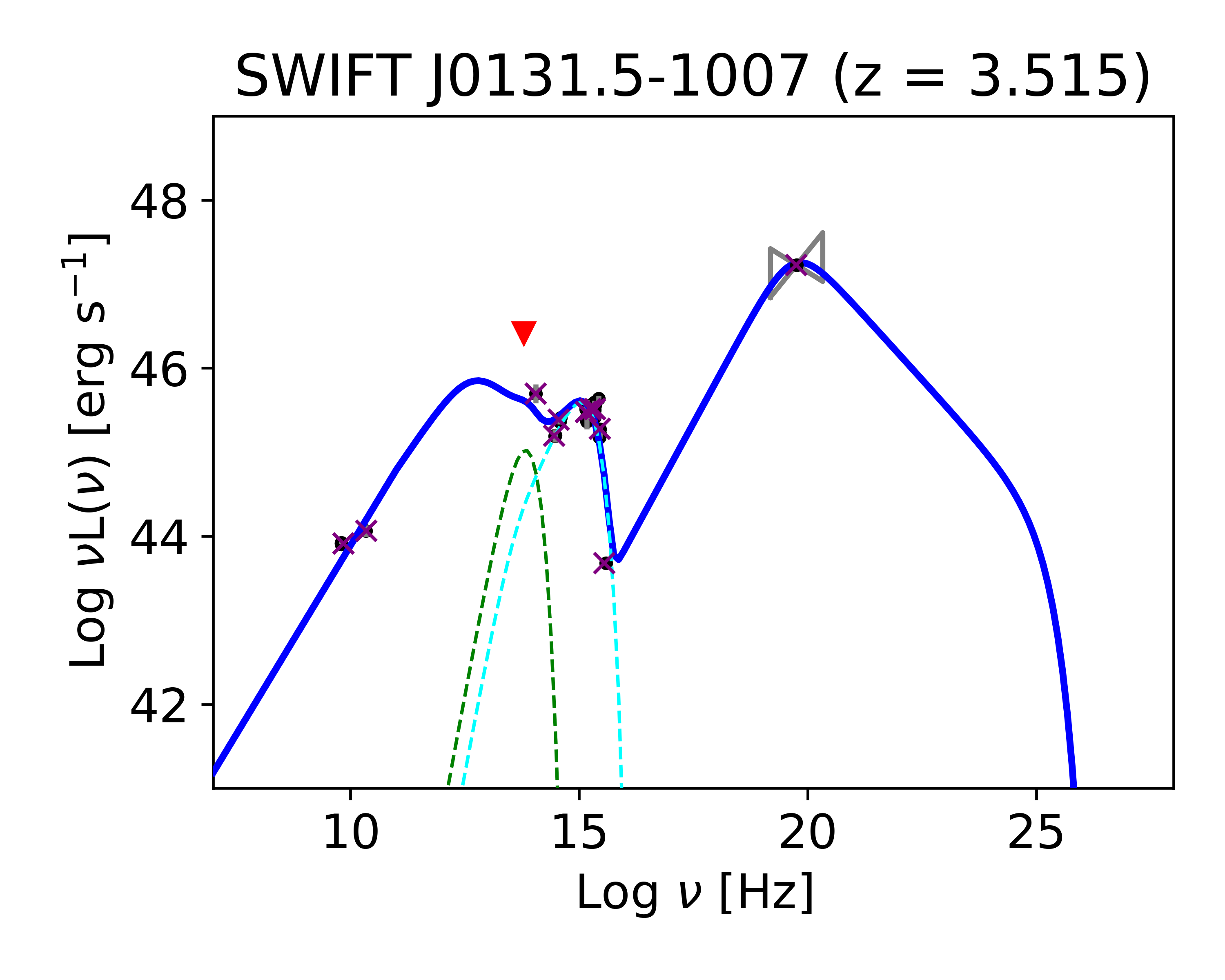}
			\includegraphics[width=.48\columnwidth]{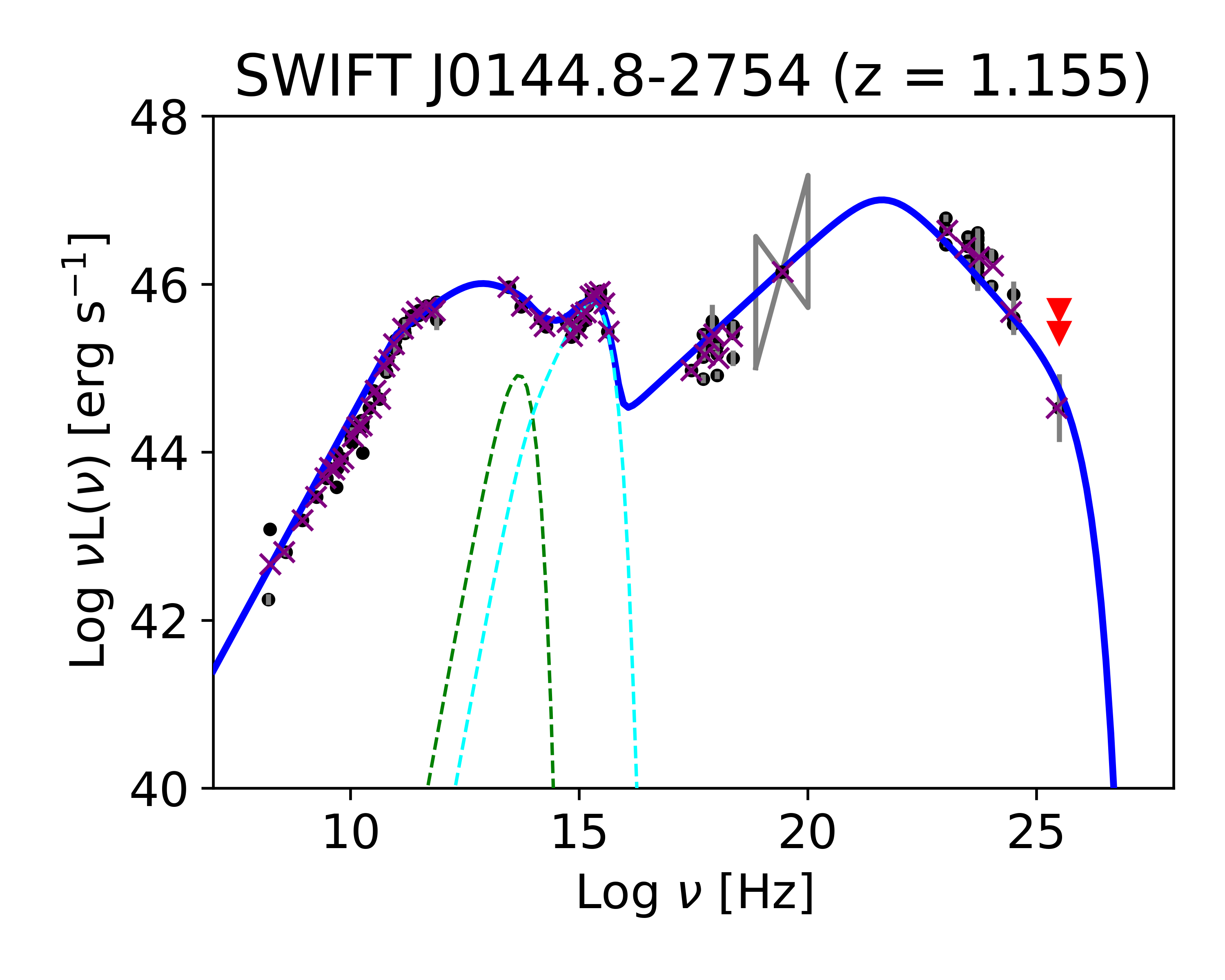}
		\end{minipage}
		\begin{minipage}{\columnwidth}
			\centering
			\includegraphics[width=.48\columnwidth]{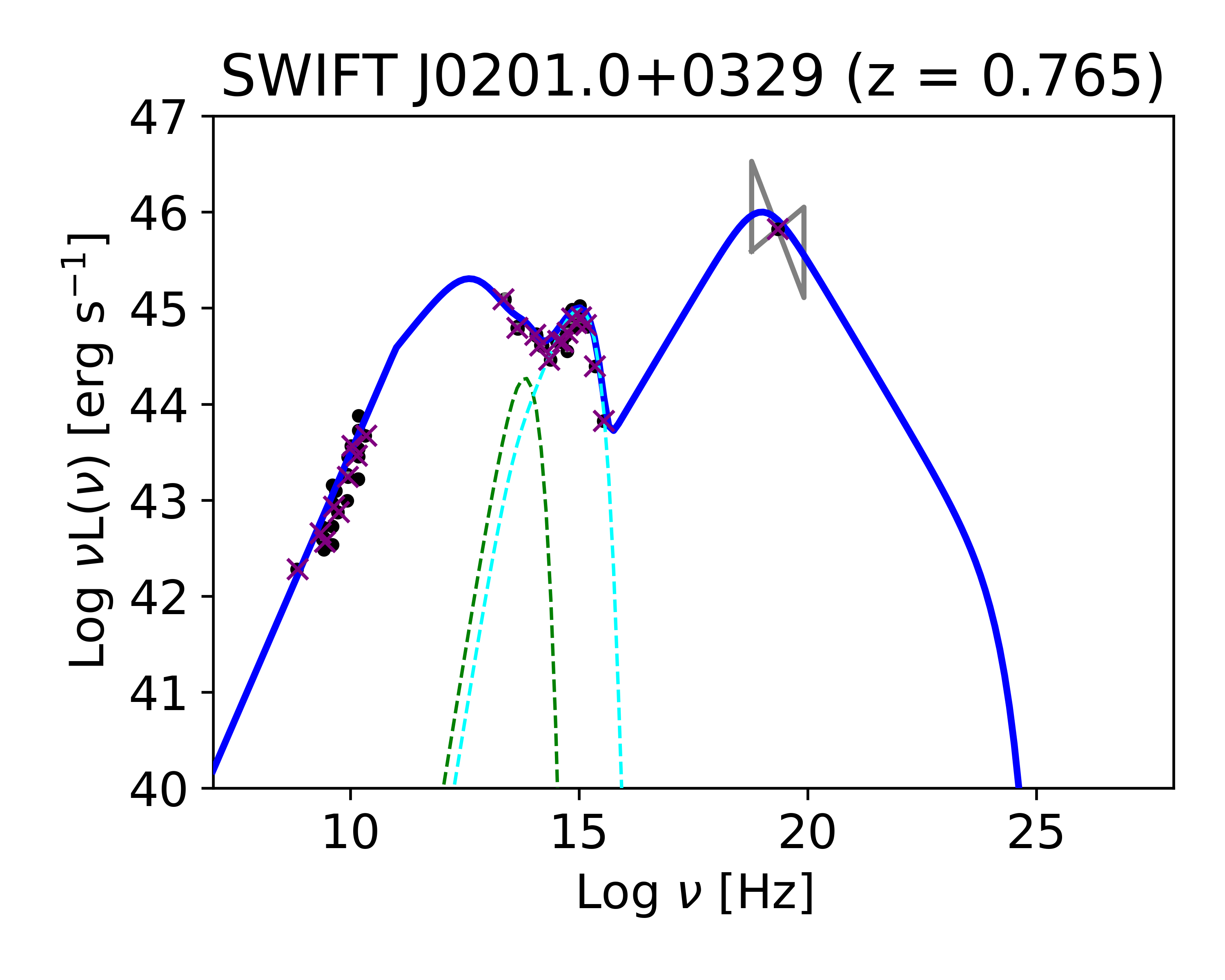}
			\includegraphics[width=.48\columnwidth]{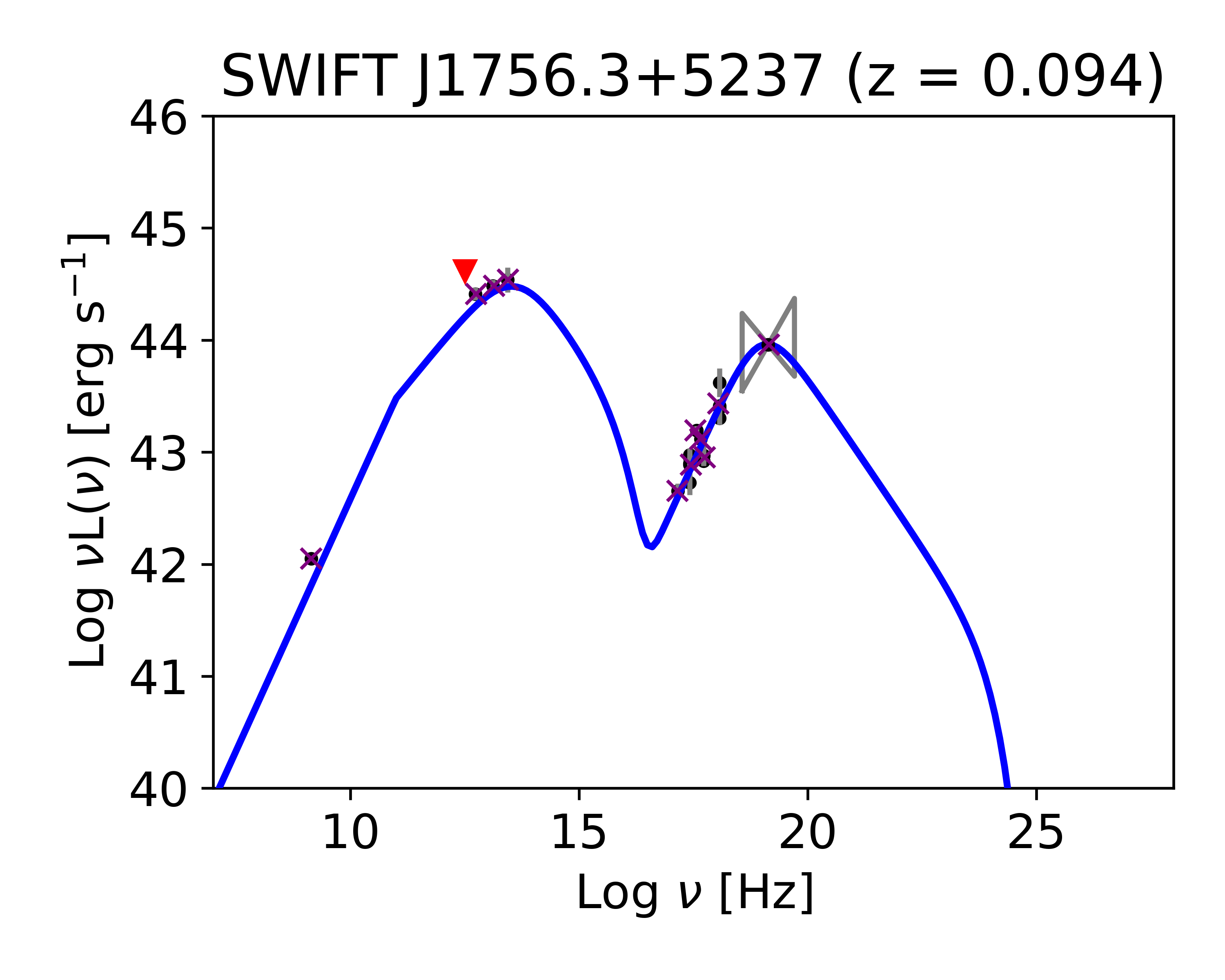}
		\end{minipage}
		\begin{minipage}{\columnwidth}
			\centering
			\includegraphics[width=.48\columnwidth]{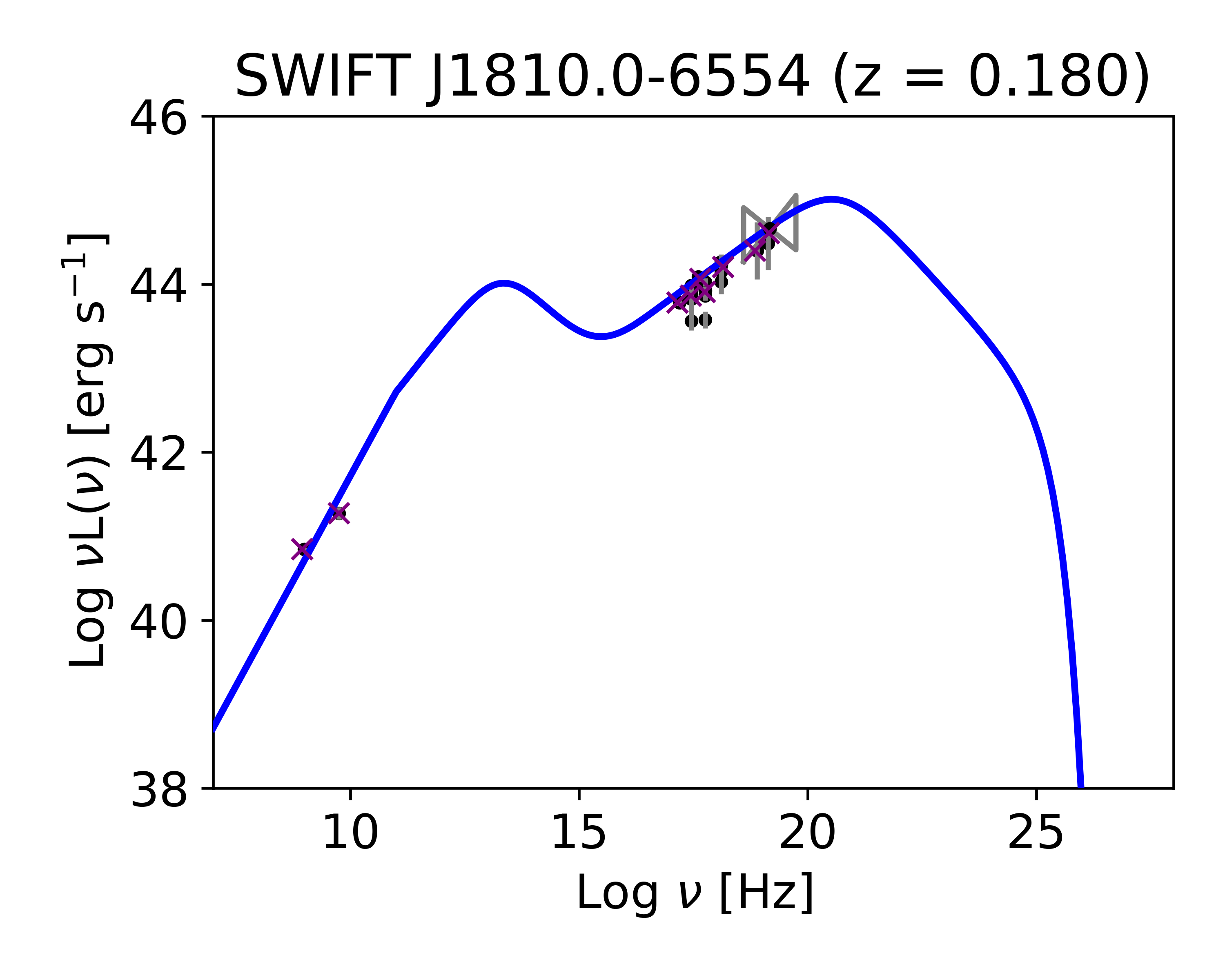}
		\end{minipage}
		\caption{Spectral energy distribution fit (blue line) for five unknown AGN sources.
			The black dots indicate archival data, the error bars are shown in gray, the red pointers represent upper limits, and the purple crosses depict the mean flux in a 0.1 width logarithmic bin of frequency.
			The model considers a phenomenological description of the jet emission for blazar-like sources with the optional addition of an emitting thin accretion disk (cyan dashed line) and an obscuring torus (green dashed line).
		}\label{es_agn_obs}
	\end{figure}

	Then we tried to apply the phenomenological jet model to the 36 unknown AGN with redshift measurements, but we found a good agreement to the data in only 5 sources, which we show in \figurename~\ref{es_agn_obs}.
	In three of these sources we found the presence of the emission from a standard accretion disk.
	In \tablename~\ref{unk_disk} the fit parameters for the five sources with a good correspondence with our phenomenological jet model are shown.
	
	\begin{table*}[tb]
		\centering
		\resizebox{\textwidth}{!}{
			\begin{tabular}{lrrrrrrrrrrrrrrrr}
				\toprule
				Source & CD & $\nu L_S$ & $a_1$ & $a_2$ & $a_3$ & $\nu_s$ & $\nu_c$ & $\nu_{cut,s}$ & $\nu_{cut,c}$ & $\nu_t$ & $a_r$ & $\dot{m}$ & $M_{BH}$ & $R_T$ & $\theta_T$ & $l_{Edd}$\\
				\midrule 
				SWIFT J0131.5-1007 & 25.00 & 7.00E+45 & 0.20 & 1.60 & 0.00 & 5.0E+12 & 5.0E+19 & 1.0E+15 & 1.0E+25 & 1.0E+11 & 0.10 & 1.00 & 2.50E+09 & 1.20 & 0.20 & 0.015\\ 
				SWIFT J0144.8-2754 & 10.00 & 1.00E+46 & 0.55 & 1.60 & 0.50 & 1.0E+13 & 5.0E+21 & 5.0E+15 & 5.0E+25 & 1.0E+11 & 0.00 & 1.50 & 1.00E+09 & 2.40 & 0.79 & 0.055\\ 
				SWIFT J0201.0+0329 & 10.00 & 2.00E+45 & 0.40 & 1.80 & 0.20 & 5.0E+12 & 2.0E+20 & 1.0E+15 & 1.0E+24 & 1.0E+11 & -0.10 & 0.25 & 1.50E+09 & 0.58 & 0.60 & 0.006\\ 
				SWIFT J1756.3+5237 & 0.30 & 3.00E+44 & 0.50 & 1.60 & 0.10 & 4.0E+13 & 1.0E+19 & 1.0E+16 & 1.0E+24 & 1.0E+11 & 0.10 & - & - & - & - & - \\ 
				SWIFT J1810.0-6554 & 10.00 & 1.00E+44 & 0.30 & 1.60 & 0.60 & 2.0E+13 & 5.0E+20 & 1.0E+16 & 1.0E+25 & 1.0E+11 & 0.00 & - & - & - & - & - \\
				\bottomrule
		\end{tabular}}
		\caption{Fit parameters for the unknown AGNs (phenomenological jet model with accretion disk) identified as blazars.
			We only found the emission signature of an accretion disk in the first three objects.
			The meanings of the parameters used are as follows: $CD$ Compton dominance ratio; $\nu L_S$ synchrotron peak luminosity in $\nu L(\nu)$ in erg s$^{-1}$; 
			$a_1$, $a_2$, $a_3$ power-law slopes; $\nu_S$ synchrotron peak frequency in Hz; $\nu_C$ inverse-Compton peak frequency in Hz; $\nu_{cut,S}$, $\nu_{cut,C}$ cutoff frequencies for low- and high-energy spectra in Hz; $\nu_T$ self-absorption frequency in Hz; $a_r$ power-law slope for $\nu<\nu_T$; $\dot{ m }$ accretion ratio in units of $M_{\odot}$ yr$^{-1}$; $M_{BH}$ black hole mass in $M_{\odot}$; $l_{Edd}$ accretion disk Eddington ratio; $R_T$ obscuring torus radius in parsec; and $\theta_T$ obscuring torus covering angle.}\label{unk_disk}
	\end{table*}
	
\begin{figure}[tb]
		\centering
		\begin{minipage}{\columnwidth}
			\centering
			\includegraphics[width=.4\columnwidth]{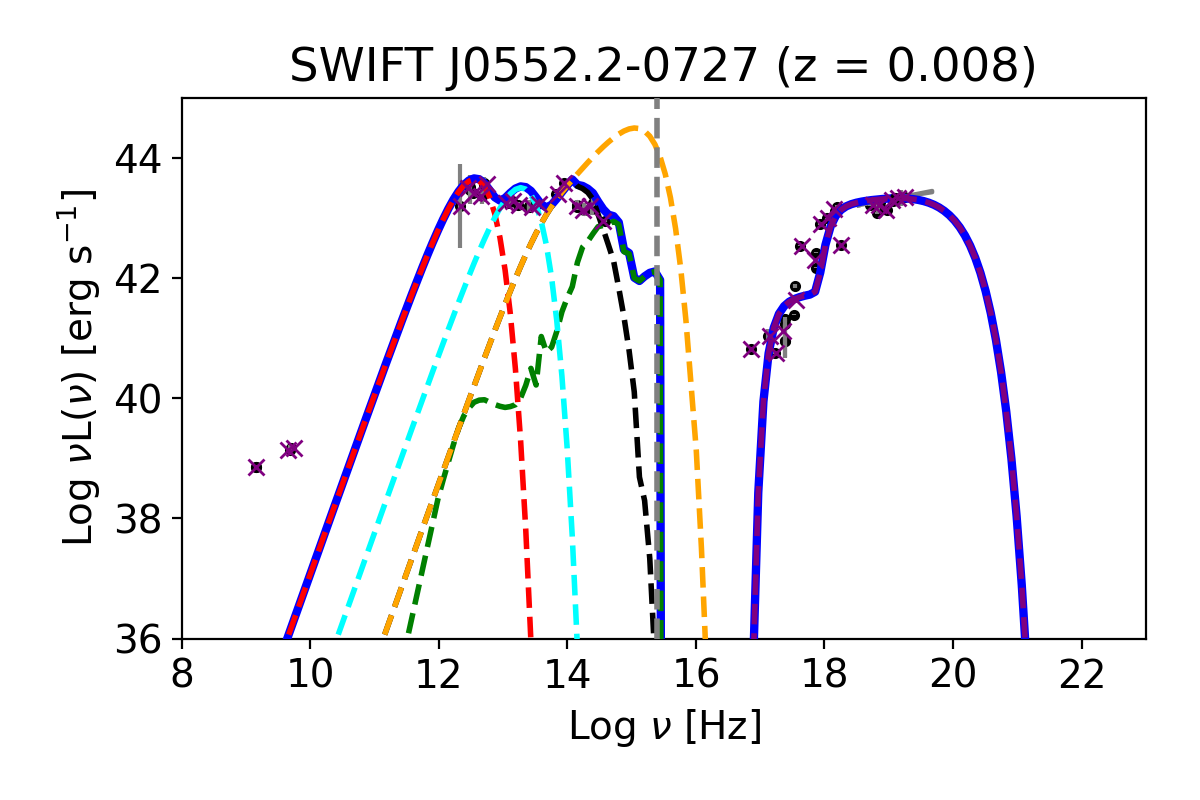}
			\includegraphics[width=.4\columnwidth]{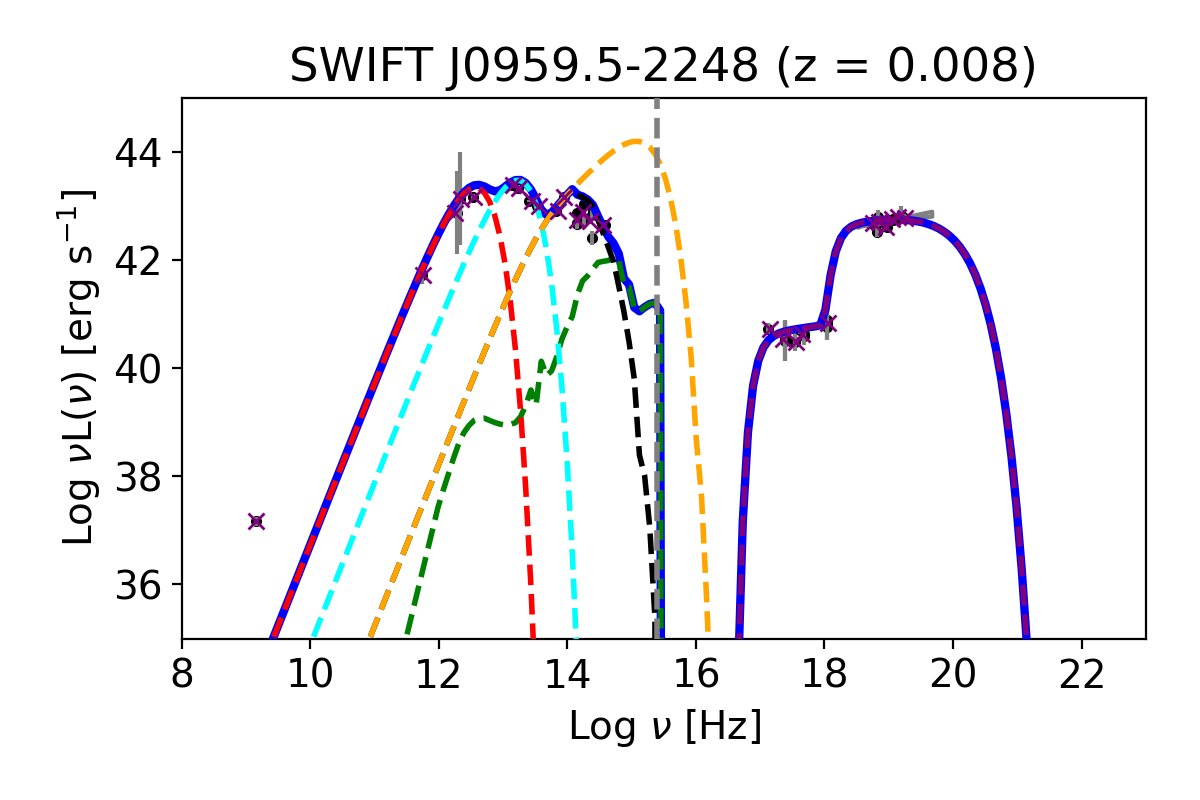}
		\end{minipage}
		\begin{minipage}{\columnwidth}
			\centering
			\includegraphics[width=.4\columnwidth]{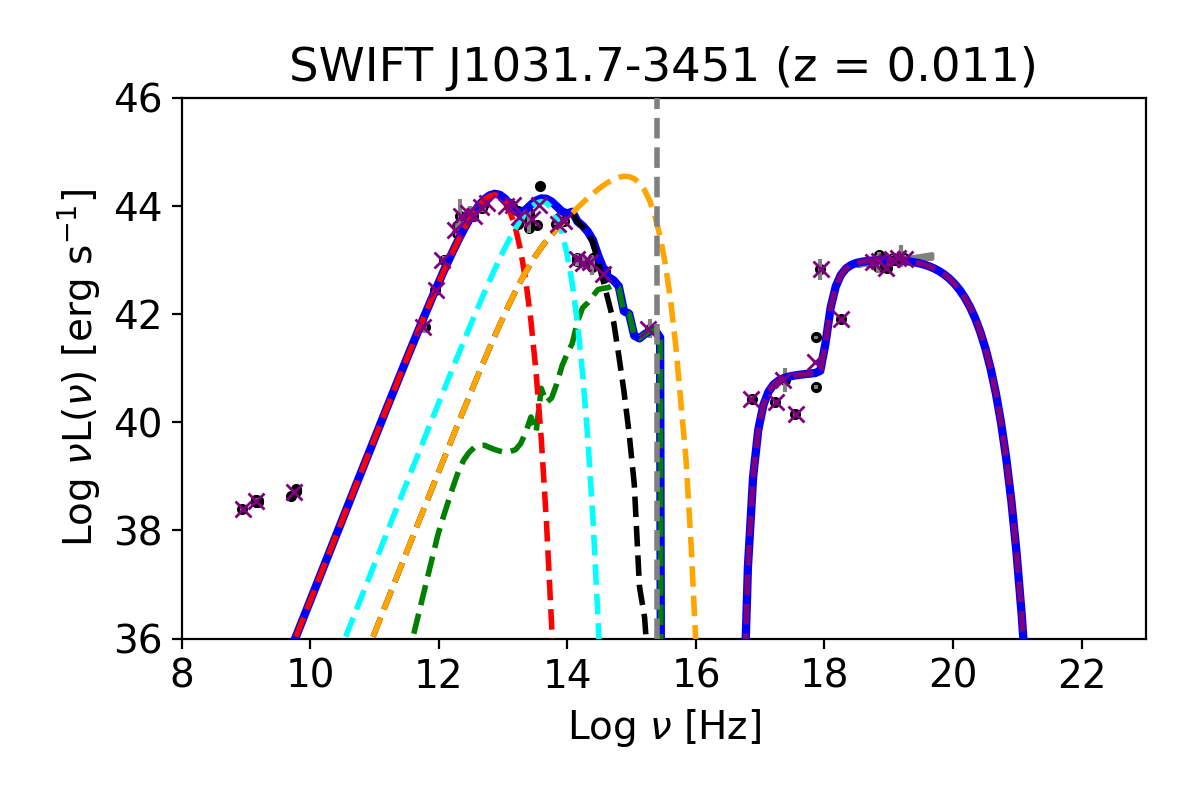}
			\includegraphics[width=.4\columnwidth]{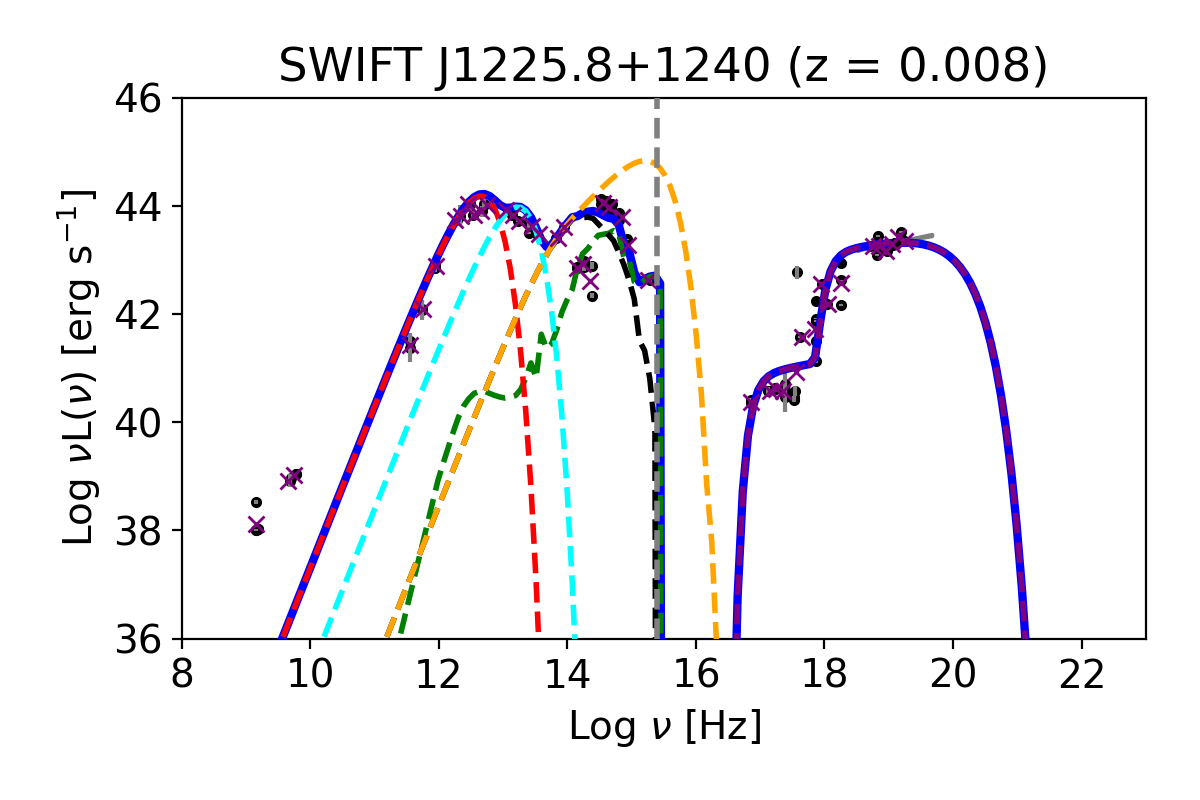}
		\end{minipage}
		\begin{minipage}{\columnwidth}
			\centering
			\includegraphics[width=.4\columnwidth]{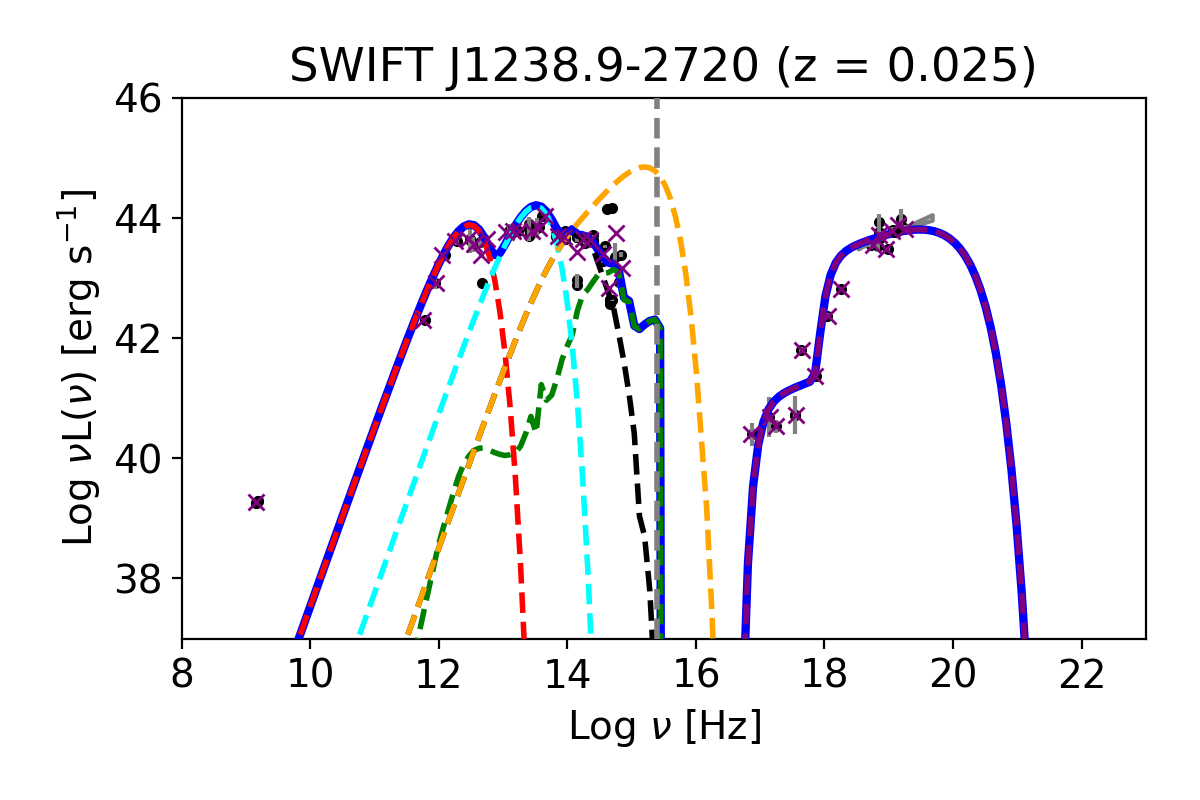}
			\includegraphics[width=.4\columnwidth]{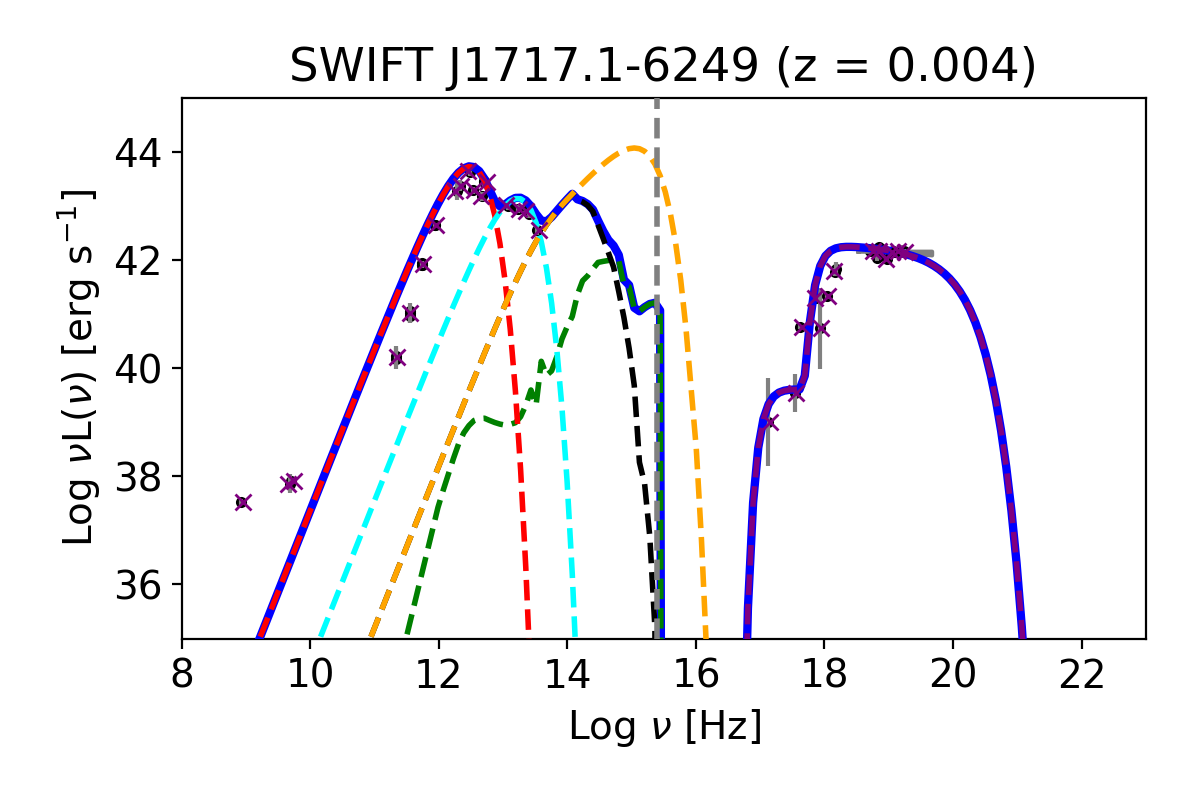}
		\end{minipage}
		\begin{minipage}{\columnwidth}
			\centering
			\includegraphics[width=.4\columnwidth]{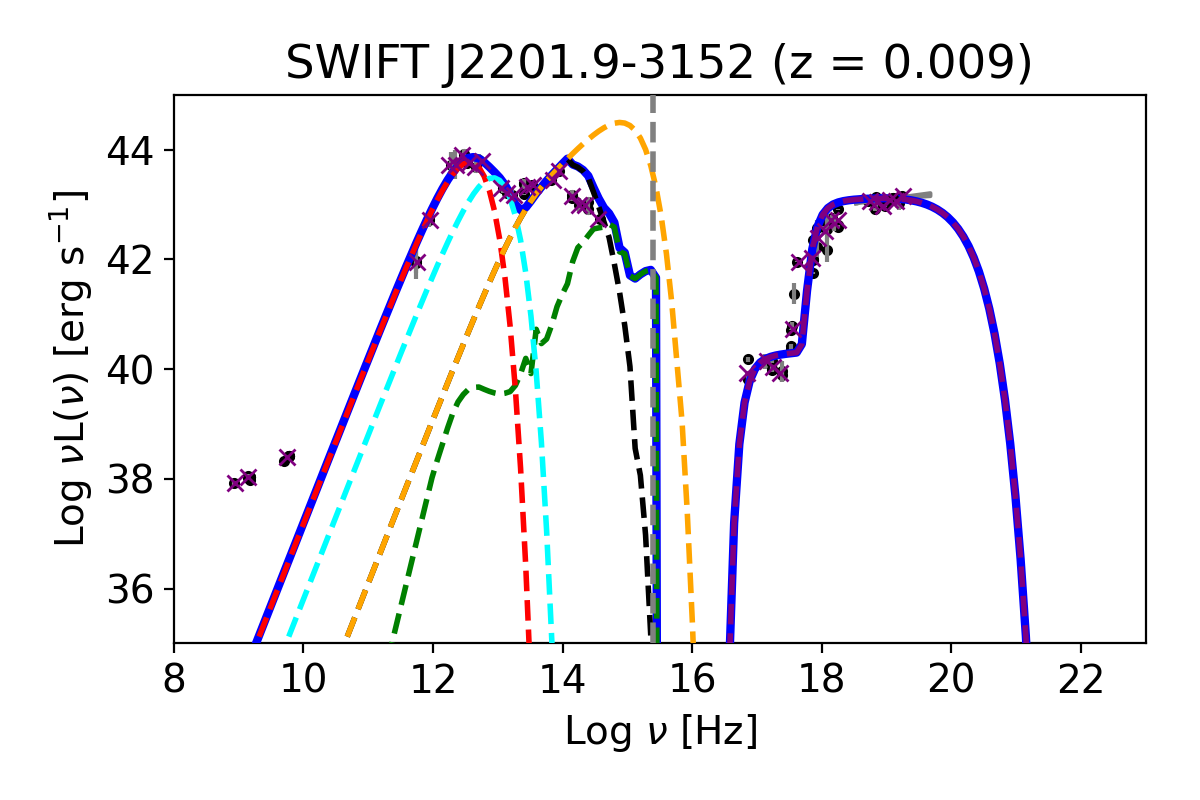}
			\includegraphics[width=.4\columnwidth]{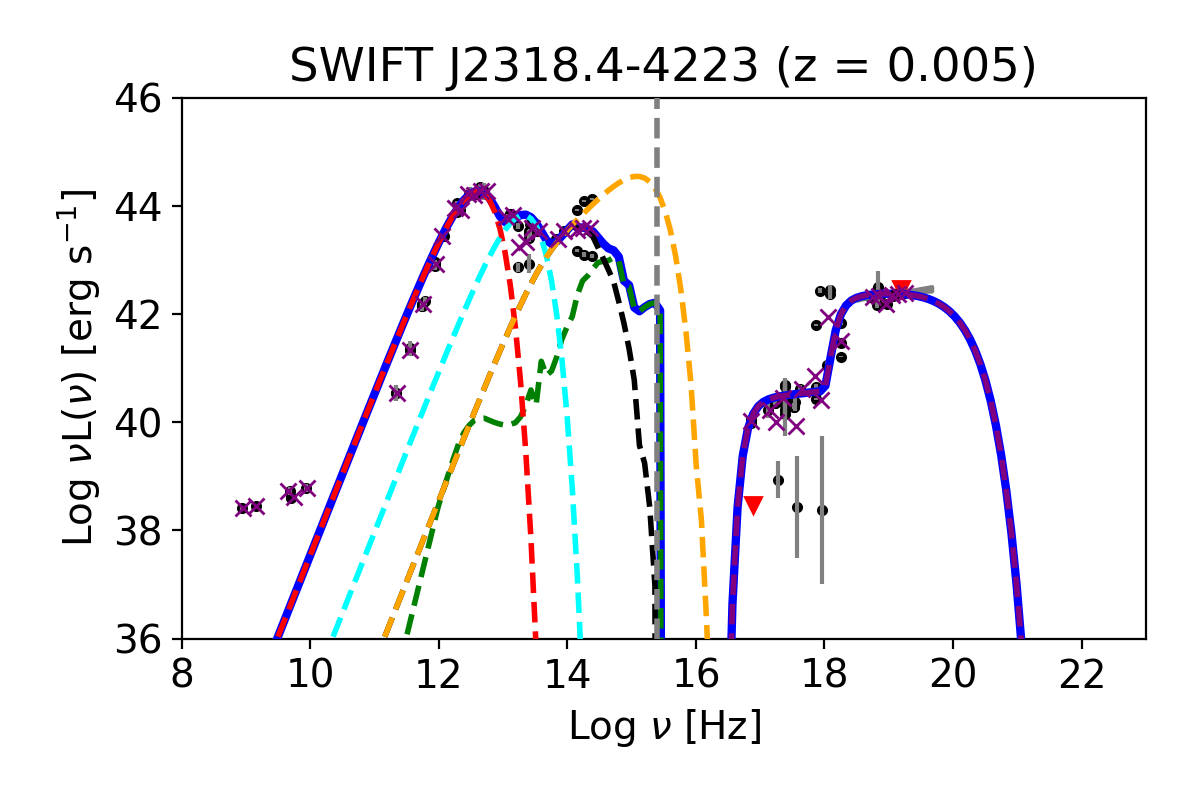}
		\end{minipage}
		\caption{Spectral energy distribution fit (blue line) for eight of the BAT Seyfert 2 sources with the highest flux. The black dots indicate archival data, the error bars are shown in gray, the red pointers represent upper limits, and the purple crosses depict the mean flux in a 0.1 width logarithmic bin of frequency. The model fit (blue line) considers absorption at kiloelectronvolt energies and in the optical band. The orange dashed line indicates the unabsorbed emission from the accretion disk, the black dashed line indicates the optical absorbed disk emission, and the green dashed line indicates the galactic template, the red dashed line indicates the emission from the dust belt, the cyan dashed line is the obscuring torus and the purple dashed line indicates the absorbed corona emission. The gray vertical dashed line indicates the hydrogen $Ly\alpha$ frequency.}\label{es_sy2}
	\end{figure}

\begin{figure}[tb]
		\centering
		\includegraphics[width=.9\columnwidth]{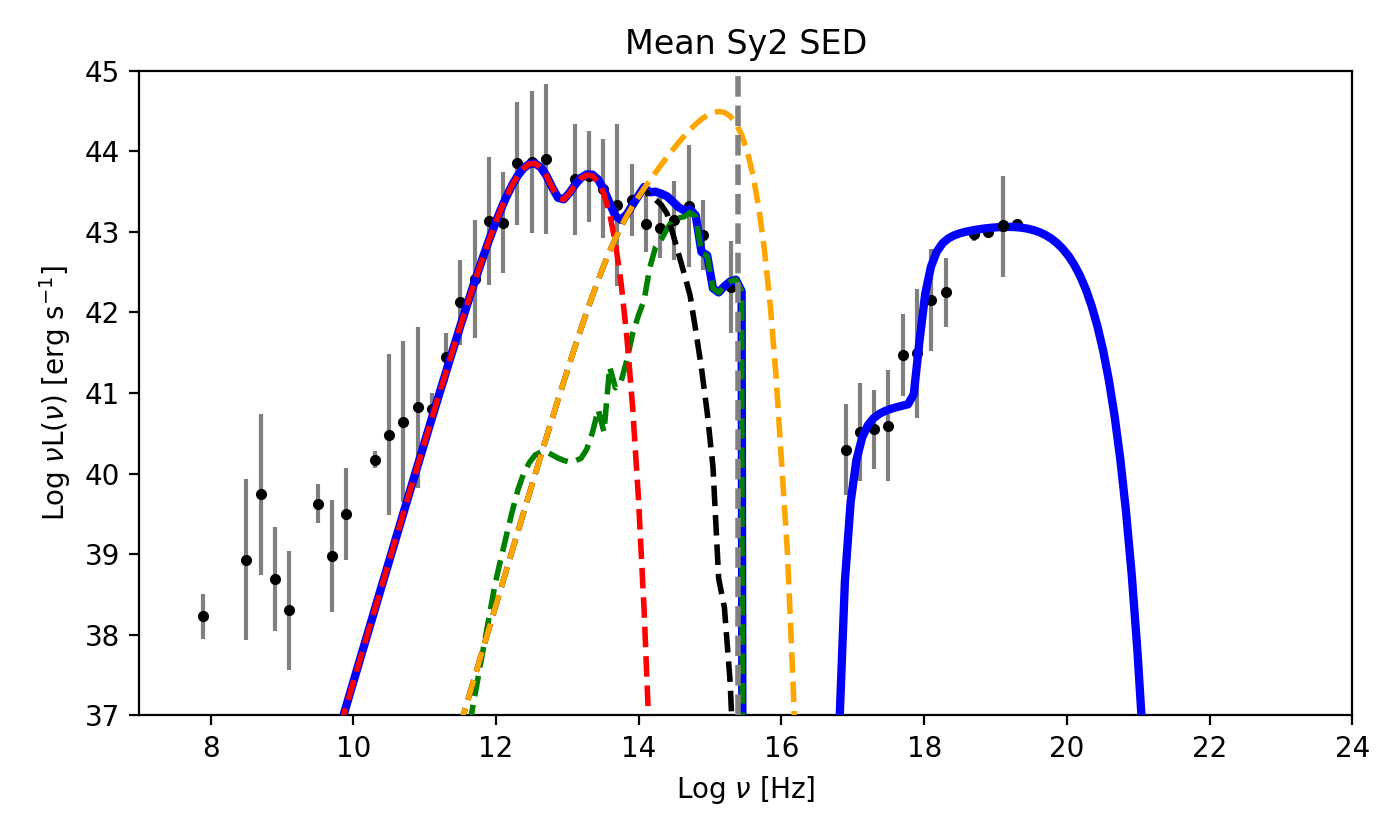}
		\caption{Spectral energy distribution fit (blue line) of the mean spectra of the 20 Seyfert 2 sources in the BAT catalog with the highest flux.
			The black dots indicate archival data, the error
			bars are shown in gray, the red pointers represent upper limits, and the
			purple crosses depict the mean flux in a 0.1 width logarithmic bin of frequency.
			The model fit (blue line) considers absorption at keV energies and in the optical band.
			The orange dashed line represents the unabsorbed emission from the accretion disk, the black dashed line indicates the optical absorbed disk emission, the green dashed line indicates the galactic template, and the red dashed line represents the sum of the emission from the dust belt and the obscuring torus.
			The gray vertical dashed line indicates the hydrogen $Ly\alpha$ frequency.
		}\label{es_sy2mean}
	\end{figure}

	Therefore, we considered the obscured AGN model.
	We tested this model on the 20 Seyfert 2 sources with the highest flux in the BAT catalog and their mean SED, obtained by normalizing the spectra on the mean BAT flux.
	We obtained a good agreement between data and the obscured AGN model for these sources, confirming that our model is a good representation of their overall sources emission.
	In \figurename~\ref{es_sy2} the fit of eight of these sources are shown, while in \figurename~\ref{es_sy2mean} we show the fit of the Seyfert 2 sample mean SED.
	
\begin{figure}[tb]
		\centering
		\includegraphics[width=\columnwidth]{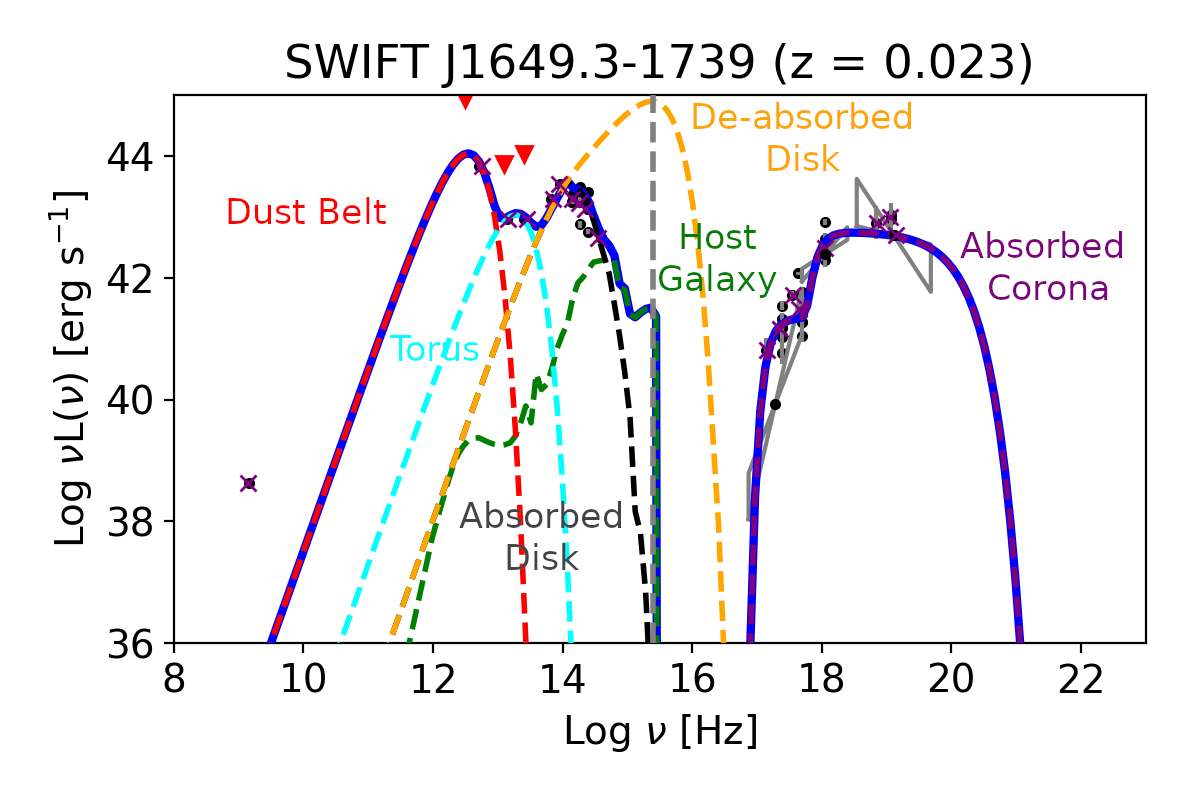}
		\caption{Example of the SED fit for obscured AGN sources.
			The black dots indicate archival data, the error
			bars are shown in gray, the red pointers represent upper limits, and the
			purple crosses depict the mean flux in a 0.1 width logarithmic bin of frequency.
			The model fit (blue line) considers absorption at keV energies and in the optical band.
			The orange dashed line represents the unabsorbed emission from the accretion disk, the black dashed line indicates the optical absorbed disk emission, the green dashed line indicates the galactic template, the red dashed line depicts the emission from the dust belt, the cyan dashed line indicates the obscuring torus, and the purple dashed line indicates the absorbed corona emission.
			The gray vertical dashed line indicates the hydrogen $Ly\alpha$ frequency.
		}\label{es_obs}
	\end{figure}

\begin{figure*}[tb]
		\centering
		\begin{minipage}{\textwidth}
			\centering
			\includegraphics[width=.23\textwidth]{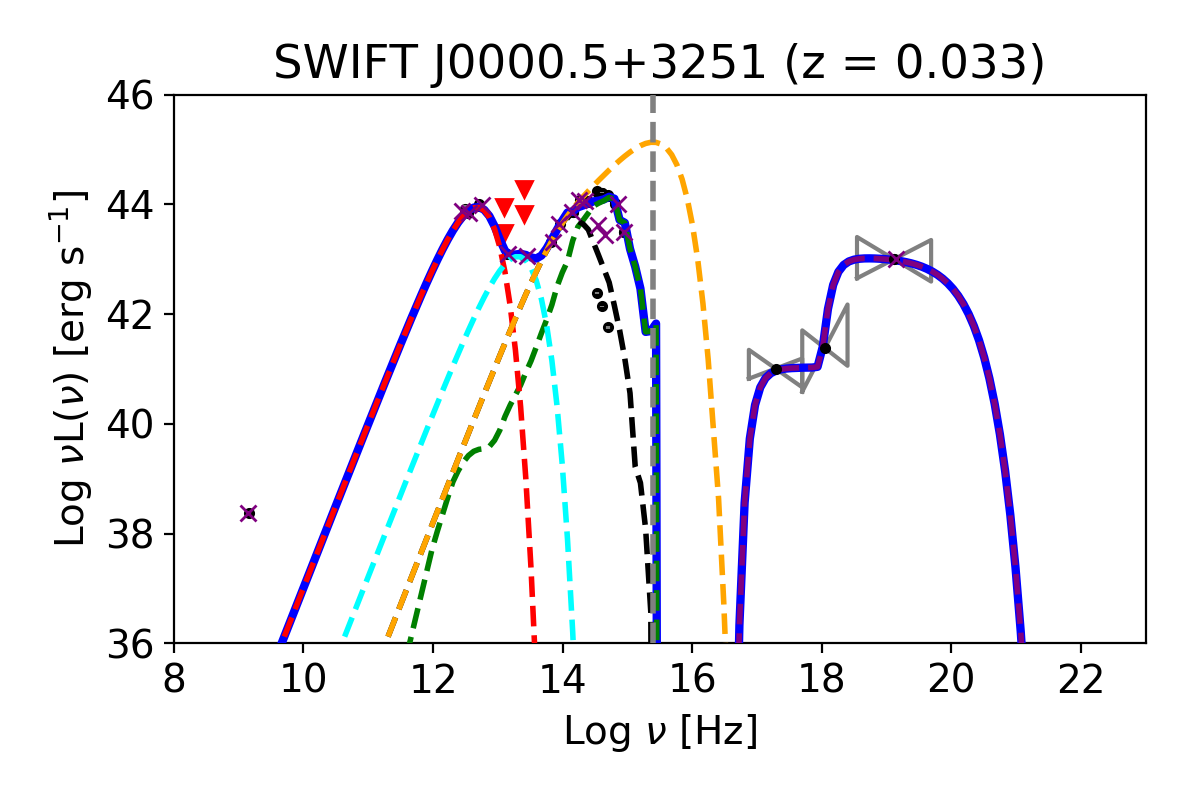}
			\includegraphics[width=.23\textwidth]{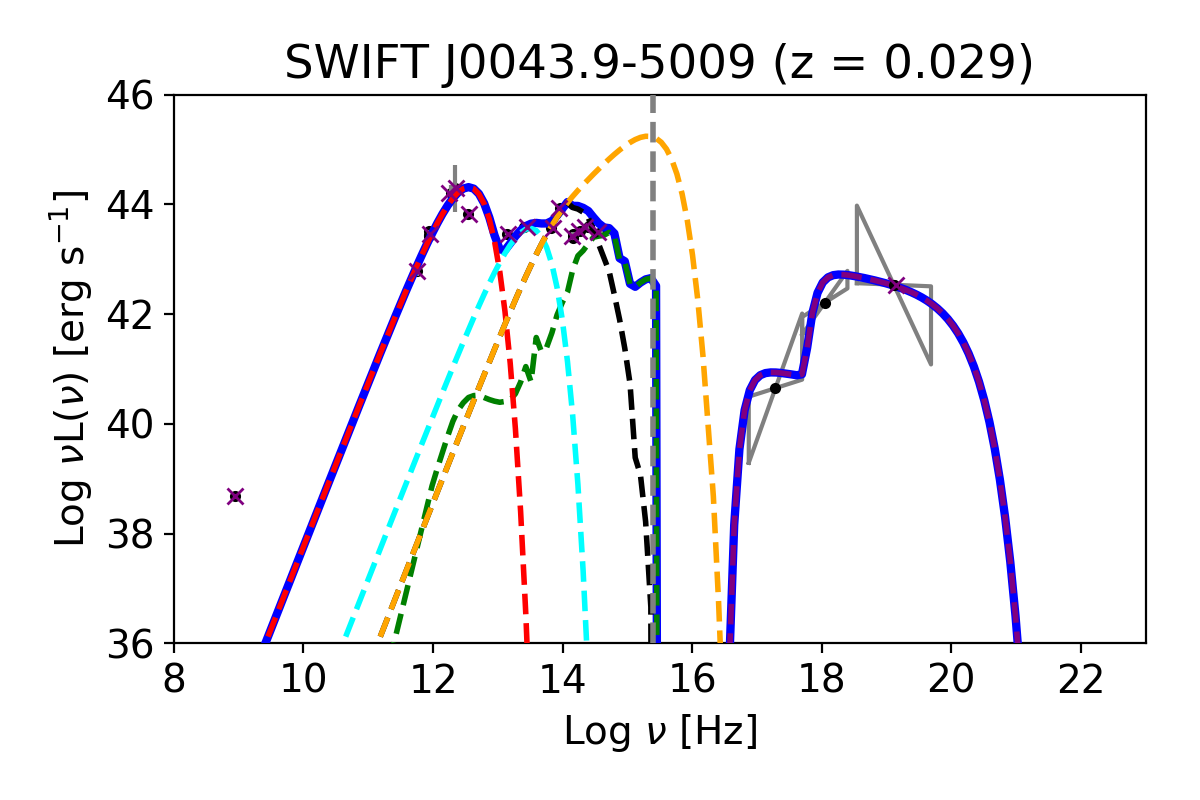}
			\includegraphics[width=.23\textwidth]{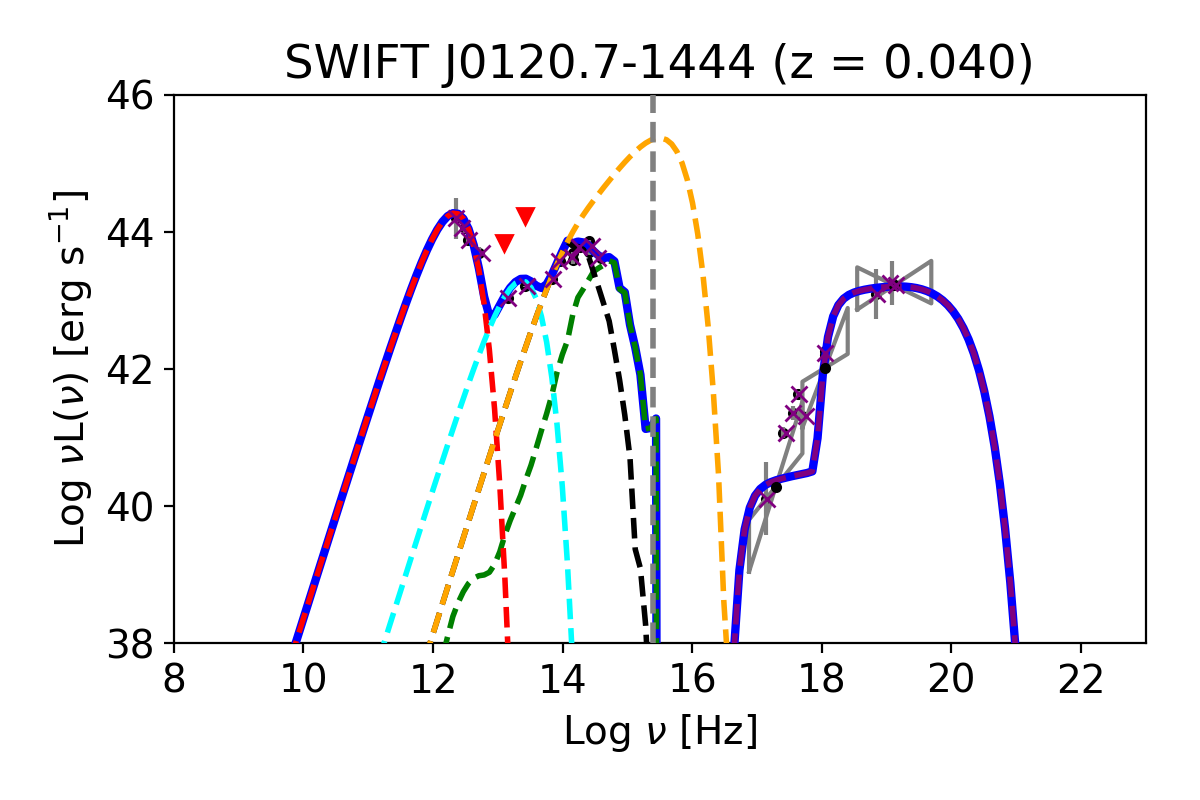}
			\includegraphics[width=.23\textwidth]{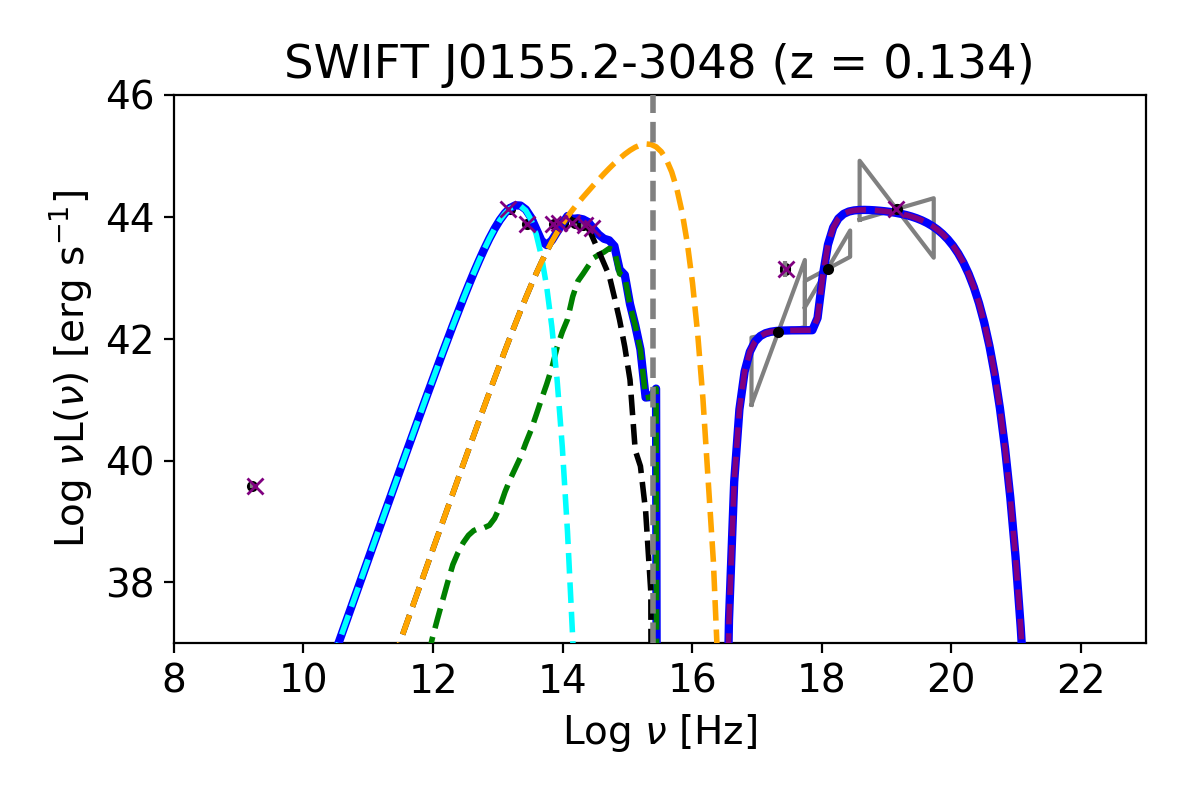}
		\end{minipage}
		\begin{minipage}{\textwidth}
			\centering
			\includegraphics[width=.23\textwidth]{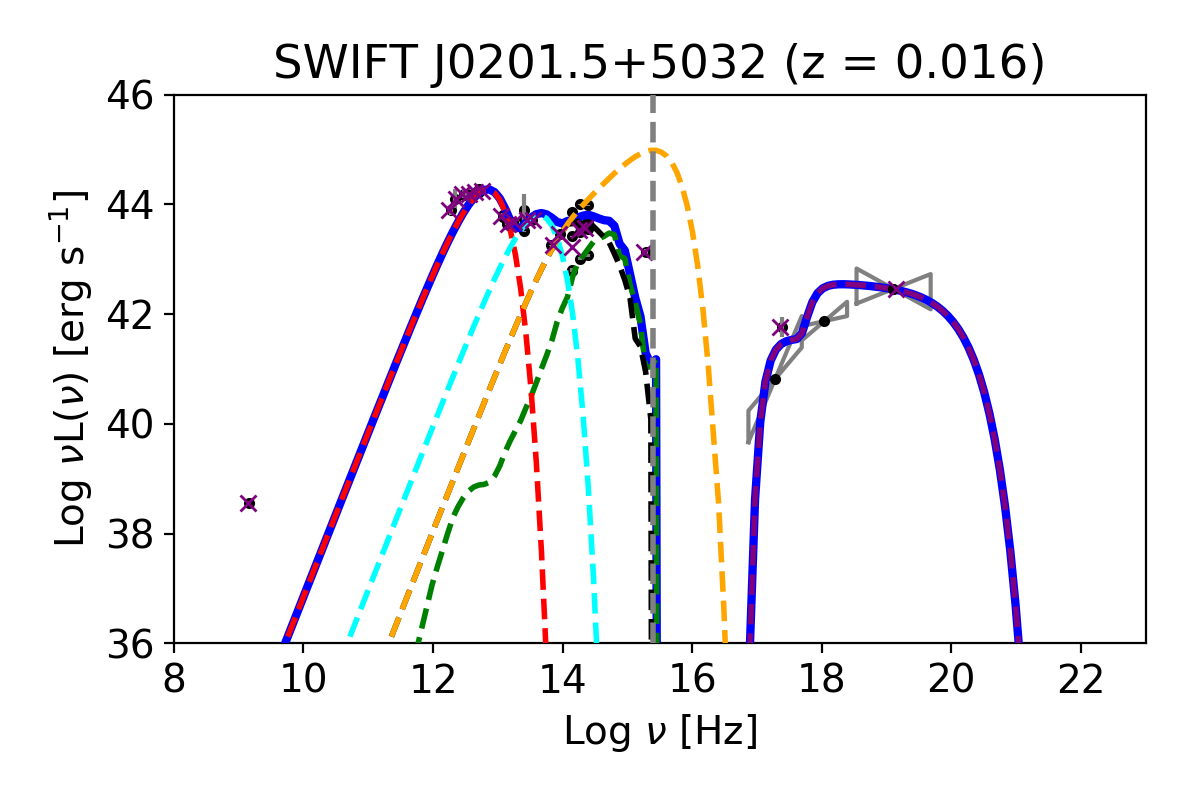}
			\includegraphics[width=.23\textwidth]{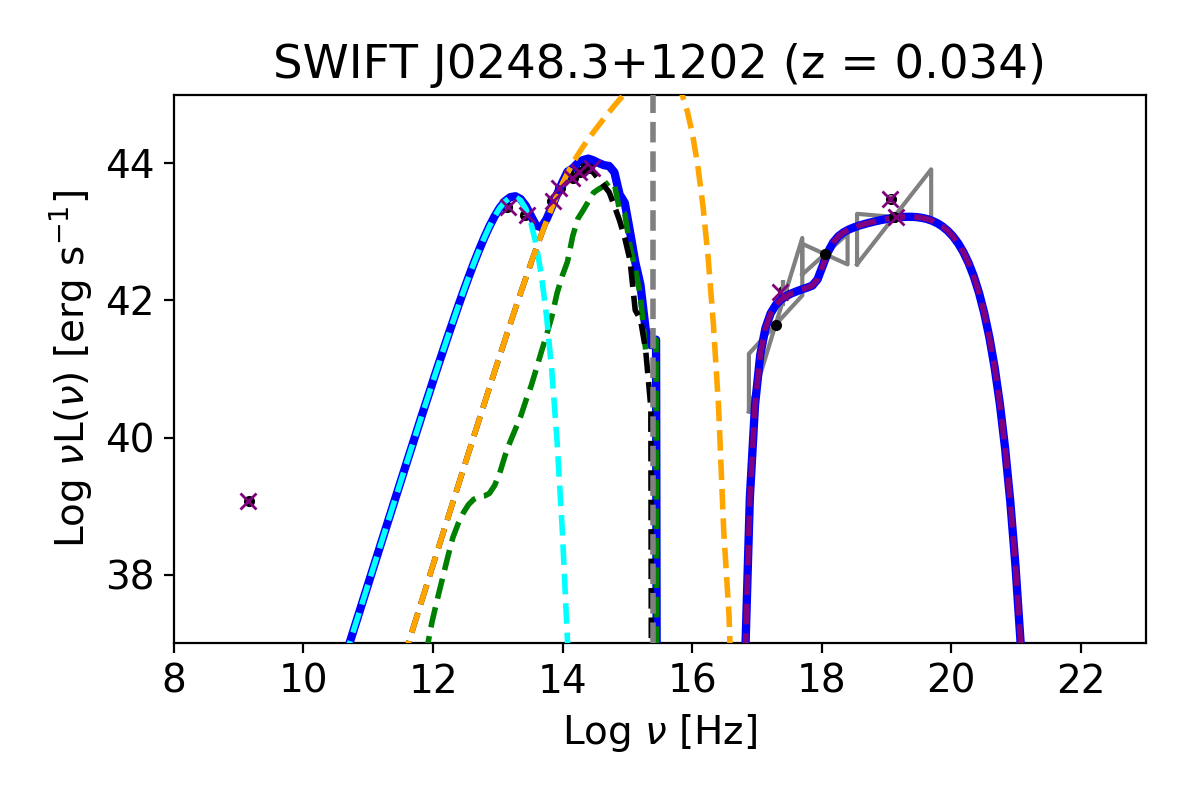}
			\includegraphics[width=.23\textwidth]{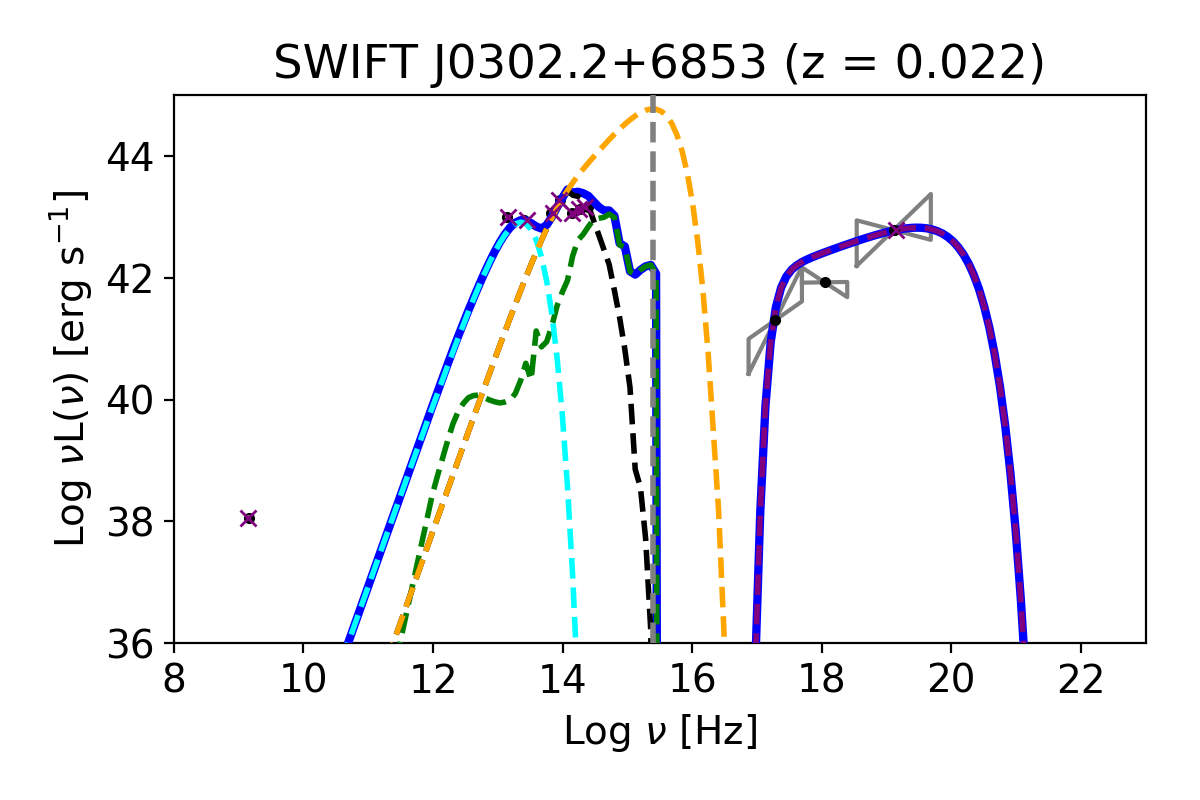}
			\includegraphics[width=.23\textwidth]{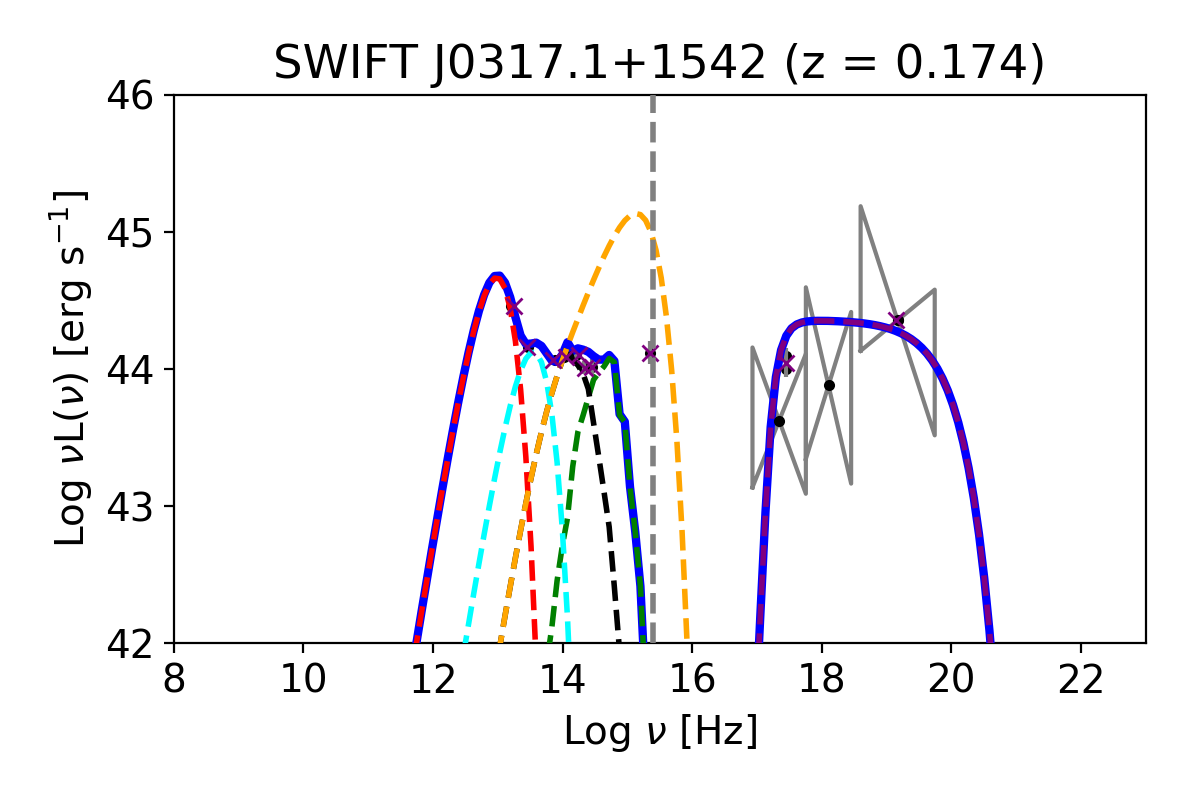}
		\end{minipage}
		\begin{minipage}{\textwidth}
			\centering
			\includegraphics[width=.23\textwidth]{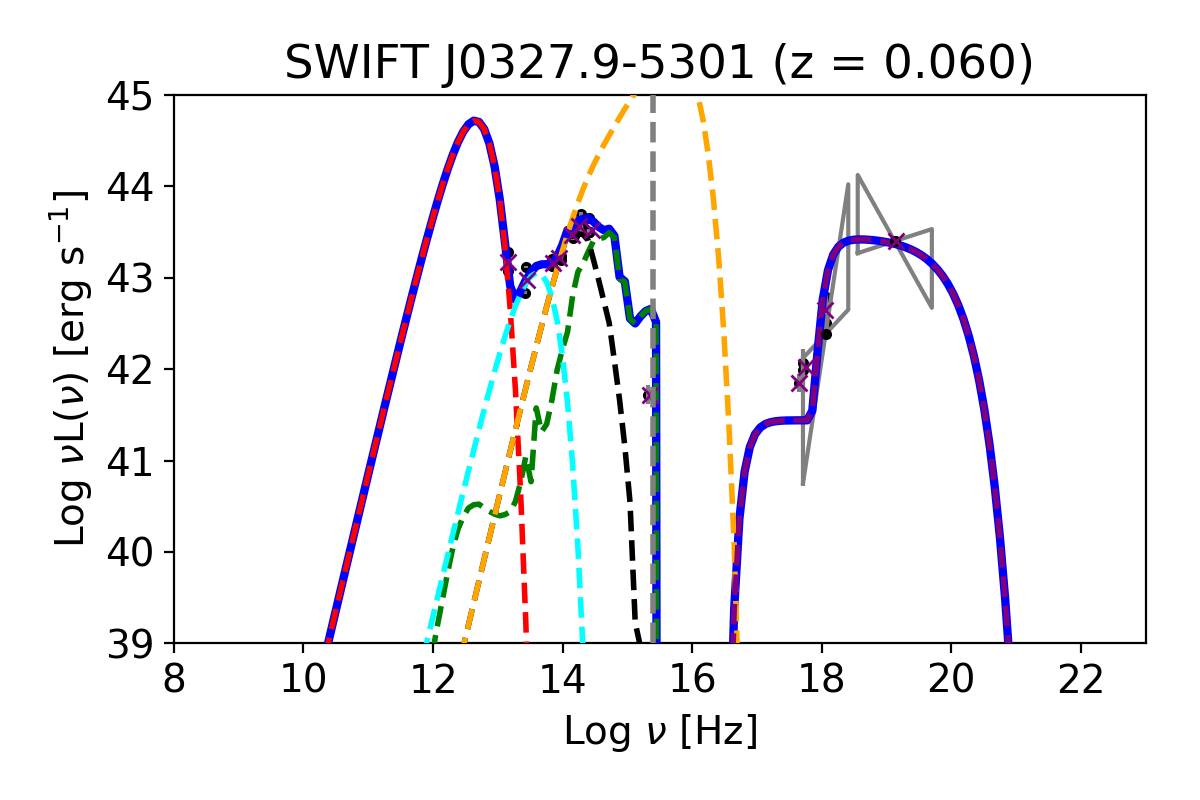}
			\includegraphics[width=.23\textwidth]{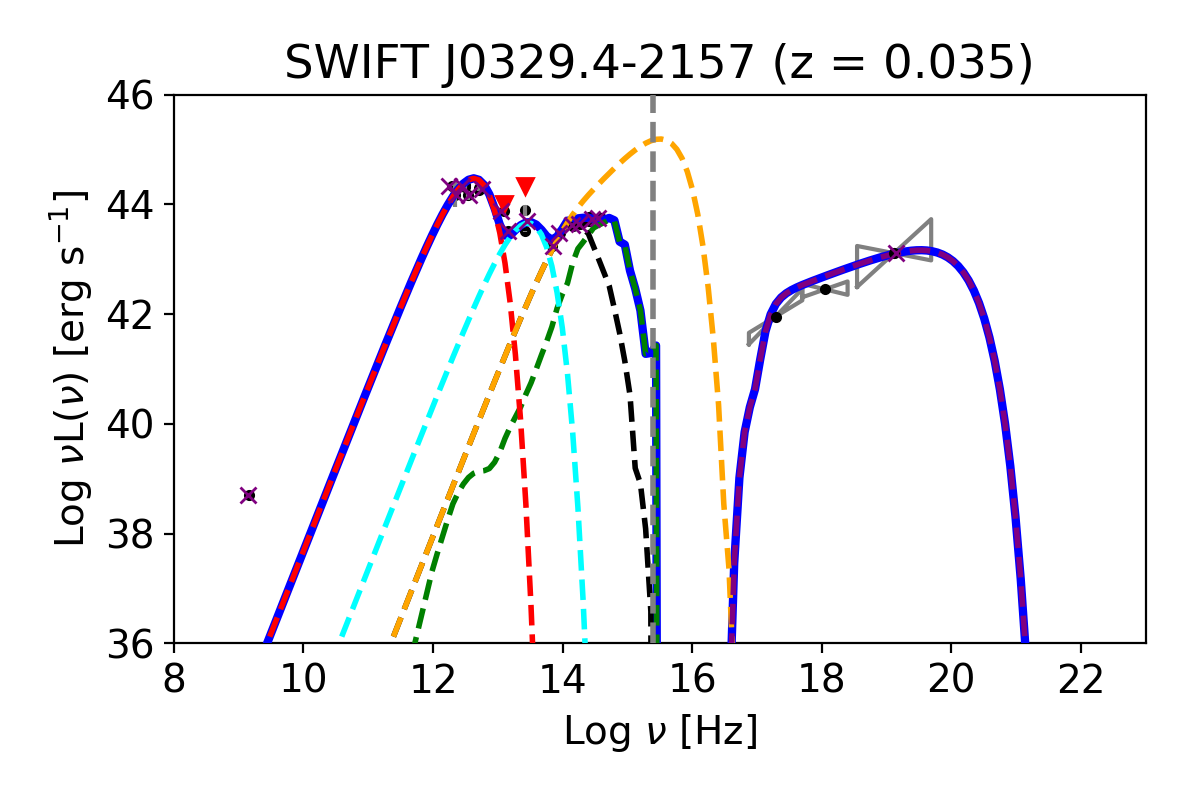}
			\includegraphics[width=.23\textwidth]{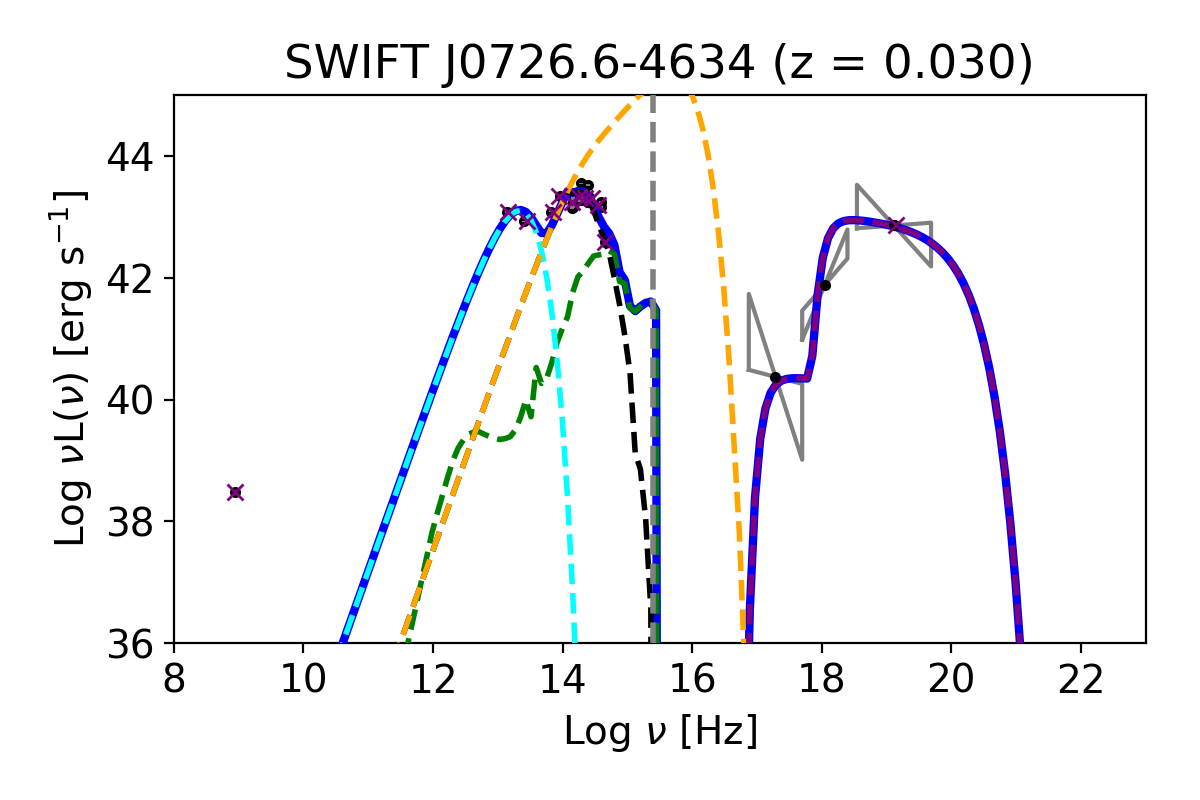}
			\includegraphics[width=.23\textwidth]{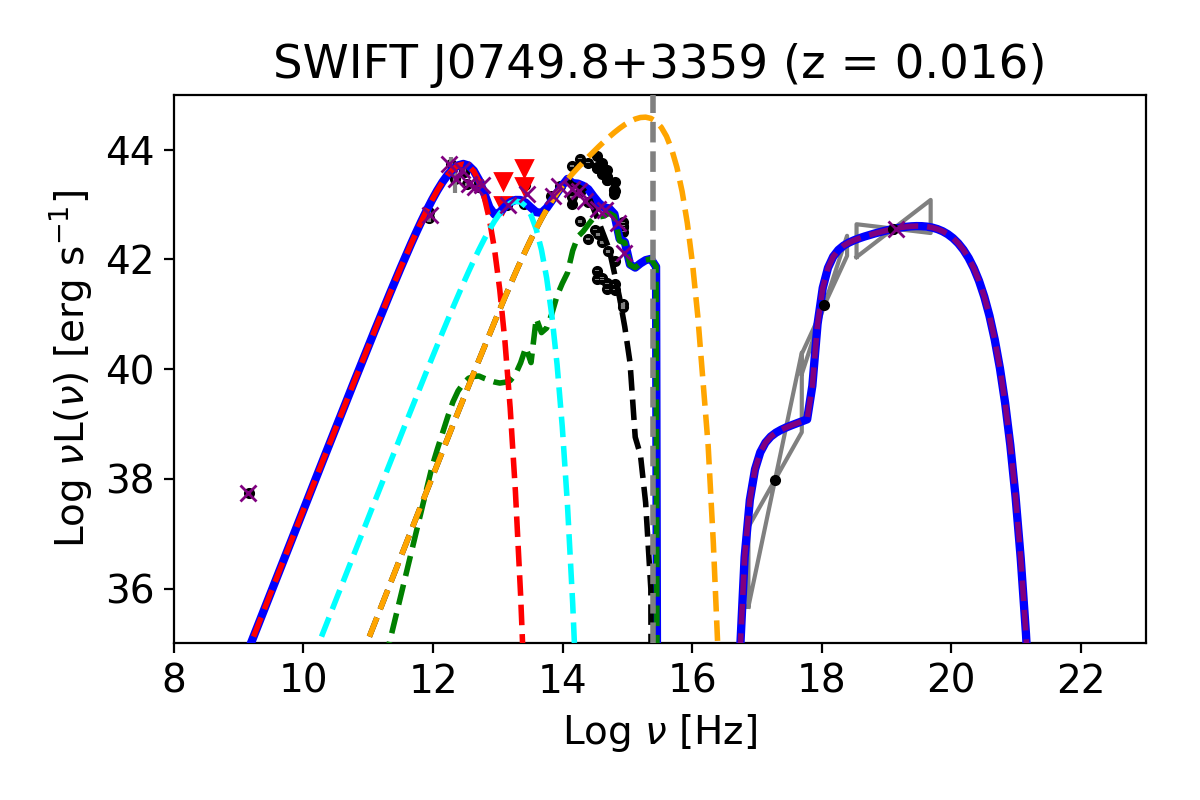}
		\end{minipage}
		\begin{minipage}{\textwidth}
			\centering
			\includegraphics[width=.23\textwidth]{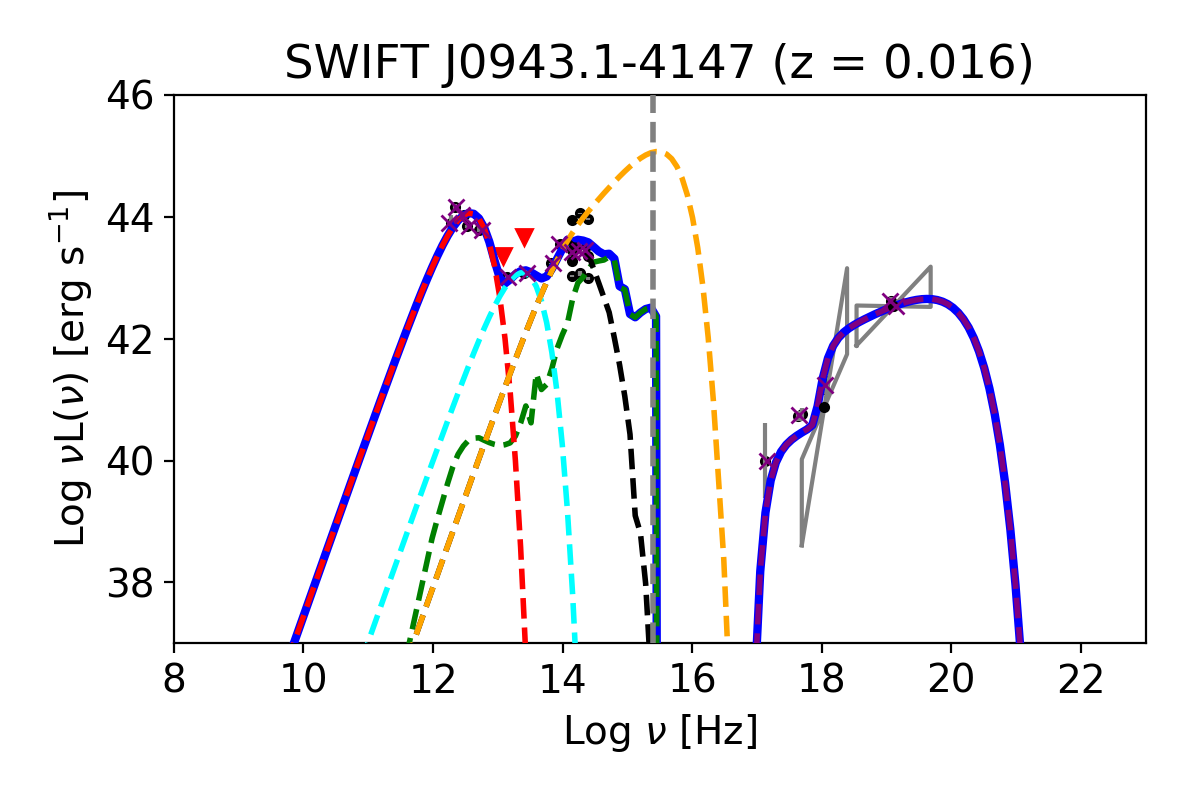}
			\includegraphics[width=.23\textwidth]{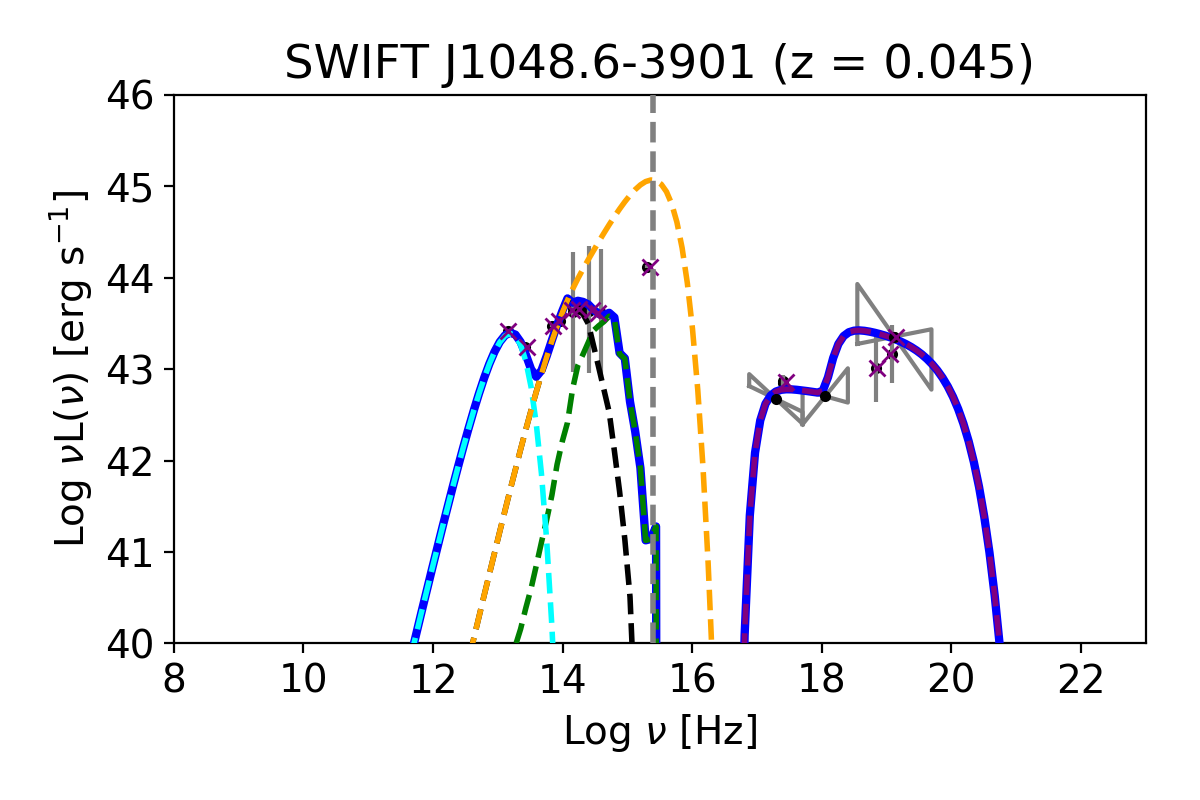}
			\includegraphics[width=.23\textwidth]{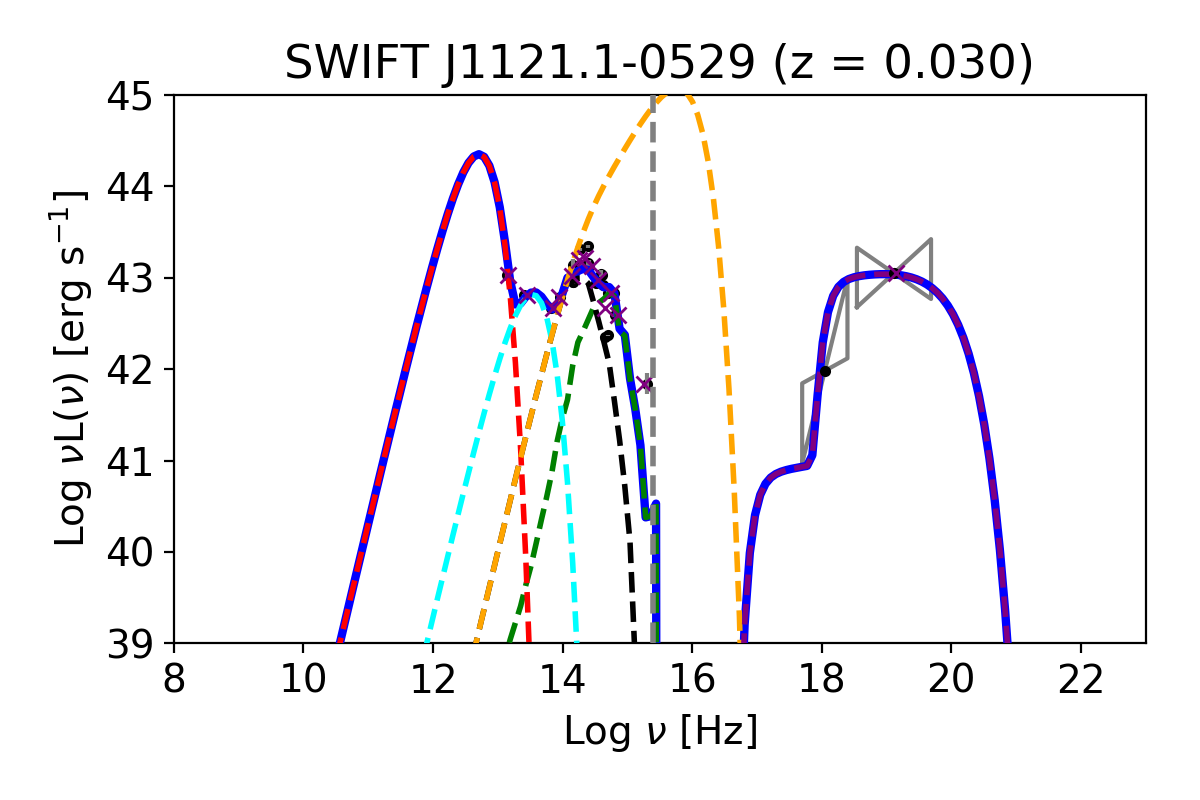}
			\includegraphics[width=.23\textwidth]{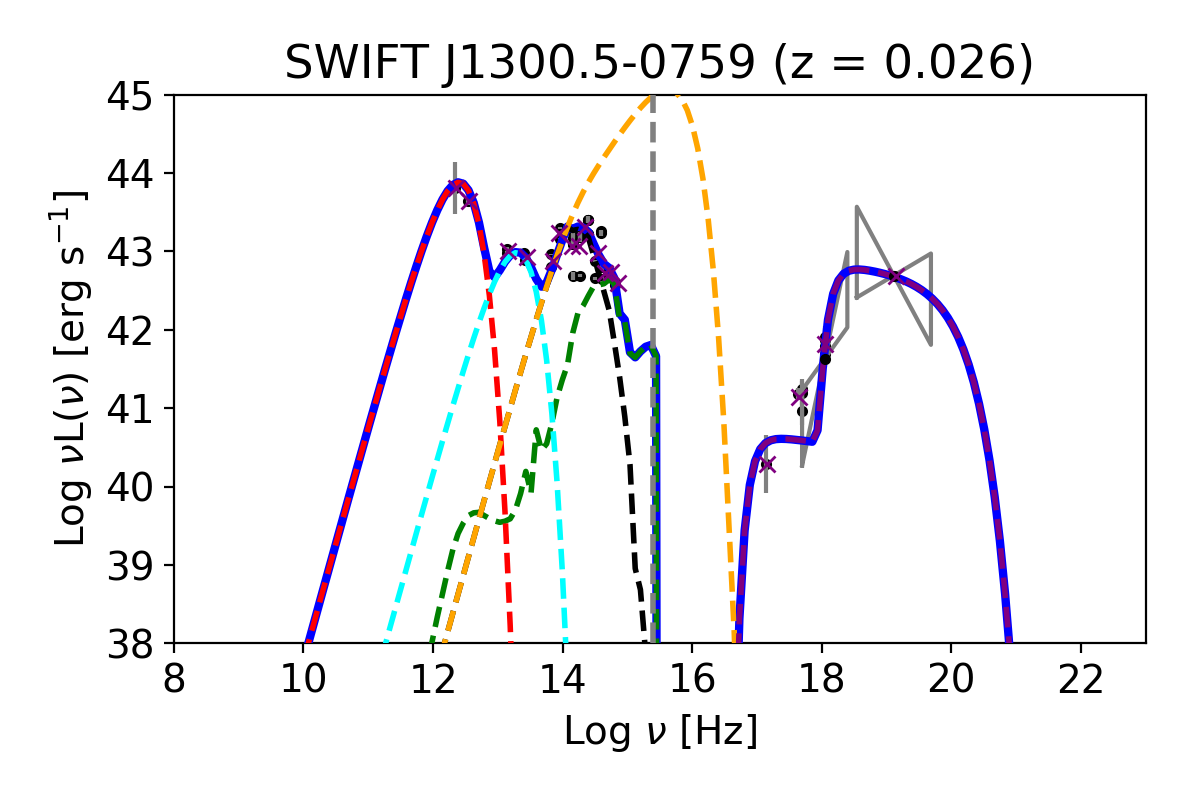}
		\end{minipage}
		\begin{minipage}{\textwidth}
			\centering
			\includegraphics[width=.23\textwidth]{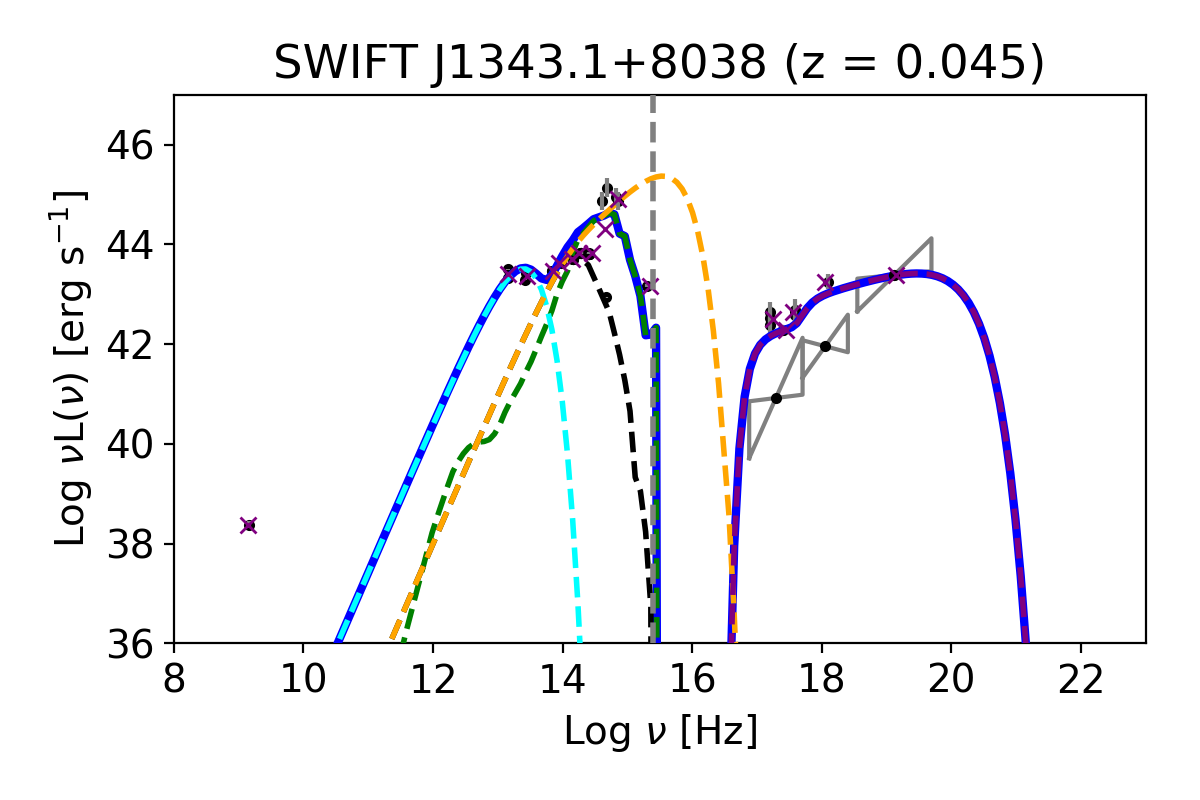}
			\includegraphics[width=.23\textwidth]{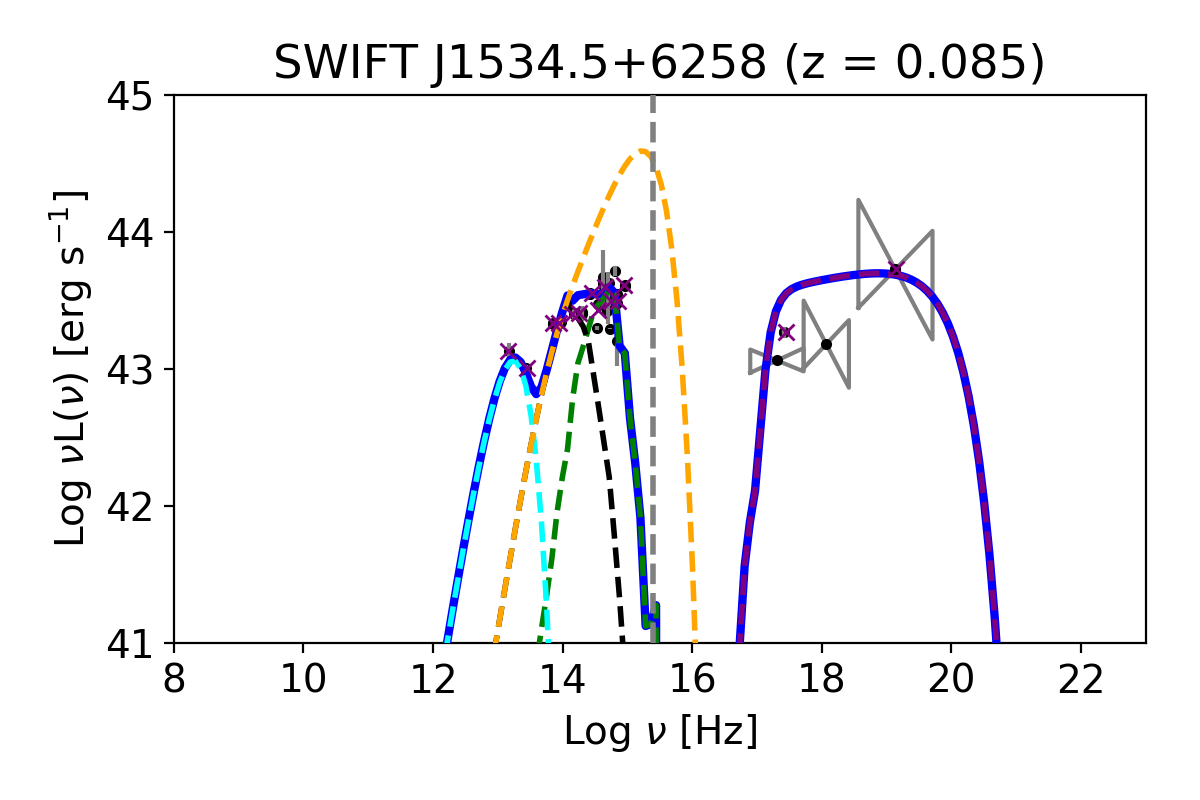}
			\includegraphics[width=.23\textwidth]{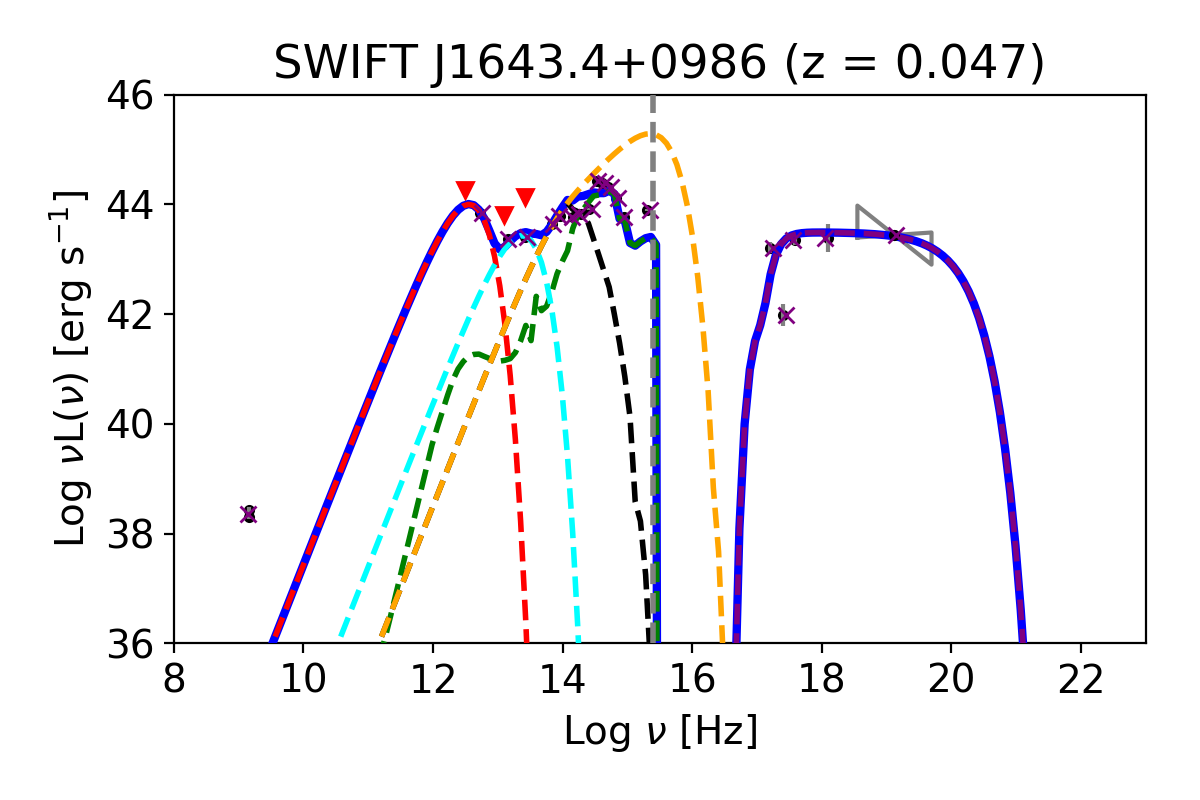}
			\includegraphics[width=.23\textwidth]{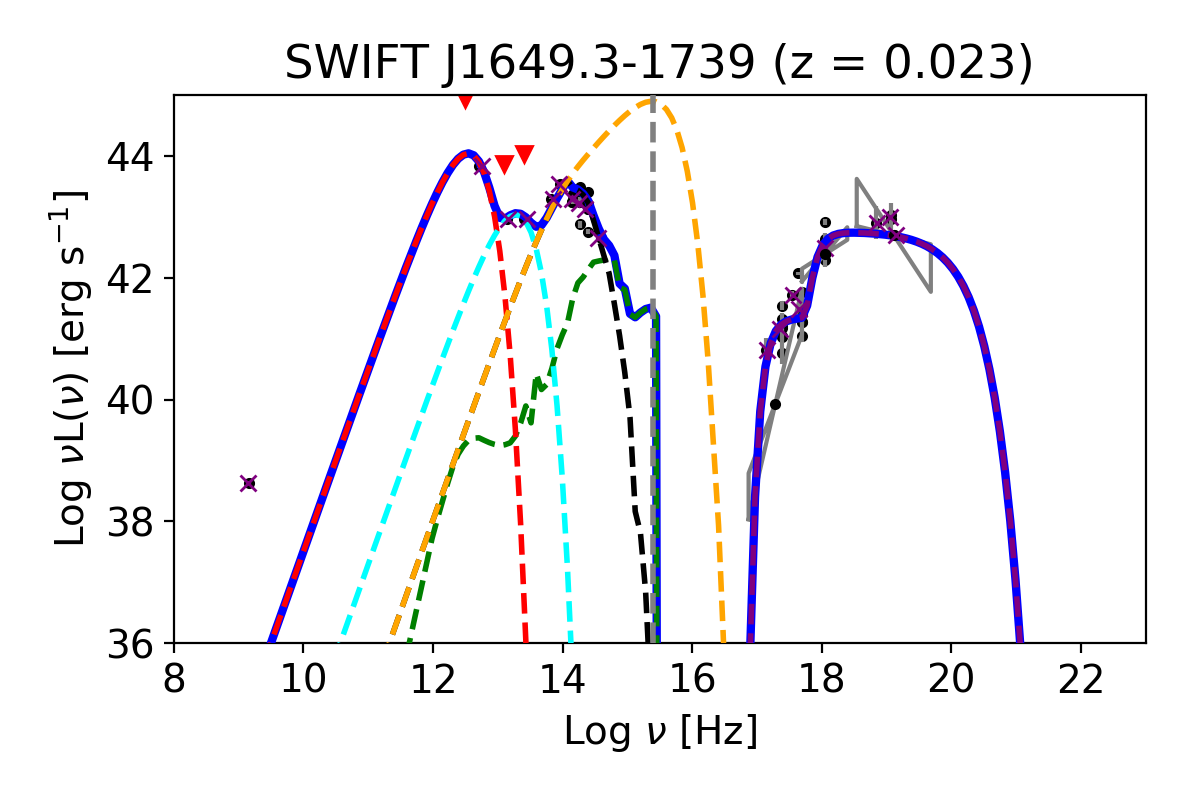}
		\end{minipage}
		\begin{minipage}{\textwidth}
			\centering
			\includegraphics[width=.23\textwidth]{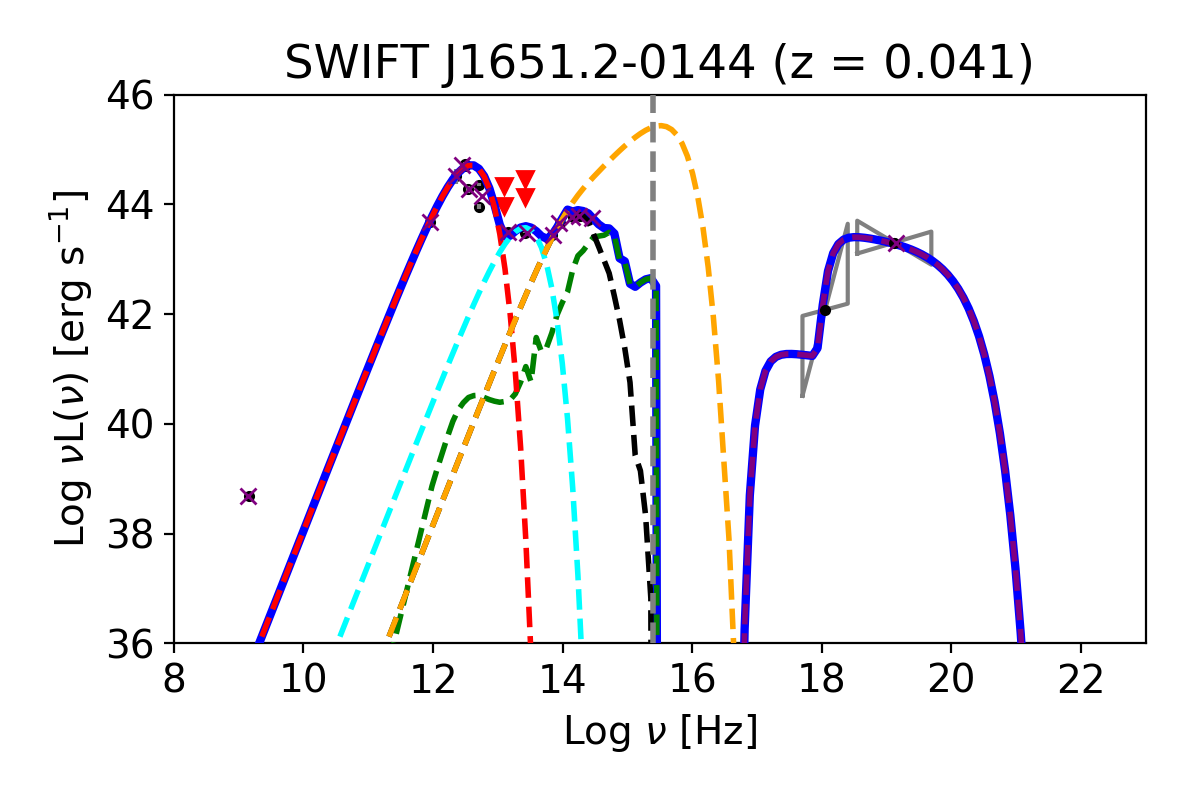}
			\includegraphics[width=.23\textwidth]{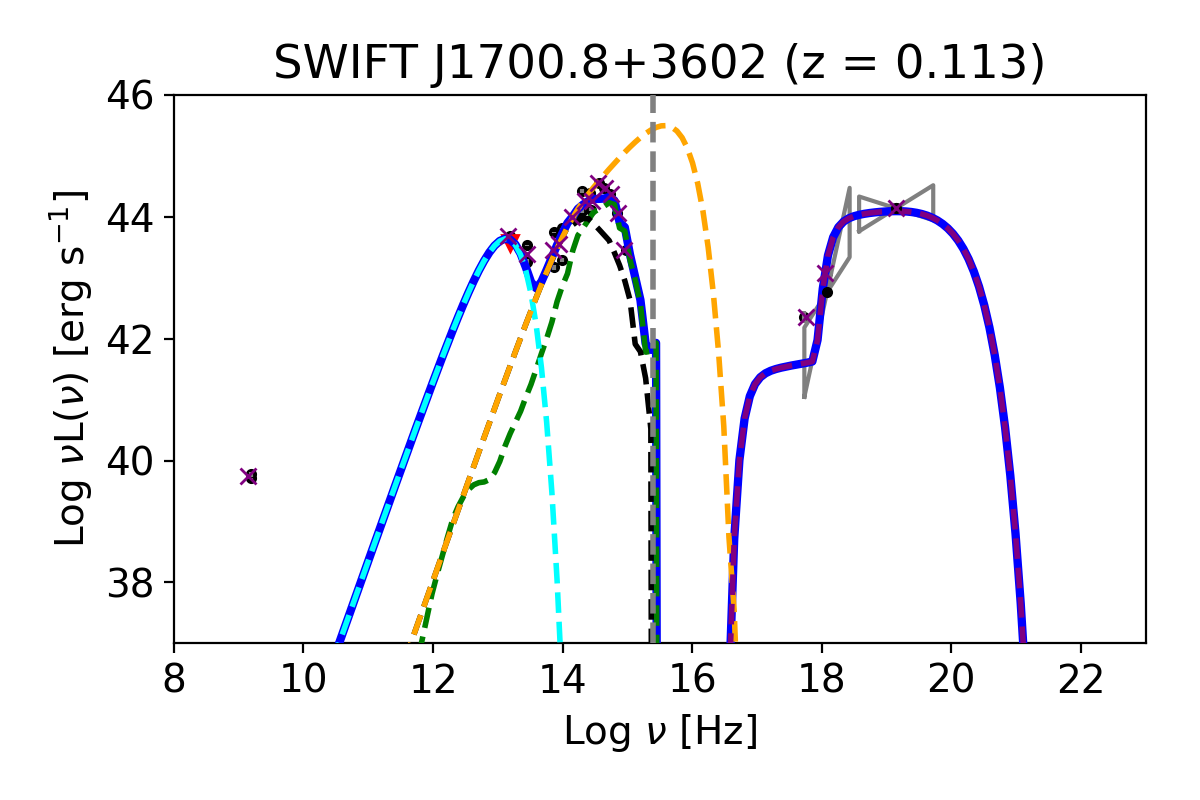}
			\includegraphics[width=.23\textwidth]{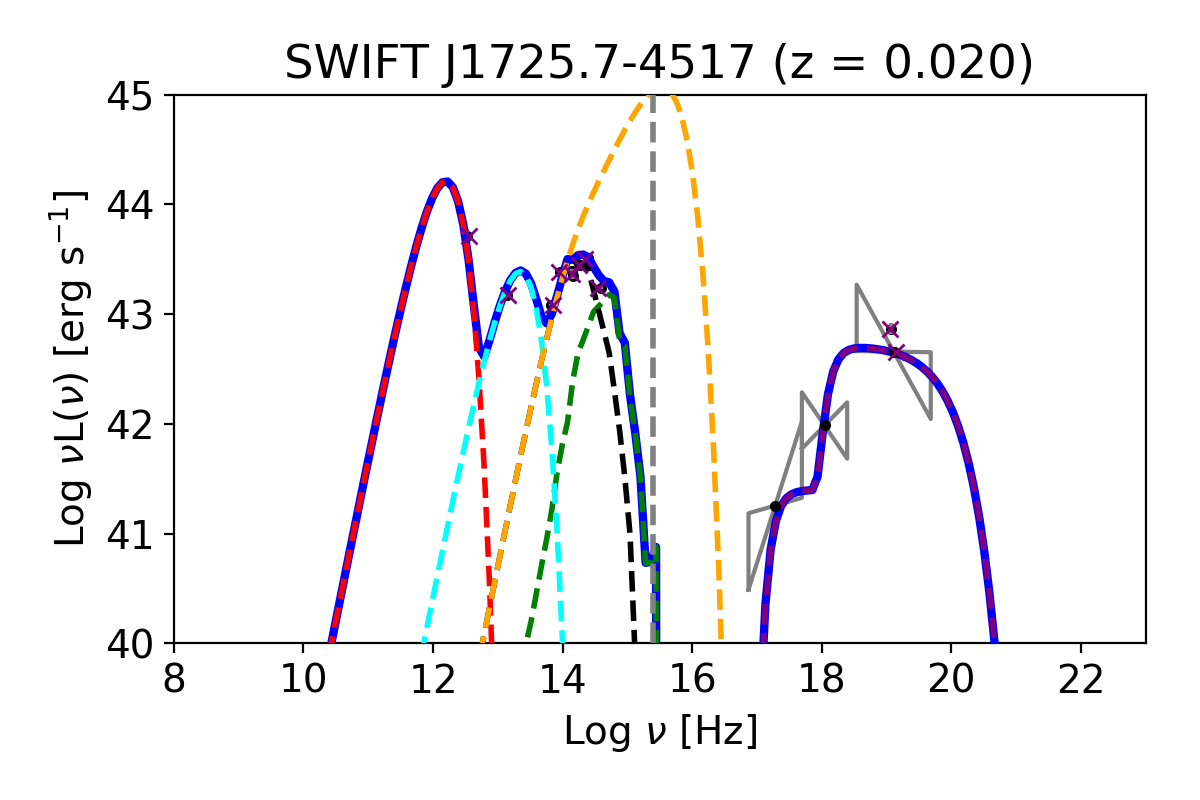}
			\includegraphics[width=.23\textwidth]{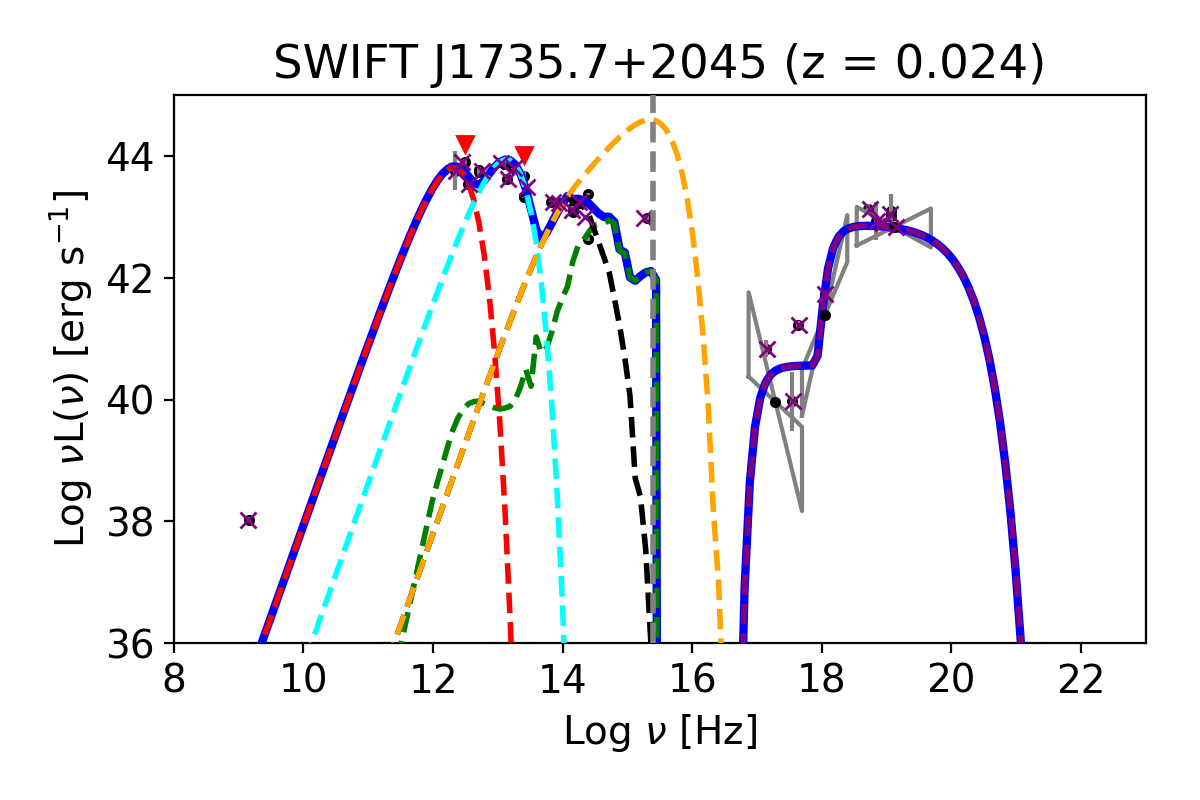}
		\end{minipage}
		\begin{minipage}{\textwidth}
			\centering
			\includegraphics[width=.23\textwidth]{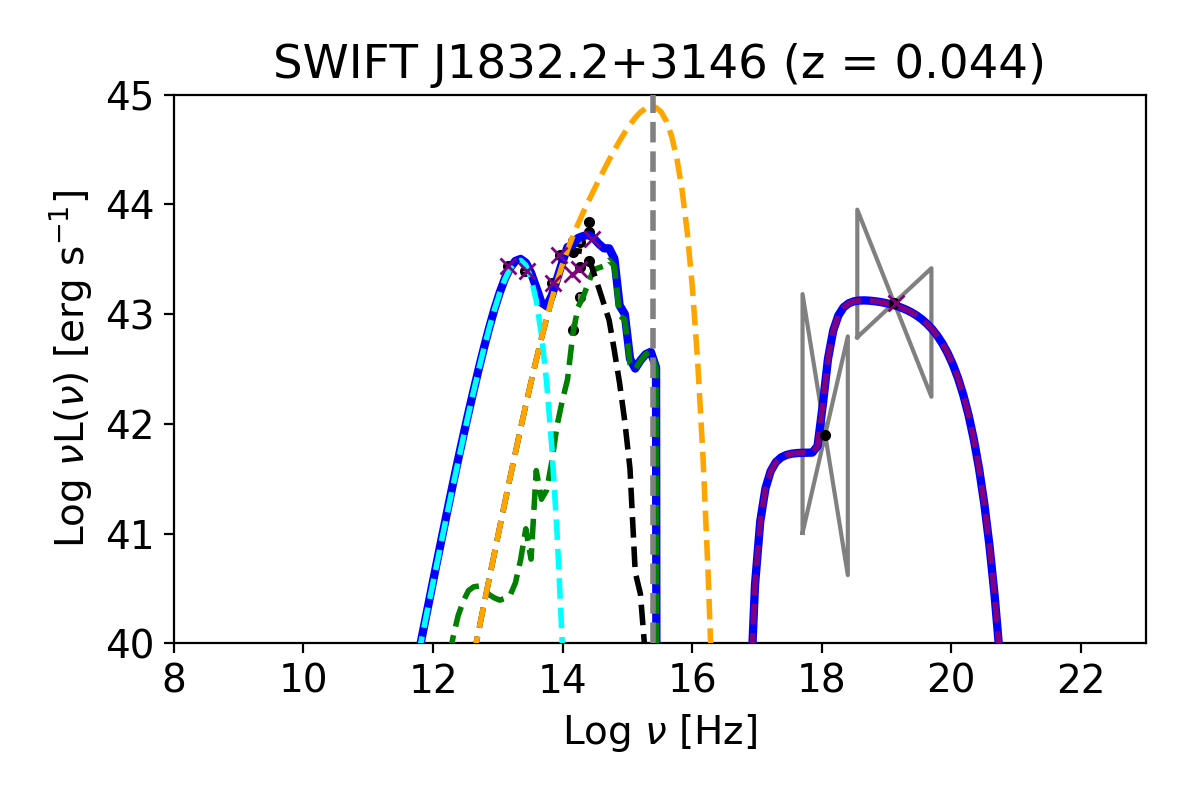}
			\includegraphics[width=.23\textwidth]{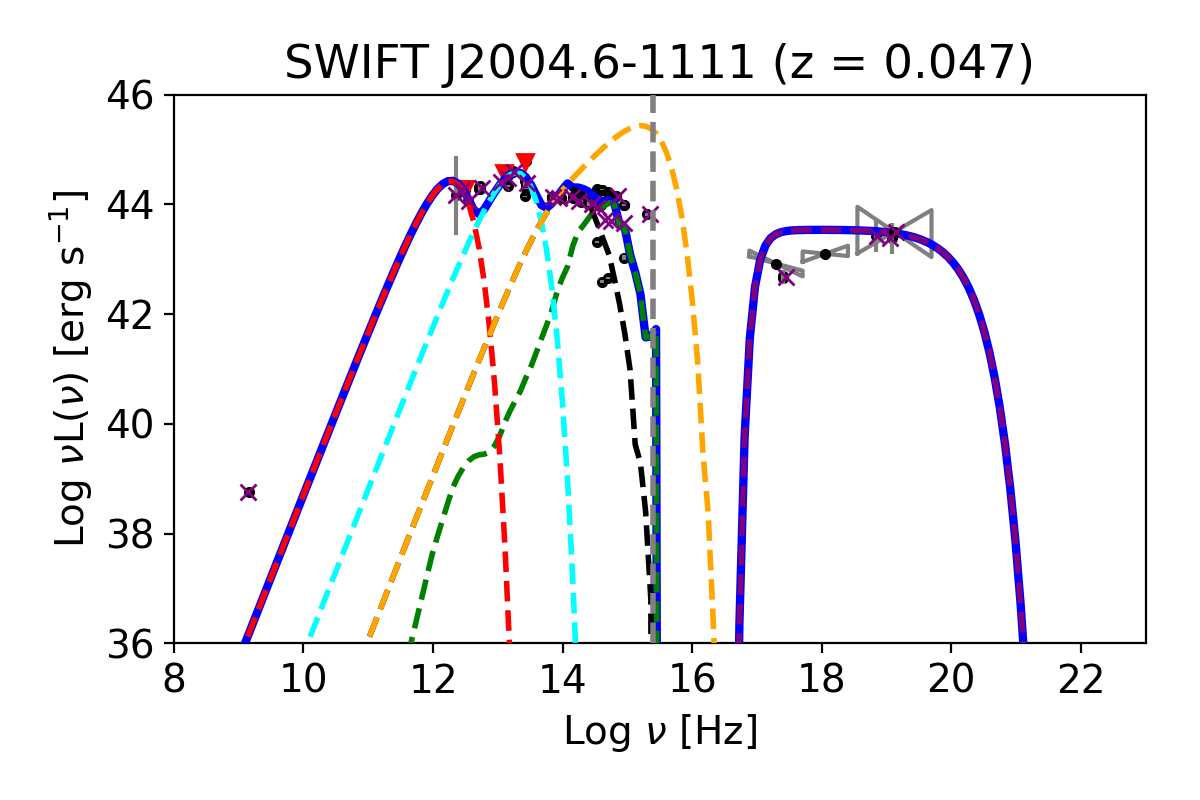}
			\includegraphics[width=.23\textwidth]{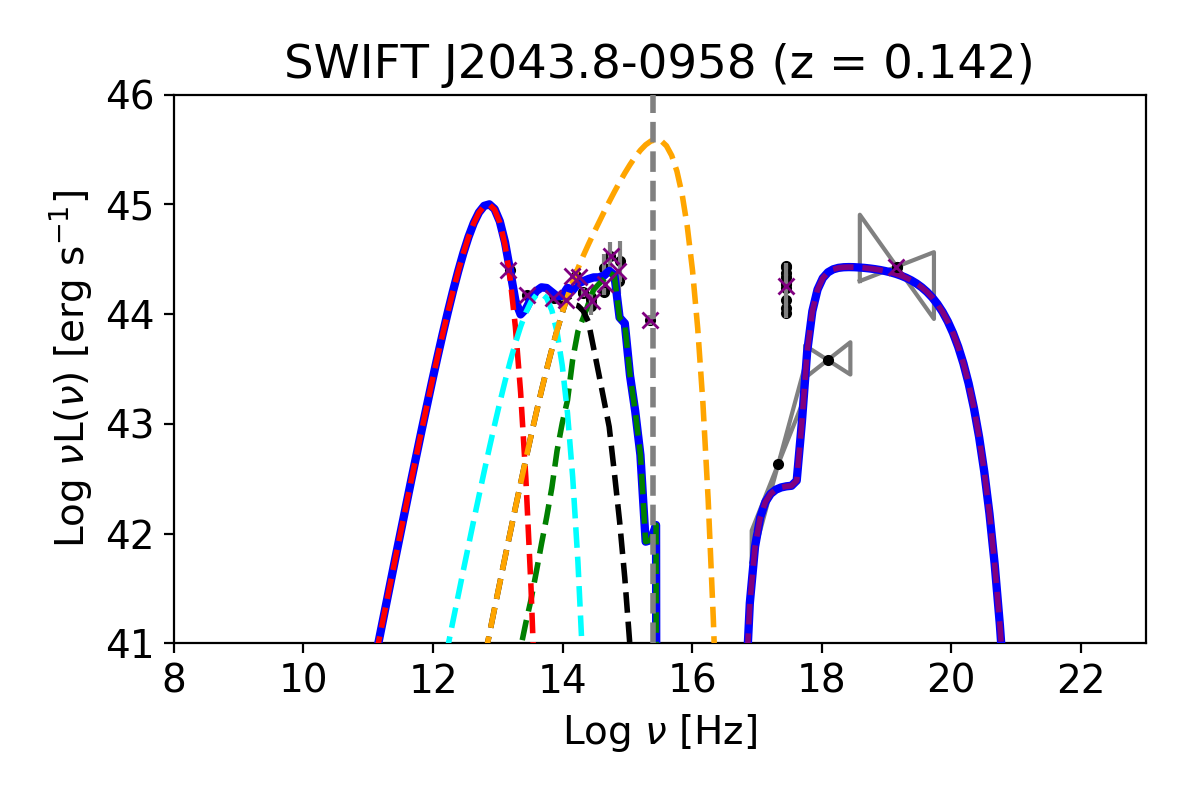}
			\includegraphics[width=.23\textwidth]{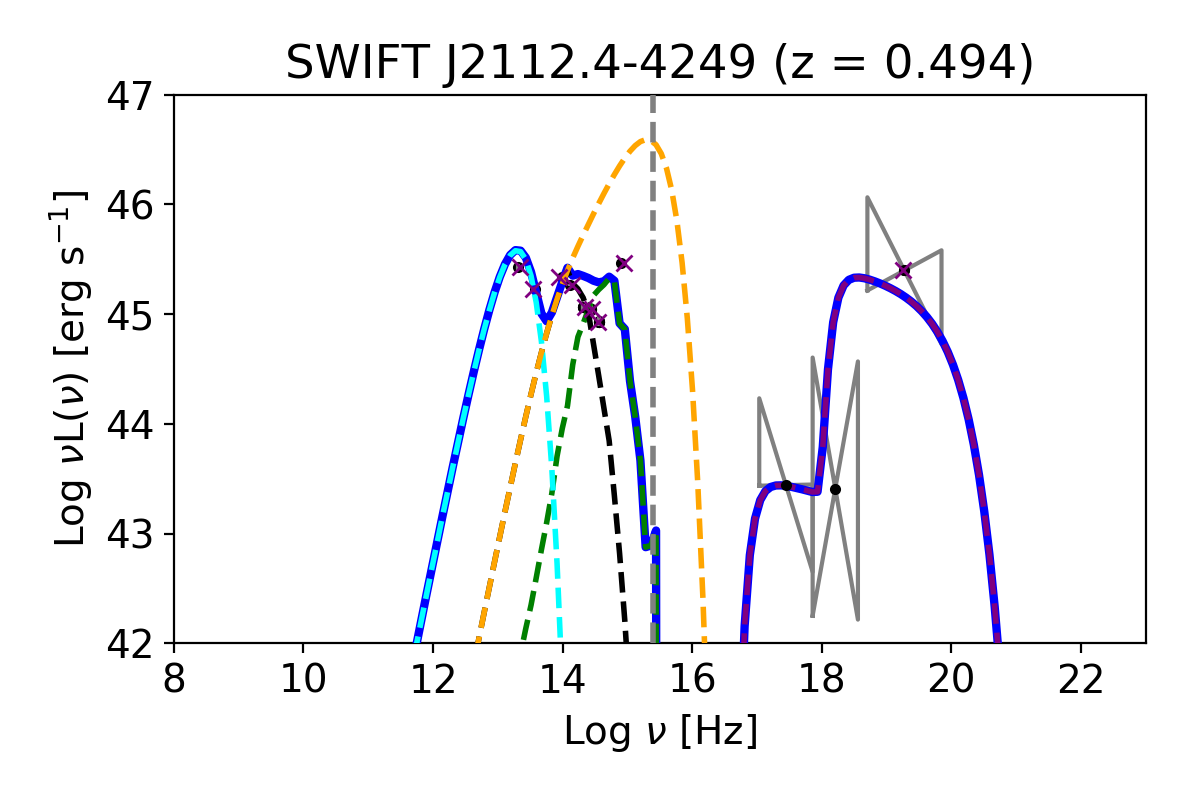}
		\end{minipage}
		\begin{minipage}{\textwidth}
			\centering
			\includegraphics[width=.23\textwidth]{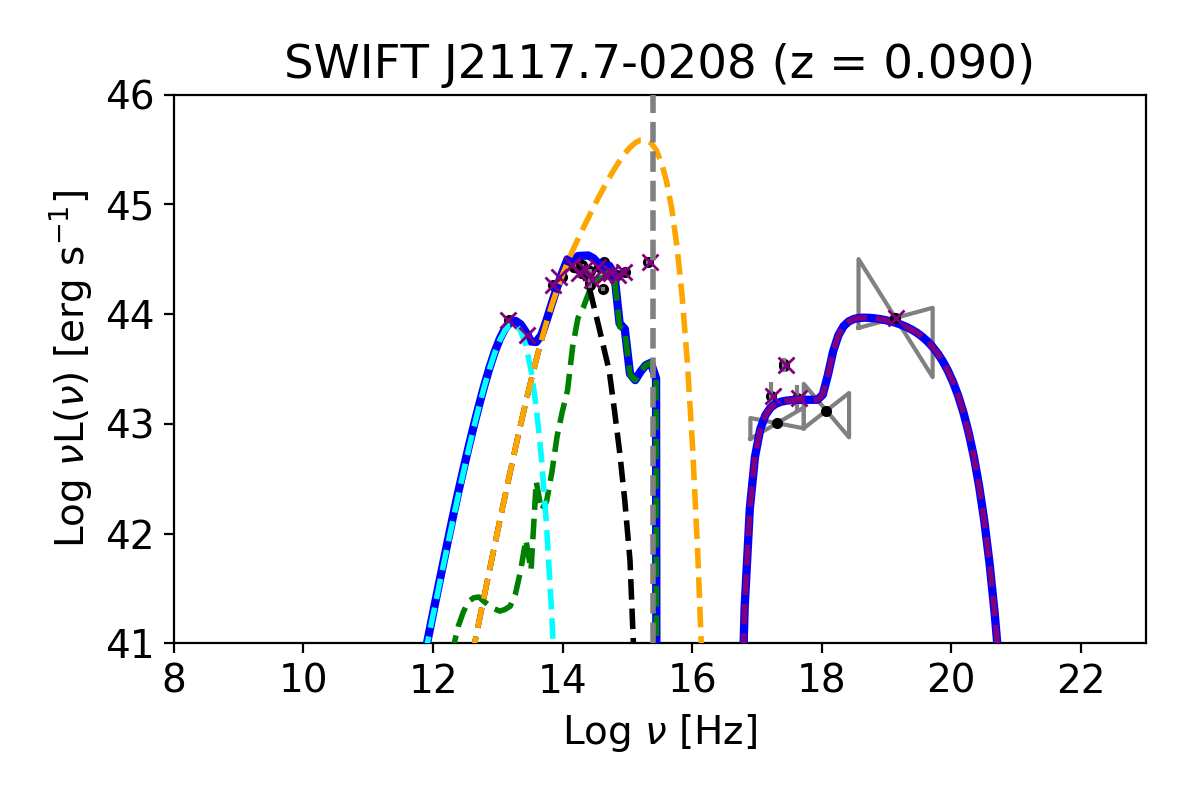}
			\includegraphics[width=.23\textwidth]{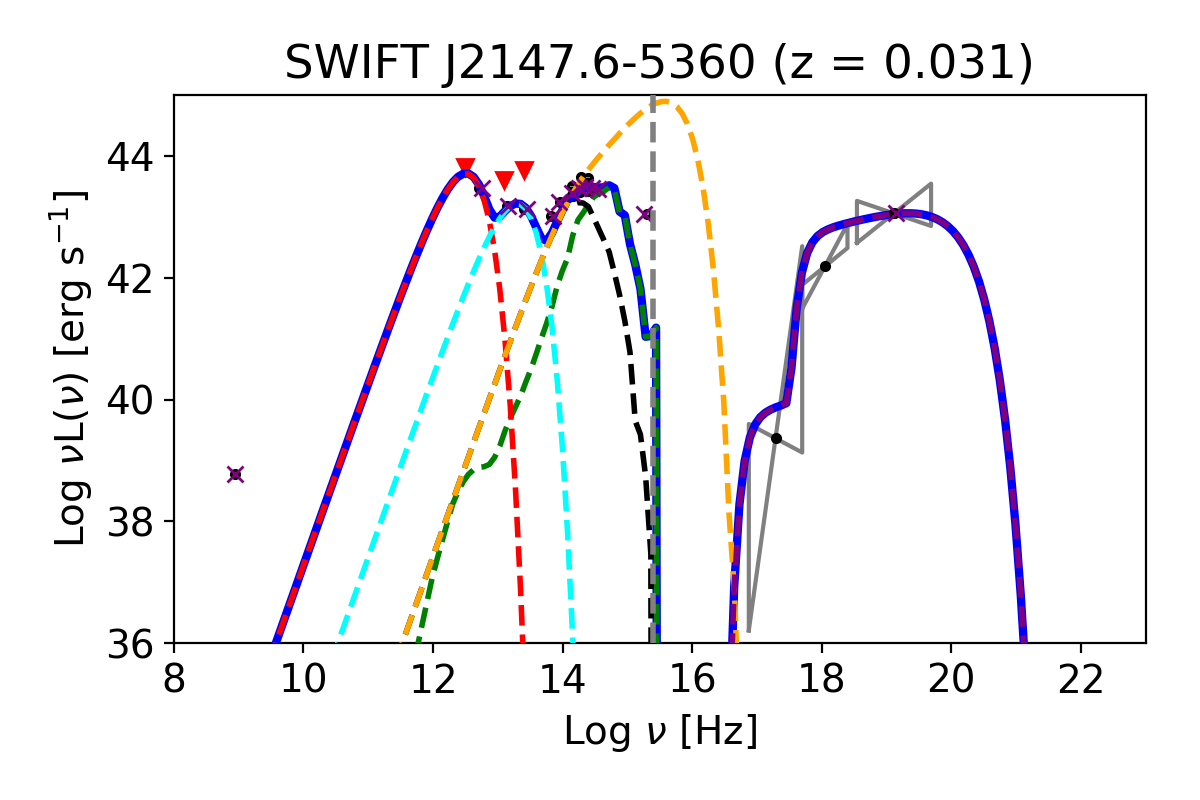}
			\includegraphics[width=.23\textwidth]{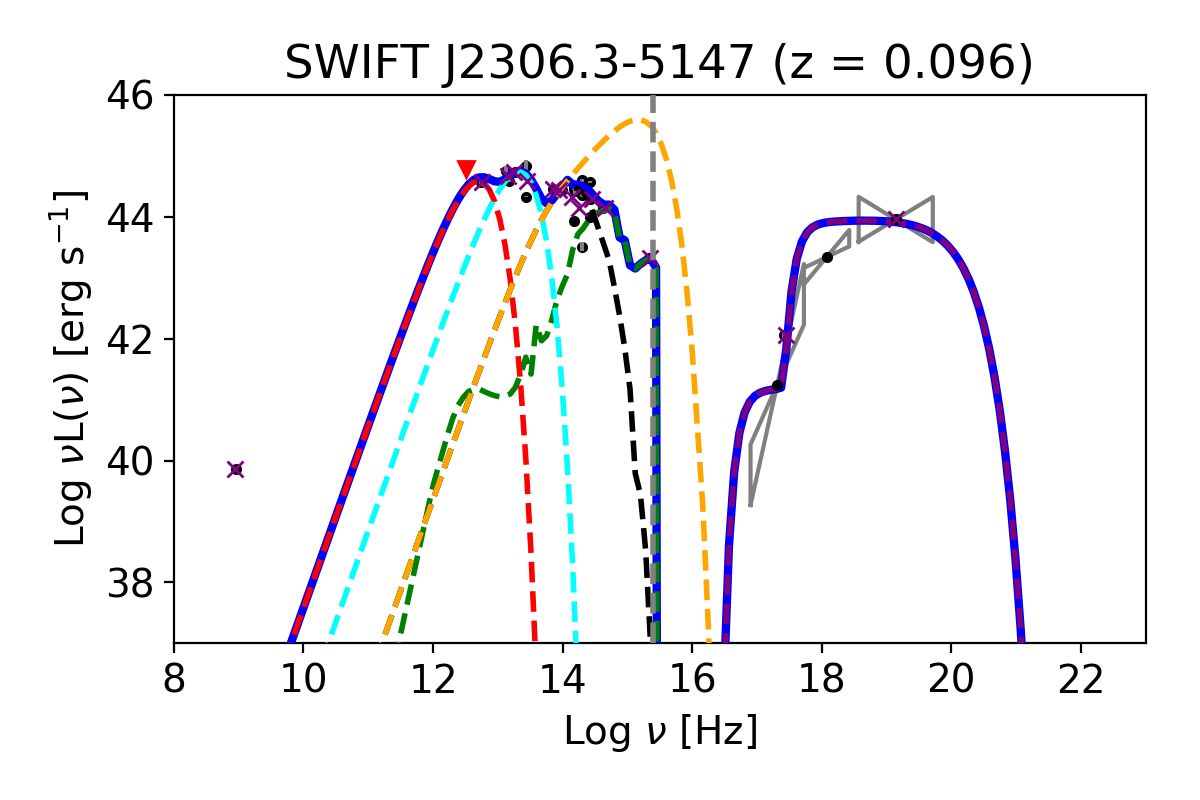}
		\end{minipage}
		\caption{Spectral energy distribution fit for obscured AGN sources.
			A complete description is given in  Figure \ref{es_obs}.
		}\label{apptest}
	\end{figure*}

	Then we applied the obscured AGN model to the remaining 31 unknown sources, which from here on we consider the main subject of our discussion, and we finally found good agreement between model and data.
	In \figurename~\ref{apptest} we show the SED fit for the source SWIFT J1649.3-1739 with all the emitting components labeled.
	We note that without a blazar-like jet contribution the radio emission at $\nu\sim 10^9$ Hz can be explained with emissions of stellar origin (e.g., HII regions and supernovae) from the host galaxy up to $\nu L(\nu)\sim10^{39-40}$ erg s$^{-1}$ \citep[e.g., see][]{normal:radio}, therefore in our model we did not take this data point into account.
	The model fits of all the 31 sources are shown in \figurename~\ref{es_obs}, while in \tablename~\ref{obs_fit_res} the fit parameters that we obtained are provided.
	We note that both $L_X$ and $\Gamma_{BAT}$ have average values compared to the non-blazar sources in the BAT AGN catalog \citep[see][]{bat:xprop}.
	
	\begin{table*}[tb]
		\centering
		\resizebox{\textwidth}{!}{
			\begin{tabular}{lrrrrrrrrrrrrr}
				\toprule
				Source & $\dot{m}$ & $M$ & $l_{Edd}$ & $R_{DB}$ & $\theta_{DB}$ & $R_{T}$ & $\theta_{T}$ & $f$ & $N_H$ & $N_H^{gal}$ & $\gamma$ & $G$ & $A_B$\\ 
				\midrule
				SWIFT J0000.5+3251 & 0.35 & 3.00E+08 & 0.043 & 97.21 & 1.10 & 3.24 & 1.40 & 0.010 & 8.00E+24 & 3.00E+21 & 0.98 & 44.10 &  7.000\\
				SWIFT J0043.9-5009 & 0.45 & 5.00E+08 & 0.033 & 226.83 & 0.95 & 1.94 & 1.30 & 0.010 & 1.00E+24 & 2.00E+21 & 1.20 & 43.45 &  7.000\\
				SWIFT J0120.7-1444 & 0.60 & 2.50E+08 & 0.088 & 648.09 & 1.05 & 3.24 & 1.40 & 0.003 & 5.00E+24 & 7.00E+21 & 0.80 & 43.55 &  7.000\\
				SWIFT J0155.2-3048 & 0.40 & 5.00E+08 & 0.029 &  - & - & 6.48 & 1.00 & 0.010 & 5.00E+24 & 5.00E+21 & 1.00 & 43.45 &  6.000\\
				SWIFT J0201.5+5032 & 0.25 & 2.50E+08 & 0.037 & 48.61 & 0.80 & 0.97 & 1.10 & 0.100 & 5.00E+23 & 4.00E+21 & 1.05 & 43.45 &  4.000\\
				SWIFT J0248.3+1202 & 0.60 & 2.50E+08 & 0.088 &  - & - & 6.48 & 1.35 & 0.260 & 3.00E+24 & 2.00E+21 & 0.70 & 43.70 &  4.000\\
				SWIFT J0302.2+6853 & 0.15 & 2.00E+08 & 0.027 &  - & - & 1.94 & 1.35 & 0.040 & 3.00E+21 & 3.00E+21 & 0.60 & 43.00 &  7.000\\
				SWIFT J0317.1+1542 & 0.35 & 1.00E+09 & 0.013 & 32.40 & 0.60 & 1.94 & 1.00 & 0.040 & 4.00E+21 & 4.00E+21 & 1.00 & 44.05 &  7.000\\
				SWIFT J0327.9-5301 & 0.70 & 1.00E+08 & 0.256 & 194.43 & 0.80 & 0.97 & 1.45 & 0.010 & 3.00E+24 & 2.00E+22 & 1.00 & 43.45 &  7.000\\
				SWIFT J0329.4-2157 & 0.40 & 2.00E+08 & 0.073 & 162.02 & 0.80 & 2.27 & 1.25 & 0.040 & 2.00E+21 & 2.00E+21 & 0.55 & 43.70 &  7.000\\
				SWIFT J0726.6-4634 & 0.50 & 1.00E+08 & 0.183 &  - & - & 3.24 & 1.42 & 0.002 & 3.00E+24 & 4.00E+21 & 1.10 & 42.40 &  7.000\\
				SWIFT J0749.8+3359 & 0.10 & 3.00E+08 & 0.012 & 162.02 & 0.90 & 2.59 & 1.25 & 0.001 & 3.00E+24 & 7.00E+21 & 0.59 & 42.80 &  7.000\\
				SWIFT J0943.1-4147 & 0.30 & 2.00E+08 & 0.055 & 162.02 & 1.00 & 2.27 & 1.38 & 0.050 & 3.00E+24 & 2.00E+21 & 0.43 & 43.30 &  7.000\\
				SWIFT J1048.6-3901 & 0.30 & 3.00E+08 & 0.037 &  - & - & 6.48 & 1.30 & 0.200 & 8.00E+24 & 2.00E+21 & 1.10 & 43.55 &  7.000\\
				SWIFT J1121.1-0529 & 0.30 & 5.00E+07 & 0.220 & 97.21 & 0.80 & 0.97 & 1.43 & 0.010 & 3.00E+24 & 5.00E+21 & 0.88 & 42.80 &  7.000\\
				SWIFT J1300.5-0759 & 0.30 & 1.00E+08 & 0.110 & 324.04 & 1.10 & 3.24 & 1.40 & 0.005 & 6.00E+24 & 6.00E+21 & 1.10 & 42.60 &  7.000\\
				SWIFT J1343.1+8038 & 0.60 & 2.00E+08 & 0.110 &  - & - & 3.24 & 1.35 & 0.500 & 2.00E+23 & 2.00E+21 & 0.60 & 44.60 &  7.000\\
				SWIFT J1534.5+6258 & 0.10 & 3.50E+08 & 0.010 &  - & - & 3.24 & 1.25 & 0.050 & 2.00E+21 & 2.00E+21 & 0.90 & 43.55 &  7.000\\
				SWIFT J1643.4+0986 & 0.50 & 4.50E+08 & 0.041 & 194.43 & 1.15 & 3.24 & 1.35 & 0.040 & 5.00E+21 & 5.00E+21 & 1.00 & 44.20 &  8.000\\
				SWIFT J1649.3-1739 & 0.20 & 2.50E+08 & 0.029 & 162.02 & 0.90 & 3.24 & 1.35 & 0.040 & 1.00E+24 & 2.00E+21 & 1.00 & 42.30 &  8.000\\
				SWIFT J1651.2-0144 & 0.70 & 2.50E+08 & 0.103 & 259.24 & 0.80 & 3.24 & 1.35 & 0.005 & 6.00E+24 & 8.00E+21 & 1.15 & 43.45 &  7.000\\
				SWIFT J1700.8+3602 & 0.80 & 2.00E+08 & 0.147 &  - & - & 12.96 & 1.35 & 0.005 & 5.00E+24 & 8.00E+21 & 0.81 & 44.20 &  4.000\\
				SWIFT J1725.7-4517 & 0.30 & 1.50E+08 & 0.073 & 972.13 & 0.90 & 3.24 & 1.30 & 0.050 & 4.00E+24 & 4.00E+21 & 1.00 & 43.15 &  6.000\\
				SWIFT J1735.7+2045 & 0.10 & 2.00E+08 & 0.018 & 324.04 & 0.85 & 8.10 & 0.75 & 0.005 & 6.00E+24 & 5.00E+21 & 0.98 & 42.90 &  7.000\\
				SWIFT J1832.2+3146 & 0.20 & 2.50E+08 & 0.029 &  - & - & 3.24 & 1.20 & 0.040 & 5.00E+24 & 2.00E+22 & 1.00 & 43.45 &  5.000\\
				SWIFT J2004.6-1111 & 0.70 & 1.00E+09 & 0.026 & 972.13 & 1.00 & 9.72 & 0.90 & 0.200 & 5.00E+19 & 5.00E+21 & 0.99 & 44.00 &  7.000\\
				SWIFT J2043.8-0958 & 1.00 & 4.00E+08 & 0.092 & 97.21 & 0.70 & 1.62 & 1.20 & 0.010 & 5.00E+23 & 2.00E+21 & 1.00 & 44.35 &  7.000\\
				SWIFT J2112.4-4249 & 10.00 & 2.50E+09 & 0.147 &  - & - & 32.40 & 1.00 & 0.008 & 8.00E+24 & 2.00E+21 & 1.15 & 45.30 &  8.000\\
				SWIFT J2117.7-0208 & 1.00 & 1.00E+09 & 0.037 &  - & - & 9.72 & 1.30 & 0.200 & 7.00E+24 & 2.00E+21 & 1.00 & 44.35 &  6.000\\
				SWIFT J2147.6-5360 & 0.20 & 1.00E+08 & 0.073 & 162.02 & 1.10 & 3.24 & 1.30 & 0.002 & 2.00E+23 & 2.00E+21 & 0.71 & 43.45 &  6.000\\
				SWIFT J2306.3-5147 & 1.00 & 1.50E+09 & 0.024 & 162.02 & 1.00 & 9.72 & 0.90 & 0.002 & 1.00E+23 & 2.00E+21 & 0.94 & 44.10 &  7.000\\
				\bottomrule
		\end{tabular}}
		\caption{Fit parameters for unknown AGNs (obscured AGN emission model).
			The meanings of the parameters are as follows: $\dot{m}$ accretion ratio in units of M$_{\odot}$ yr$^{-1}$; $M$ black hole mass in M$_{\odot}$; $l_{Edd}$ accretion disk Eddington ratio; $R_{DB}$ dust belt radius in parsec; $\theta_{DB}$ dust belt covering angle; $R_{T}$ obscuring torus radius in parsec; $\theta_{T}$ obscuring torus covering angle; $f$ intrinsic corona scattered component (as fraction of the total flux); $N_H^{gal}$ hydrogen column density from host galactic gas in cm$^{-2}$; $N_H$ hydrogen column density in cm$^{-2}$; $\gamma$ X-ray corona power-law index; $G$ normalization coefficient for the galaxy emission template; and $A_B$ extinction in magnitudes at wavelength $\lambda_B=0.44$ $\mu$m.}\label{obs_fit_res}
	\end{table*}
	
	For all sources with an inferred Eddington ratio $l\sim10^{-2}$ we tried to apply an ADAF model instead of a standard accretion disk, but we chose not to consider it because to reach the high X-ray luminosity of our sources (up to $\nu L_X \sim 10^{44}$ erg s$^{-1}$) a high accretion rate would be needed, which would flatten the shape of the resulting SED. Thus the result would be inconsistent with the data, especially in the infrared band, even after accounting for absorption and the host galaxy contribution.
	
	Even if we cannot extract precise physical parameters from our work, we can make some considerations concerning the average behavior of the sources.
	We found that the majority of the 31 spectra are consistent with a black hole mass between $10^8$ and $10^9$ solar masses and we obtained a mean unabsorbed luminosity for the accretion disk $L_{\mathit{disk}}\sim 10^{45.6}$ erg s$^{-1}$.
	Therefore, all sources have an inferred Eddington ratio of 0.01 to 0.3, with a mean $l_{\mathit{Edd}}=0.07$.
	
	The main feature characterizing obscured AGNs is the presence of heavy absorption both in optical and X-ray band.
	At soft-X-ray energies the main contributing factor is the photoelectric absorption.
	Out of 31 sources, 30 have data in the soft-X-ray band, which is important to study the amount of absorption, but for many of these sources we have just a few measurements in this energy band, not allowing for a good sampling of the luminosity behavior.
	However, we can use the data to constrain the hydrogen column density along the line of sight, since the absorption scales as $F(E)\propto e^{-\sigma(E)N_H}$, where $\sigma$ is the cross section of photoelectric absorption.
	We found that $N^{gal}_H\sim10^{20-21}$ cm$^{-2}$, assigned from archival data by the source position in the sky, can always properly describe the Galactic absorption for all sources with data around 1 keV.
	Regarding the torus contribution, we found that the SED of 20 out of 30 objects with data at $\nu\sim 10^{19}$ Hz are consistent with $N_H>10^{24}$ cm$^{-2}$.
	This ratio of two-thirds is higher than the intrinsic $\sim0.5$ expected in the BAT survey of non-blazar AGNs \citep[e.g., see][]{bat:xprop}.
	This is most likely as consequence of our source selection. We are studying objects that are more likely to be highly obscured since they lack optical lines observations.
	Even if the source of the obscuration at X-ray energies (gas) is different from the obscuring source at longer wavelengths (dust) and may occur on different physical size scales, there is usually good agreement between the X-ray signatures of absorption and the optical spectral types of AGNs if $N_ H = 10^{22}$ cm$^{-2}$ is taken as the threshold between X-ray absorbed and unabsorbed AGNs \citep[e.g., see][]{xopt:abs1,xopt:abs2}.
	We found no relevant differences in the $N_H$ distribution between spiral and elliptic galaxies, with the type assigned by visual recognition from archival optical images.
	
	For all 31 sources we found the spectral signature of the re-emission by an obscuring torus.
	From the peak frequency and luminosity we can derive  a mean dust temperature of $T_T \sim 300$ K and a mean covering factor of $q_T=\cos\theta_{T}=0.3$; most of the sources have $q_T<0.4$.
	Therefore, from equation \ref{eq_T_torus}, we obtain a mean radius of $R_T\sim 5$ parsec.
	
\begin{figure}[tb]
		\centering
		\includegraphics[width=\columnwidth]{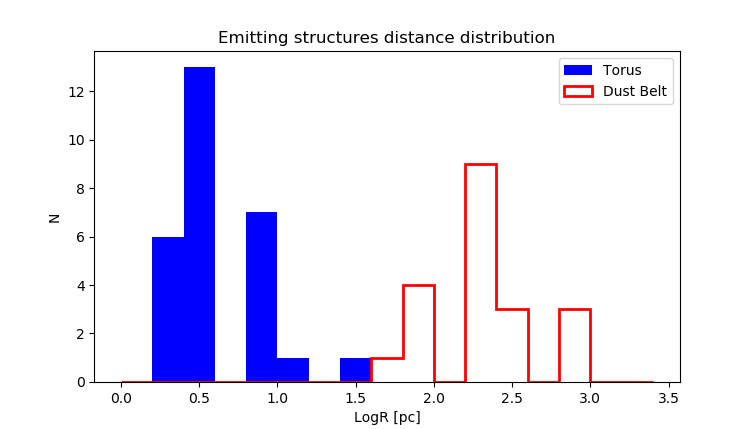}
		\caption{Distance distribution from the black hole of the IR and submillimeter emitting structures found in our work.
			The blue columns represent the radius of the classical obscuring torus in the AGN inner region; the red line indicates the radius of an emitting dust belt, of which we have found traces in 20 out of 31 sources.
			The distances are derived from equation \ref{eq_T_torus}.
		}\label{distR}
	\end{figure}

	To best describe the SED of all the 20 unknown AGN sources with data in the millimeter and submillimeter wavelength band, we had to consider a second emitting toroidal structure at a larger distance from the black hole.
	It is reasonable to interpret this component as a dust belt inside the galaxy rather than a proper dusty torus surrounding the black hole.
	The mean obtained value for the temperature distribution is  $T_{DB}\sim50$ K, the mean covering factor $q_{DB}=0.6,$ and the distance distribution shows a peak at $\sim150$ pc, where some sources reach $\sim$1 kpc.
	In \figurename~\ref{distR} we show the distribution of the emitting structures distance.
	In evaluating the absorption and consequent re-emission on this scale, we must take into account the contribution of stars. It is reasonable to think that a large fraction of the radiation heating the dust does not come from the AGN, but from nearby stars.
	Nonetheless, since the dust belt blackbody has a peak luminosity that is, on average, equal or larger than the host galaxy peak luminosity, we can comfortably assume that the stars alone cannot provide enough energy to explain the measured emission.
	Furthermore, a heavily obscured accretion disk, whose presence is highly suggested by the X-ray spectral emission, could easily provide the necessary energy to heat the dust emitting this spectral feature. 
	This is a clear clue that the host galaxy environment plays a key role in obscuring the AGN.
	
	Furthermore, we can make two considerations about the general slope of the SED in these 31 sources.
	First, all but five sources show a clear sign of absorption in the soft-X ray energy range.
	Those that do not are SWIFT J0317.1+1542, SWIFT J0329.4-2157, SWIFT J1643.4+0986, SWIFT J2004.6-1111, and SWIFT J0302.2+6853. Of these, only the last source\ shows no signs of re-emission in the submillimeter band, which in our model is interpreted as the emission from a dust belt.
	These sources need more in-depth studies, in particular obtaining optical spectra could be very useful.
	Also, regarding the low-energy emission, most of the sources with data at submillimeter wavelengths show an IR-to-optical luminosity ratio of about one, with a few notable exceptions.
	For example in SWIFT J0317.1+1542, SWIFT J1121.1-0529, and SWIFT J1300.5-0759, our model shows a far-IR peak that is nearly ten times in luminosity than the optical data.
	This is due to the luminosity measurements from AKARI and WISE being much higher than the one from, for example, DENIS or 2MASS.
	Furthermore, even in the WISE data we see a high luminosity difference between different energy bands.
	Consequently, to fit the data with the sum of two blackbody emissions, this forces our model to have a very marked peak in the far-infrared.
	
	With the data at our disposal, we assigned each of the selected sources to one of three classes a) Blazar if the obtained SED is well reproduced by a phenomenological blazar model; b) Obscured AGN when our obscured model is in good agreement with the data and the following spectral features are present in the SED: i) infrared luminosity similar to the optical emission; ii) an exponential cutoff in the soft-X band due to photoelectric absorption; and iii) a hard-X luminosity about the same intensity as in the optical band; or c) Obscured AGN candidate when the SED can be reproduced satisfactorily with our obscured model but at least one of the characterizing spectral features, listed in the obscured AGN category, is not present.
	These sources require a more in-depth spectroscopic study.
	Thank to this work we can classify the 36 new sources with redshift in the unknown AGN class in BAT 105 as follows: 5 are blazars, 16 are heavily obscured AGNs, and 15 are obscured AGNs candidates (see \tablename~\ref{classi}).
	
	\begin{table}[tb]
		\centering
		\begin{tabular}{lr}
			\toprule
			Source name & Classification \\
			\midrule
			SWIFT J0000.5+3251 & Obscured AGN \\
			SWIFT J0043.9-5009 & Obscured AGN \\
			SWIFT J0120.7-1444 & Obscured AGN \\
			SWIFT J0131.5-1007 & Blazar \\
			SWIFT J0144.8-2754 & Blazar \\
			SWIFT J0155.2-3048 & Obscured Candidate \\
			SWIFT J0201.0+0329 & Blazar \\
			SWIFT J0201.5+5032 & Obscured AGN \\
			SWIFT J0248.3+1202 & Obscured Candidate \\
			SWIFT J0302.2+6853 & Obscured Candidate \\
			SWIFT J0317.1+1542 & Obscured Candidate \\
			SWIFT J0327.9-5301 & Obscured AGN \\
			SWIFT J0329.4-2157 & Obscured Candidate \\
			SWIFT J0726.6-4634 & Obscured Candidate \\
			SWIFT J0749.8+3359 & Obscured AGN \\
			SWIFT J0943.1-4147 & Obscured AGN \\
			SWIFT J1048.6-3901 & Obscured Candidate \\
			SWIFT J1121.1-0529 & Obscured AGN \\
			SWIFT J1300.5-0759 & Obscured AGN \\
			SWIFT J1343.1+8038 & Obscured Candidate \\
			SWIFT J1534.5+6258 & Obscured Candidate \\
			SWIFT J1643.4+0986 & Obscured Candidate \\
			SWIFT J1649.3-1739 & Obscured AGN \\
			SWIFT J1651.2-0144 & Obscured AGN \\
			SWIFT J1700.8+3602 & Obscured Candidate \\
			SWIFT J1725.7-4517 & Obscured AGN \\
			SWIFT J1735.7+2045 & Obscured AGN \\
			SWIFT J1756.3+5237 & Blazar \\
			SWIFT J1810.0-6554 & Blazar \\
			SWIFT J1832.2+3146 & Obscured Candidate \\
			SWIFT J2004.6-1111 & Obscured Candidate \\
			SWIFT J2043.8-0958 & Obscured Candidate \\
			SWIFT J2112.4-4249 & Obscured AGN \\
			SWIFT J2117.7-0208 & Obscured Candidate \\
			SWIFT J2147.6-5360 & Obscured AGN \\
			SWIFT J2306.3-5147 & Obscured AGN \\
			\bottomrule
		\end{tabular}
		\caption{BAT 105 unknown AGNs classification as results from our work.}\label{classi}
	\end{table}
	
	\section{Conclusions}
	
	This is one of the first studies aimed at classifying hard X-ray extra-galactic sources on the basis of their overall emission properties and spectrum.
	Through our work we found that a multiwavelength model proved to be efficient to quickly characterize obscured AGNs, which is a procedure that usually requires high spatial resolution and spectroscopic observations.
	Furthermore, it can give useful insight into the structure of the region surrounding the AGN.
	
	Our model is simple in its application to the data, nonetheless it takes into account a wide variety of physical processes responsible for the whole AGN electromagnetic spectrum.
	We recognize that this approach is affected by the lack of data and that good coverage of the whole spectrum is essential to properly describe a source.
	Nonetheless, when spectroscopic observations are not available or the SED is under-sampled, this is the only method that is efficient in characterizing heavily obscured AGNs.
	Our approach is limited, but makes the best use possible of the available data.
	
	A method that is capable of easily identifying obscured AGNs is extremely important in relation to our ability to understand the CXB and the accretion history in our Universe. This is especially important considering that, according to many current models, around two-thirds of all AGNs should be obscured in order to explain the CXB above 30 keV, which is a population that we do not fully understand yet.
	We expect that an approach such as ours will be very useful to quickly identify these types of sources.
	
	\section{Acknowledgements}
	This work made use of data supplied by the UK Swift Science Data Center at the University of Leicester.
	We would like to express appreciation to Claudio Ricci for his help in the spectral analysis, to Michael Koss for his valuable suggestions during the writing of this paper, and to our anonymous referee for his/her constructive criticism and useful comments. 
	
	\bibliography{giuliani.bib}

\begin{thebibliography}{32}
\expandafter\ifx\csname natexlab\endcsname\relax\def\natexlab#1{#1}\fi

\bibitem[{Antonucci(1993)}]{uni_antonucci}
Antonucci, R. 1993, Annu. Rev. Astron. Astrophys., 31, 473

\bibitem[{Barger {et~al.}(2005)Barger, Cowie, Mushotzky, Yang, Wang, Steffen,
  \& Capak}]{bat_vs_disk}
Barger, A., Cowie, L., Mushotzky, R., {et~al.} 2005, ApJ, 129, 578

\bibitem[{Barthelmy {et~al.}(2005)Barthelmy, Barbier, Cummings, Fenimore,
  Gehrels, Hullinger, Krimm, Markwardt, Palmer, Parsons, Sato, Suzuki,
  Takahashi, Tashiro, \& Tueller}]{bat_art}
Barthelmy, S., Barbier, L., Cummings, J., {et~al.} 2005, SSRv, 120, 143

\bibitem[{Baumgartner {et~al.}(2013)Baumgartner, Tueller, Markwardt, Skinner,
  Barthelmy, Mushotzky, Evans, \& Gehrels}]{bat_70}
Baumgartner, W., Tueller, J., Markwardt, C.~B., {et~al.} 2013, ApJS, 207, 12

\bibitem[{Boettcher {et~al.}(2013)Boettcher, Reimer, Sweeney, \&
  Prakash}]{lepto:hadro}
Boettcher, M., Reimer, A., Sweeney, K., \& Prakash, A. 2013, ApJ, 389, A54

\bibitem[{Condon(1992)}]{normal:radio}
Condon, J. 1992, ARA\&A, 30, 575

\bibitem[{Croswell(1996)}]{z_gal}
Croswell, K. 1996, The alchemy of the heavens (Oxford University Press)

\bibitem[{Delvecchio {et~al.}(2014)Delvecchio, Gruppioni, Pozzi, Berta,
  Zamorani, Cimatti, Lutz, Scott, Vignali, Cresci, Feltre, Cooray, Vaccari,
  Fritz, Le~Floc'h, Magnelli, Popesso, Oliver, Bock, Carollo, Contini,
  Le~Fevre, Lilly, Mainieri, Renzini, \& Scodeggio}]{obs_ir}
Delvecchio, I., Gruppioni, C., Pozzi, F., {et~al.} 2014, MNRAS, 3, 2736

\bibitem[{Fornalski(2018)}]{photoelectric}
Fornalski, K. 2018, J.Phys.Commun., 2

\bibitem[{Ghisellini {et~al.}(2017)Ghisellini, Righi, Costamante, \&
  Tavecchio}]{blazar_sequence}
Ghisellini, G., Righi, C., Costamante, L., \& Tavecchio, F. 2017, MNRAS, 469,
  255

\bibitem[{Mahadevan(1997)}]{adaf_scaling}
Mahadevan, R. 1997, ApJ, 477, 585

\bibitem[{Malizia {et~al.}(2012)Malizia, Bassani, Bazzano, Bird, Masetti,
  Panessa, Stephen, \& Ubertini}]{xopt:abs1}
Malizia, A., Bassani, L., Bazzano, A., {et~al.} 2012, MNRAS, 426, 1750

\bibitem[{Massaro {et~al.}(2016)Massaro, Crespo, D'Abrusco, Landoni, Masetti,
  Ricci, Milisavljevic, Paggi, \& Smith}]{class:gamma}
Massaro, F., Crespo, N.~A., D'Abrusco, R., {et~al.} 2016, AAS/High Energy
  Astrophysics Division, 15, 106.20

\bibitem[{Merloni {et~al.}(2014)Merloni, Bongiorno, Brusa, Iwasawa, Mainieri,
  Magnelli, Salvato, Berta, Cappelluti, Comastri, Fiore, Gilli, Koekemoer,
  Floc’h, Lusso, Lutz, Miyaji, Pozzi, Riguccini, Rosario, Silverman,
  Symeonidis, Treister, Vignali, \& Zamorani}]{xopt:abs2}
Merloni, A., Bongiorno, A., Brusa, M., {et~al.} 2014, MNRAS, 437, 3550

\bibitem[{Mushotzy {et~al.}(2008)Mushotzy, Winter, McIntosh, \&
  Tueller}]{bat_edd}
Mushotzy, R., Winter, L., McIntosh, D., \& Tueller, J. 2008, ApJ, 684, L65

\bibitem[{Narayan \& Yi(1994)}]{adaf_letter}
Narayan, R. \& Yi, I. 1994, ApJ, 428, L13

\bibitem[{Narayan \& Yi(1995)}]{adaf_article}
Narayan, R. \& Yi, I. 1995, ApJ, 452, 710

\bibitem[{Oh {et~al.}(2018)Oh, Koss, Markwardt, Schawinski, Baumgartner,
  Barthelmy, Cenko, Gehrels, Mushotzky, Petulante, Ricci, Lien, \&
  Trakhtenbrot}]{bat_survey}
Oh, K., Koss, M., Markwardt, C., {et~al.} 2018, ApJS, 235, 4

\bibitem[{Paiano {et~al.}(2017)Paiano, Franceschini, \&
  Stamerra}]{class:paiano}
Paiano, S., Franceschini, A., \& Stamerra, A. 2017, MNRAS, 468, 4902

\bibitem[{Parisi {et~al.}(2012)Parisi, Masetti, Jiménez-Bailón, Chavushyan,
  Palazzi, Landi, Malizia, Bassani, Bazzano, Bird, Charles, Galaz, Mason,
  McBride, Minniti, Morelli, Schiavone, \& Ubertini}]{class:parisi}
Parisi, P., Masetti, N., Jiménez-Bailón, E., {et~al.} 2012, A\&A, 545, A101

\bibitem[{Pei(1992)}]{abs_pei}
Pei, Y. 1992, ApJ, 395, 130

\bibitem[{Perola {et~al.}(2002)Perola, Matt, Cappi, Fiore, Guainazzi, Maraschi,
  Petrucci, \& Piro}]{cutoff:perola}
Perola, G., Matt, G., Cappi, M., {et~al.} 2002, A\&A, 389, 802

\bibitem[{Polletta {et~al.}(2007)Polletta, Tajer, Maraschi, Trinchieri,
  Lonsdale, Chiappetti, Andreon, Pierre, Fèvre, Zamorani, Maccagni, Garcet,
  Surdej, Franceschini, Alloin, Shupe, Surace, Fang, Rowan-Robinson, Smith, \&
  Tresse}]{swire_temp}
Polletta, M., Tajer, M., Maraschi, L., {et~al.} 2007, ApJ, 663, 81

\bibitem[{Ricci {et~al.}(2017)Ricci, Trakhtenbrot, Koss, Ueda, Vecchio,
  Treister, Schawinski, Paltani, Oh, \& Lamperti}]{bat:xprop}
Ricci, C., Trakhtenbrot, B., Koss, M.~J., {et~al.} 2017, ApJS, 233, 17

\bibitem[{Risaliti {et~al.}(1999)Risaliti, Maiolino, \& Salvati}]{obs_local}
Risaliti, G., Maiolino, R., \& Salvati, M. 1999, ApJ, 522, 157

\bibitem[{Rojas {et~al.}(2017)Rojas, Masetti, Minniti, Jiménez-Bailón,
  Chavushyan, Hau, McBride, Bassani, Bazzano, Bird, Galaz, Gavignaud, Landi,
  Malizia, Morelli, Palazzi, Patiño-Álvarez, Stephen, \&
  Ubertini}]{class:rojas}
Rojas, A., Masetti, N., Minniti, D., {et~al.} 2017, A\&A, 602, A124

\bibitem[{Schnorr-Muller {et~al.}(2016)Schnorr-Muller, Davies, Korista,
  Burtscher, Rosario, Storchi-Bergmann, Contursi, Genzel, Gracia-Carpio, Hicks,
  Janssen, Koss, Lin, Lutz, Maciejewski, Muller-Sanchez, de~Xivry, Riffel,
  Riffel, Schartmann, Sternberg, Sturm, Tacconi, Veilleux, \& Ulrich}]{sy:av}
Schnorr-Muller, A., Davies, R., Korista, K., {et~al.} 2016, MNRAS, 462, 3570

\bibitem[{Setti \& Woltjer(1989)}]{agn_xrb}
Setti, G. \& Woltjer, L. 1989, A\&A, 224, L21

\bibitem[{Shakura \& Sunyaev(1973)}]{ss_acc_disk}
Shakura, N. \& Sunyaev, R. 1973, A\&A, 24, 337

\bibitem[{Ueda {et~al.}(2014)Ueda, Akiyama, Hasinger, Miyaji, \&
  Watson}]{ueda_2014}
Ueda, Y., Akiyama, M., Hasinger, G., Miyaji, T., \& Watson, M. 2014, ApJ, 786,
  104

\bibitem[{Ueda {et~al.}(2007)Ueda, Eguchi, Terashima, Mushotzky, Tueller,
  Markwardt, Gehrels, Hashimoto, \& Potter}]{suzaku_obs}
Ueda, Y., Eguchi, S., Terashima, Y., {et~al.} 2007, ApJ, 664, L79

\bibitem[{Urry \& Padovani(1995)}]{uni_urry}
Urry, C. \& Padovani, P. 1995, PASP, 107, 803

\end{thebibliography}
	
\end{document}